\documentclass[onecolumn,pre,superscriptaddress,notitlepage]{revtex4-2}

\usepackage{amsmath, amssymb}
\usepackage[caption=false]{subfig}
\usepackage{graphicx}
\usepackage{textcomp}
\usepackage{float}
\usepackage{nicefrac}
\usepackage{color}
\usepackage{booktabs} 
\usepackage{chngcntr}

\newcommand{\cor}{{\rm{cor}}}
\newcommand{\one}{\mathbf{1}}
\DeclareMathOperator{\Tr}{Tr}
\newcommand{\idt}{\mathbb{I}}
\newcommand{\jdt}{\mathbb{J}}

\newcommand{\Din}{D_\mathrm{in}}
\newcommand{\Dout}{D_\mathrm{out}}
\newcommand{\din}{d_\mathrm{in}}
\newcommand{\dout}{d_\mathrm{out}}

\begin{document}
\title{The Enmity Paradox}

\author{Amir Ghasemian}
\email[]{amir.ghasemian@yale.edu}
\affiliation{Yale Institute for Network Science, Yale University, New Haven, CT 06511, USA}

\author{Nicholas A. Christakis}
\affiliation{Yale Institute for Network Science, Yale University, New Haven, CT 06511, USA}

\begin{abstract}
The ``friendship paradox'' of social networks states that, on average, ``your friends have more friends than you do.'' 
Here, we theoretically and empirically explore a related and overlooked paradox we refer to as the ``enmity paradox.'' We use empirical data from 24,687 people living in 176 villages in rural Honduras.
We show that, for a real negative undirected network (created by symmetrizing antagonistic interactions), the paradox exists as it does in the positive world. Specifically, a person’s enemies have more enemies, on average, than a person does. Furthermore, in a mixed world of positive and negative ties, we study the conditions for the existence of the paradox, both theoretically and empirically, finding that, for instance, a person's friends typically have more enemies than a person does. 
We also confirm the ``generalized'' enmity paradox for non-topological attributes in real data, analogous to the generalized friendship paradox (e.g., the claim that a person’s enemies are richer, on average, than a person is). As a consequence, the naturally occurring variance in the degree distribution of both friendship and antagonism in social networks can skew people’s perceptions of the social world.
\end{abstract}

\maketitle
The empirical observation that a person's friends have, on average, more friends than they do is called the ``friendship paradox''~\cite{feld1991your}.
The friendship paradox can be explained by the fact that, in computing the average degree of individuals' friends, high-degree individuals are counted more than individuals with low degree. It is a kind of sampling bias. Computing the average degree based on a person’s network neighbors' perspective is biased towards a higher mean value than computing it from a person’s own perspective. 

The friendship paradox has been generalized to non-topological characteristics in social networks, such as happiness and wealth (people’s friends are happier and richer than they are, on average)~\cite{eom2014generalized}. The positive correlation between network degree and various characteristics is the origin of this generalization. The friendship paradox has also been used in social network polling and estimating power-law degree distributions~\cite{nettasinghe2019friendship}, and it has been studied in relation to some other topological properties of networks, such as betweenness, closeness, eigenvector, and Katz centrality~\cite{grund2014your, higham2019centrality}, as well as extensions to directed networks~\cite{higham2019centrality, alipourfard2020friendship}.
Moreover, the friendship paradox can be considered a special case of human social sensing~\cite{galesic2021human}.
Furthermore, individuals with more social connections are more indicative of early trends than the average member of the population for many societal phenomena, from the spread of disease~\cite{christakis2010social} to the spread of information~\cite{garcia2014using}. It is even possible to exploit the friendship paradox to develop effective strategies to intervene in networks~\cite{kim2015social, shakya2017exploiting, kumar2021interventions, alexander2022algorithms}.

Researchers have also recently examined what topological features influence the strength of friendship paradox. 
For the local formulation of the friendship paradox, for three classes of network models (the Poisson random graph, the configuration model, and a model of a random degree-assortative network), it has been shown that networks with more heterogeneous degree distributions and with negative assortativity tend to have the strongest friendship paradox~\cite{cantwell2021friendship}.
Two formulations of friendship paradox have been examined based on the global and local structure of a network~\cite{feld1991your}. By exploiting a topological property called ``inversity,'' i.e., the Pearson correlation between a node's degree and the inverse degree of its neighbors, it is possible to evaluate the relationship between these two formulations and the strength of the friendship paradox~\cite{kumar2021interventions,shirado2019resource}. Similar discussions for the generalized friendship paradox in directed networks are possible~\cite{alipourfard2020friendship}. 

To date, the friendship paradox has been investigated primarily from the perspective of positive networks, and little attention has been paid to questions regarding a world in which solely negative ties exist or a world in which both positive and negative ties exist.
As a result, it is unclear whether we can still observe these paradoxes in the face of antagonistic ties. Does a negative world manifest the same paradoxes? If so, what mechanisms would cause such paradoxes to occur in a negative world?

Thus, here, we explore the ``enmity paradox,'' by which we mean, most generally, that the mean number of enemies a person has is \textit{lower} than the mean number of enemies their enemies have. Furthermore, we examine whether this paradox is observed in a mixed world of both positive and negative ties by comparing a person's average number of friends with their enemies' average number of friends, and by comparing a person's average number of enemies with their friends' average number of enemies. 
 
\section*{The enmity paradox}
\label{sec:enm_paradox}
It is a fact of mathematics that, in a population with variance in the degree distribution (and subject to certain provisos), the friendship paradox exists. But its extent is very much a result of underlying social factors (such as variation across people in the number of friends they want, or whether popular people are preferred as friends). Some aspects of this phenomenon can also be self-reinforcing as a consequence of positive feedback that results from the biased perception~\cite{jackson2019friendship}. 

However, even though there is an intrinsic mathematical symmetry between positive and negative network objects (as instantiated by adjacency matrices), the empirical existence of an enmity paradox is not assured. We mathematically clarify and then empirically investigate the existence, origins, and manifestations of the enmity paradox. 

First, we derive the equations for the enmity paradox, which are similar to the equations derived for the friendship paradox~\cite{feld1991your}. Based on our comparison of the mean number of enemies of individuals with the mean number of enemies' enemies in the negative world, we theoretically assert that the enmity paradox should indeed arise in a similar manner to the friendship paradox.

Consider a simple signed network $G=(V, E_{(+)}, E_{(-)})$, where $V$ is the vertex set with $n$ nodes, and $E_{(+)}$ and $E_{(-)}$ are the set of positive and negative edges revealing two not-necessarily-dependent worlds. The adjacency matrices corresponding to the positive and negative interactions are denoted as $A_{(+)}$ and $A_{(-)}$, respectively. 
In order to create undirected networks, we either (1) remove unreciprocated edges from the network, or (2) symmetrize the network by removing the edges' direction.
For example, the node $j$ is considered a neighbor of the node $i$ if, in the former, both edges $(i,j)$ and $(j,i)$ are included in the set $E$, while, in the latter, if at least one of these edges exist in the edge set. An edge is referred to as a ``friend'' if it belongs to $E_{(+)}$, and as an ``enemy'' if it belongs to $E_{(-)}$. In addition, we also refer to a node $j$ as an $\ell$-hop neighbor of node $i$ if there is a walk with length $\ell$ between $i$ and $j$; a walk can include repeated nodes. 
Here, we focus solely on the enmity and friendship paradoxes for undirected networks. The enmity paradox for directed networks are provided in the Appendix D. 

There are two types of degrees for each node $i$ in these two parallel worlds, $k_{(+),i} = \sum_j A_{(+),ij}$ corresponding to positive network, and $k_{(-),i}= \sum_j A_{(-),ij}$ corresponding to negative network. The probability of a node with degree $k_{(+)}$ ($k_{(-)}$) is denoted as $p_{k_{(+)}}^{(0)}$ ($p_{k_{(-)}}^{(0)}$) or, more simply, $p_{k_{(+)}}$ ($p_{k_{(-)}}$); and the probability of a node's friend with positive degree $k_{(+)}$ and negative degree $k_{(-)}$ is denoted as $p_{k_{(+)}}^{(1)}$ and $p_{k_{(-)}}^{(1)}$, respectively. 
Similarly, the probability of a node's enemy with positive degree $k_{(+)}$ and negative degree $k_{(-)}$ is denoted as $q_{k_{(+)}}^{(1)}$ and $q_{k_{(-)}}^{(1)}$, respectively.  
For simplicity, whenever it is clear from context, we denote the degree, the degree distribution, and the degree distributions of neighboring nodes corresponding to the friendship and enmity networks using $k$, $p_{k}$, and $p_{k}^{(1)}$ and $q_{k}^{(1)}$. 

To establish the enmity paradox, we need to compare the average negative degree of a random node with the average negative degree of a random enemy of a random node. The average negative degree of a random node can be written as $\langle k_{(-)} \rangle_{p_{k_{(-)}}} = \sum_{k_{(-)}} k_{(-)} p_{k_{(-)}}$.
Going along a random negative edge to one of its enemies leads to a node with negative degree $k_{(-)}$ with probability $q_{k_{(-)}}^{(1)}$ proportional to $k_{(-)} p_{k_{(-)}}$, i.e., $q_{k_{(-)}}^{(1)} = \nicefrac{k_{(-)} p_{k_{(-)}}}{\langle k_{(-)} \rangle_{p_{k_{(-)}}}}$. 
Therefore, the average degree of a random enemy can be written as $\langle k_{(-)} \rangle_{q_{k_{(-)}}^{(1)}} = \sum_{k_{(-)}} k_{(-)} q_{k_{(-)}}^{(1)} =  \sum_{k_{(-)}} \nicefrac{{k_{(-)}}^2 p_{k_{(-)}}}{\langle k_{(-)} \rangle_{p_{k_{(-)}}}}$ for enmity networks. Using Jenson’s inequality, it can be shown that $\langle k_{(-)} \rangle_{q_{k_{(-)}}^{(1)}} \geq \langle k_{(-)} \rangle_{p_{k_{(-)}}}$. This inequality can be called the ``enmity paradox'' for the negative world, and it follows from the same mathematical facts as the friendship paradox in the positive world.

Social networks that simultaneously involve both positive and negative ties are theoretically more complicated. That is, 
analytically, we have shown that people have fewer enemies on average than their enemies; however, the result of mixing positive and negative worlds is not obvious. Do people have more or fewer enemies than their friends, or do they have more or fewer friends than their enemies?

Considering a dependency between positive and negative degrees with the correlation denoted by $\rho_{k_{(+)},k_{(-)}} = \nicefrac{\left(E\left[{k_{(+)}k_{(-)}}\right] - E{k_{(+)}}E{k_{(-)}}\right)}{\sigma_{k_{(+)}}\sigma_{k_{(-)}}}$, we have the joint probability of degrees $k_{(+)}$ and $k_{(-)}$ as $p_{k_{(+)},k_{(-)}} = f(\sigma_{k_{(+)},k_{(-)}})$. For this scenario, the average degree of a person's enemies would be denoted by $\langle k_{(-)}\rangle_{p_{k_{(-)}}}$; the average degree of a person's friends would be denoted by $\langle k_{(+)}\rangle_{p_{k_{(+)}}}$; the average degree of a person's friends' enemies would be denoted by $\langle k_{(-)} \rangle_{p_{k_{(+)},k_{(-)}}^{(1)}}$; and the average degree of a person's enemies' friends would be denoted by $\langle k_{(+)} \rangle_{q_{k_{(+)},k_{(-)}}^{(1)}}$, where the first two are driven before and the last two are as follows.
\begin{subequations}
\label{eq:ave_deg_neighbor}
\begin{align}
\label{eq:mix_worlds_ave_enm_deg_fr}
\langle {k_{(-)}} \rangle_{p_{k_{(+)},k_{(-)}}^{(1)}} =& \sum_{k_{(+)}, k_{(-)}} \nicefrac{k_{(-)}k_{(+)}p_{k_{(+)},k_{(-)}}}{\langle k_{(+)}\rangle}_{p_{k_{(+)}}}\, \\
\label{eq:mix_worlds_ave_fr_deg_enm}
\langle {k_{(+)}} \rangle_{q_{k_{(+)},k_{(-)}}^{(1)}} =& \sum_{k_{(+)}, k_{(-)}} \nicefrac{k_{(+)}k_{(-)}p_{k_{(+)},k_{(-)}}}{\langle k_{(-)}\rangle}_{p_{k_{(-)}}}\, . 
\end{align}
\end{subequations}
Therefore, we have three different regimes as follows: (1) If we have independent positive and negative worlds, i.e., $p_{k_{(+)},k_{(-)}} = p_{k_{(+)}}p_{k_{(-)}}$, then $\langle k_{(-)} \rangle_{p_{k_{(+)},k_{(-)}}^{(1)}} = \langle k_{(-)}\rangle_{p_{k_{(-)}}}$ and there is no difference between the average number of a person's enemies and average number of enemies of a person's friends, and similarly $\langle k_{(+)} \rangle_{q_{k_{(+)},k_{(-)}}^{(1)}} = \langle k_{(+)}\rangle_{p_{k_{(+)}}}$, which means no difference between the average number of a person's friends and the average number of a person's enemies' friends. (2) If there is a positive correlation between the two worlds, i.e., $\rho_{k_{(+)},k_{(-)}} > 0$ (or when $\sigma_{k_{(+)},k_{(-)}} > 0$), then we have paradoxes in the mixed world in the same direction as the enmity and friendship paradoxes, as a person's friends have more enemies and a person's enemies have more friends compared to a person. (3) Finally, if there is a negative correlation between the two worlds, i.e., $\rho_{k_{(+)},k_{(-)}} < 0$ (or when $\sigma_{k_{(+)},k_{(-)}} < 0$), then we have paradoxes in the mixed world in the opposite direction with the enmity and friendship paradoxes, as a person's friends have a smaller number of enemies and a person's enemies have a smaller number of friends compared to the person. 

\subsection*{Computation and representation of enmity paradox for empirical data}
\label{sub_sec:comp_and_rep}
\begin{figure}[t!]
\begin{center}
  \includegraphics[width=0.75\columnwidth]{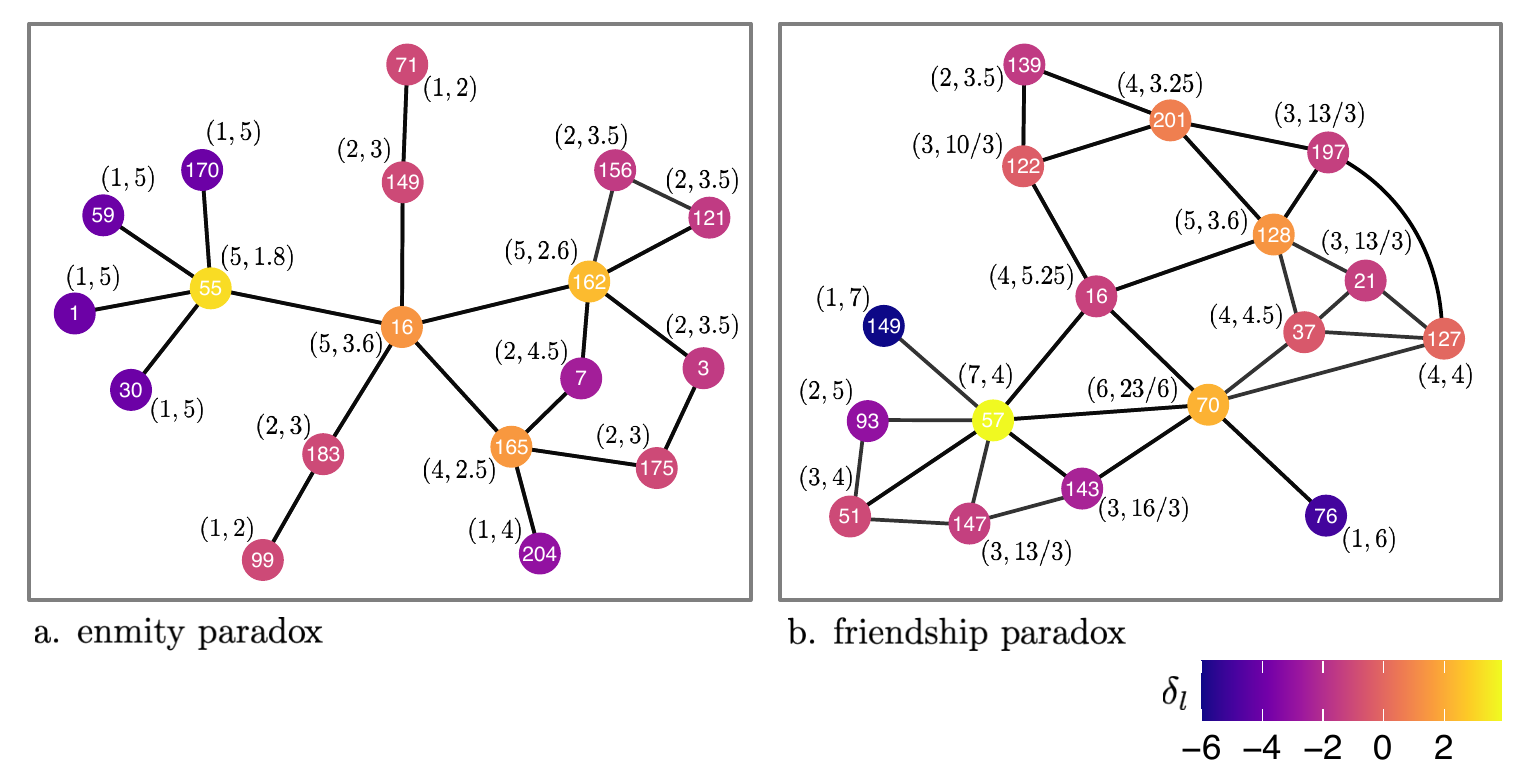}
  \end{center}
  \vspace{-8mm}
\caption{Global versus local enmity and friendship paradoxes. The networks are induced graphs for vertices at distance 2 from one vertex in one village in Honduras. On the left is the enmity network constructed from enmity interactions, and on the right is the friendship network constructed from friendship interactions. Darker blue represents nodes with maximum paradox, while red represents nodes with balance, and yellow represents nodes with the opposite circumstance. The $(a, b)$ value for each node represents the (mean degree, mean neighbors' degree) and the difference, i.e., $a-b$, represents the individual local paradox. It is the mean of all individual local paradoxes that constitutes the ``local paradox.'' In the case of the enmity paradox, the local paradox is equal to $-1.25$, while in the case of the friendship paradox, it is equal to $-1.04$. The ``global paradox'' is the difference between the mean degree of a random node and the mean degree of a random endpoint on a random edge. Accordingly, the global enmity paradox is equal to $-0.93$ and the global friendship paradox is equal to $-0.69$ here.}
\label{fig:global_local_enm_fr_paradox}
\end{figure}

Investigating the enmity and friendship paradoxes in practice requires writing the equations using the data as we are given neither the generative probability of the ties nor the joint probability of negative and positive ties.
The friendship paradox has previously been formulated in two different variations which relate to global bias and local bias~\cite{feld1991your, alipourfard2020friendship}. In the global formulation, we compare the average degree of a random node $i$ with the average degree of a random neighbor of a random node $j$---or a random end of a random edge. In the local formulation, we compare the difference between the average of a random node's degree and the average of that node's neighbors' degrees locally (Fig.~\ref{fig:global_local_enm_fr_paradox}). From now on, by ``paradox strength,'' we mean the magnitude of the global difference $\delta_g$ or of local difference $\delta_l$.

The paradox in the global variation is the result of the oversampling of high-degree nodes. In the local variation, the paradox can be intensified locally if there is a positive correlation between a node's degree and the inverse degree of its neighbors, or it can be attenuated if there is a negative correlation between these variables. 
This correlation measure is known as ``inversity,'' and although it is related to degree assortativity, it is not the same~\cite{kumar2021interventions}; the relevance of such metrics has also been previously explored empirically~\cite{shirado2019resource}.

The enmity paradox can similarly be defined for local and global formulations.
To illustrate the differences between these two mathematical formulations of the enmity and friendship paradoxes, we first redo the empirical computations for the enmity and friendship paradoxes using empirical data, followed by a matrix representation of the enmity and friendship paradoxes. 

The global formulation of the enmity paradox is the comparison between the average degree of a random node $i$, i.e., $\sum_i \nicefrac{k_i}{n}$ and the average degree of its neighbors' degrees, i.e., $\nicefrac{\sum_i k_i^2}{\sum_i k_i}$, where the difference can be written using matrix formulation as Eq.~\ref{eq:dif_global},
\begin{align}
\label{eq:dif_global}
\delta_{g,-w}(-) =& \frac{(\one^TA_{(-)}\one)^2 - \one^TA_{(-)}^2\one\cdot\one^T\one}{\one^TA_{(-)}\one \cdot\one^T\one} \, ,
\end{align}
where $\one$ is a vector of ones with length $n$ ($\one^T\one=n$). In this notation, the $-w$ and $+w$ indicate the type of one's neighbor, as one's enemy and friend, respectively. The $(-)$ and $(+)$ denote the type of comparison as enemies or friends.
This global quantity is a non-positive due to the Cauchy–Schwarz inequality ($|\langle \mathbf{u},\mathbf{v} \rangle|^2\leq \langle \mathbf{u}, \mathbf{u} \rangle \langle \mathbf{v}, \mathbf{v} \rangle$) with $\mathbf{u} = A\one$ and $\mathbf{v} = \one$, where the equality happens when $\mathbf{u} = k \mathbf{v} $, i.e., when we have a $k$-regular network. 
As a result, degree heterogeneity is directly related to the enmity ``paradox strength,'' since the numerator of the right-hand side in Eq.~\ref{eq:dif_global} is the negative variance of the negative degree distribution. 
These results are similar to the results for the friendship paradox.

For the local formulation, we have the difference between each individual's degree and the average degree of its neighbors, which is reminiscent of a Laplacian matrix acting on the degree vector $d$, where $i$-th outcome of this operation can be written as $\Delta_i = k_i - \nicefrac{\sum_j A_{ij} k_j}{k_i}$; and, by averaging out these outcomes, we have the local formulation of enmity paradox as Eq.~\ref{eq:dif_local},
\begin{align}
\label{eq:dif_local}
\delta_{l,-w}(-) =&  
\frac{\one^TA_{(-)}\one - \one^TD_{(-)}^{-1}A_{(-)}D_{(-)}\one}{\one^T\one} \nonumber\\
=&\frac{\one^TA_{(-)}\one - \one^TD_{(-)}A_{(-)}D_{(-)}^{-1}\one}{\one^T\one}\, ,
\end{align}
where $D_{(-)}$ is the diagonal matrix of negative degrees. Eq.~\ref{eq:dif_local} can also be written
as $\delta_{l,-w}(-) = \nicefrac{\one^T L_{(-)} k_{(-)}}{\one^T\one}$, where, $L_{(-)} = \idt - D_{(-)}^{-1} A_{(-)}$ is the Laplacian matrix of the negative world acting on the negative degree vector, $k_{(-)} = D_{(-)}\one$. Our derivation relies on the fact that $A$ is symmetric since we only consider undirected networks here (the formulations for directed networks are in the Appendix D). We can also demonstrate that this quantity is always non-positive. Since, for any two matrices $A$ and $B$, we can write $\Tr(AB) = \Tr(BA)$, thus Eq.~\ref{eq:dif_local} can be written as follows,
\begin{align}
\label{eq:dif_local_proof_npos}
\delta_{l,-w}(-) 
=& \frac{\Tr{\left[A_{(-)}(2\jdt - D_{(-)}\jdt D_{(-)}^{-1} - D_{(-)}^{-1}\jdt D_{(-)})\right]}}{2\one^T\one}\, ,
\end{align}
where, $\jdt$ is the matrix of all ones, and, due to the non-positive entries of the matrix $2\jdt - D_{(-)}\jdt D_{(-)}^{-1} - D_{(-)}^{-1}\jdt D_{(-)}$ for each pair $(i,j)$ (see Ref.~\cite{kumar2021interventions}), the resultant trace should also be non-positive.

The difference between $\delta_{g,-w}(-)$ and $\delta_{l,-w}(-)$ can be written as Eq.~\ref{eq:delta_local_global}, where 
$\sigma_{D,(-)}^2$ is the variance of the negative degree of an endpoint of a random edge; $\sigma_{ID,(-)}^2$ is the variance of the inverse negative degree of an endpoint of a random edge; and $\overline{ k_{(-)}}$ is the average negative degree. This quantity is negative ($\delta_{g,-w}(-) < \delta_{l,-w}(-)$) if the correlation between the degree of one endpoint $i$ and the inverse degree of another endpoint $j$ on a random edge $(i,j)\in E_{(-)}$, i.e., $\rho_{(-)} = \cor(k_{(-),i},\nicefrac{1}{k_{(-),j}}|(i,j)\in E_{(-)})$ is negative; it is positive ($\delta_{g,-w}(-) > \delta_{l,-w}(-)$) if the correlation is positive; and it is zero if the correlation is zero~\cite{kumar2021interventions}.  
\begin{align}
\label{eq:delta_local_global}
\delta_{g,-w}(-) - \delta_{l,-w}(-)
=& \frac{\one^TD_{(-)}^{-1}A_{(-)}D_{(-)}\one}{\one^T\one} - \frac{\one^TA_{(-)}^2\one}{\one^TA_{(-)}\one} \nonumber\\
=&\rho_{(-)}\overline{ k_{(-)}}\sigma_{D,(-)}\sigma_{ID,(-)}\, 
\end{align}

For a mixed world of positive and negative edges, the equations can be written as follows. The global formulation of the difference between number of friends of a random node with the number of friends of enemies of a random node can be formulated as Eq.~\ref{eq:dif_fr_mixed_global}. And the global formulation of the difference between the number of enemies of a random node with the number of enemies of friends of a random node can be formulated as Eq.~\ref{eq:dif_en_mixed_global}.
\begin{subequations}
\label{eq:dif_mixed_global}
\begin{align}
\label{eq:dif_fr_mixed_global}
\delta_{g,-w}(+) =& \frac{\one^TA_{(+)}\one\cdot\one^TA_{(-)}\one - \one^TA_{(+)}A_{(-)}\one\cdot\one^T\one}{\one^TA_{(-)}\one \cdot\one^T\one}\, , \\
\label{eq:dif_en_mixed_global}
\delta_{g,+w}(-) =& \frac{\one^TA_{(+)}\one\cdot\one^TA_{(-)}\one - \one^TA_{(+)}A_{(-)}\one\cdot\one^T\one}{\one^TA_{(+)}\one \cdot\one^T\one} \, .
\end{align}
\end{subequations}
The numerator of these equations can be reduced to $\sum_i k_{(-),i} \sum_i k_{(+),i} - n\sum_i k_{(-),i}k_{(+),i}$ which is negative when the correlation between the positive and negative degrees is positive.

For the local formulation, these equations reduce to Eq.~\ref{eq:dif_mixed_local}.
\begin{subequations}
\label{eq:dif_mixed_local}
\begin{align}
\label{eq:dif_fr_mixed_local}
\delta_{l,-w}(+) =& \frac{\one^T A_{(+)}\one - \one^T D_{(-)}^{-1} A_{(-)} D_{(+)}\one}{\one^T\one} \, , \\
\label{eq:dif_en_mixed_local}
\delta_{l,+w}(-) =& \frac{\one^T A_{(-)}\one - \one^T D_{(+)}^{-1} A_{(+)} D_{(-)}\one}{\one^T\one} \, . 
\end{align}
\end{subequations}
Eq.~\ref{eq:dif_fr_mixed_local} can be also written as the operation of the Laplacian matrix of the negative world on the positive degree vector $k_{(+)}$, i.e., $\delta_l = \one^T L_{(-)}k_{(+)}$, where, $L_{(-)} = \idt - D_{(-)}^{-1} A_{(-)}$. Similar statements can be shown regarding the relationships between $\delta_l$ and $\delta_g$ in a mixed world. The difference between $\delta_{g,-w}(+)$ (\ref{eq:dif_fr_mixed_global}) / $\delta_{g,+w}(-)$ (\ref{eq:dif_en_mixed_global}) and $\delta_{l,-w}(+)$ (\ref{eq:dif_fr_mixed_local}) / $\delta_{l,+w}(-)$ (\ref{eq:dif_en_mixed_local}) can be written by introducing the ``generailized inversity'' measure in the mixed world, i.e., the correlation between the positive degree / negative degree of one endpoint $i$ and the inverse negative/positive
degree of another endpoint $j$ on a random negative/positive edge $(i,j)\in E_{(-)}/E_{(+)}$ (see Eqs. [S32] and [S35] in Appendix G). 

\section*{Generalized enmity paradox}
\label{sec:gen_enm_paradox}
In this section, we use a similar representation to the one used in the previous section to explore when the enmity paradox can be generalized for non-topological attributes such as happiness, wealth, and health (e.g., on average, a person’s friends are richer or happier than they are, as has previously been shown). Due to the similarity between the enmity and friendship paradox, we expect that the condition for generalization of the global definition is similar to the condition previously studied for the generalized friendship paradox~\cite{eom2014generalized}. Similarly, the condition for the relationship between the local and global formulation of the generalized enmity paradox should be similar to the condition of their relationships for the generalized friendship paradox~\cite{alipourfard2020friendship}. Given the vector of non-topological characteristic denoted by $x$, we define a diagonal matrix $D_x$ with diagonal entries $x_i, i\in\{1,...,n\}$. Thus, the generalized enmity paradox for the global definition of the difference $\delta_{g,-w}(x)$ can be written as Eq.~\ref{eq:dif_global_gen}.
\begin{align}
\label{eq:dif_global_gen}
\delta_{g,-w}(x) =& \frac{\one^TD_x\one\one^TA_{(-)}\one - \one^TA_{(-)}D_x\one\cdot\one^T\one}{\one^TA_{(-)}\one \cdot\one^T\one}  
\end{align}
The numerator can be reduced to $\Tr{\left[D_{(-)}\left(\jdt D_x\jdt - D_x\jdt^2\right)\right]}=\one^T D_{(-)}(J-n\idt)D_x\one = \sum_{i,j}k_{(-), i}x_j - n \sum_i k_{(-), i} \sum_i x_i$, which is negative only if there is a negative correlation between $k$ and $x$.

The local definition of enmity paradox can also be written as Eq.~\ref{eq:dif_local_gen}.
\begin{align}
\label{eq:dif_local_gen}
\delta_{l,-w}(x) =& 
\frac{\one^TD_x\one - \one^TD_{(-)}^{-1}A_{(-)}D_x\one}{\one^T\one}\nonumber \\
=&\frac{2\one^TD_x\one - \one^TD_{(-)}^{-1}A_{(-)}D_x\one - \one^TD_xA_{(-)}D_{(-)}^{-1}\one}{2\one^T\one} \, 
\end{align}
Using $\idt = D_{(-)}^{-1}D_{(-)}$ and $D_{(-)}\one = A_{(-)}\one$, the numerator can be reduced to $$\Tr{\left[A_{(-)}\left(2\jdt D_xD_{(-)}^{-1} - D_x \jdt D_{(-)}^{-1} - D_{(-)}^{-1}\jdt D_x \right)\right]}\, .$$ Here, there are a variety of possibilities; for example, if $x_i<k_i$, this quantity is always negative and all local differences are also negative, then this is merely an uninteresting sufficient condition. The relationship between $\delta_{g,-w}(x)$ and $\delta_{l,-w}(x)$ can be formalized using the edge-based correlation between $x_i$ on one endpoint of edge $(i,j)$ and the inverse degree of another endpoint of that edge, $\nicefrac{1}{k_{(-),j}}$, i.e., $\delta_{g,-w}(x)-\delta_{l,-w}(x) \propto \rho_{(x)}$, where 
$\rho_{(x)} = \cor(x_i,\nicefrac{1}{k_{(-),j}}|(i,j)\in E_{(-)})$~\cite{alipourfard2020friendship}. If the aforementioned edge-based correlation is positive, we have $\delta_{l,-w}(x)<\delta_{g,-w}(x)$; if it is negative, we have $\delta_{g,-w}(x)<\delta_{l,-w}(x)$; and if there is no correlation, then two measures are equivalent. Using this relationship we have four possibilities of (positive, positive), (positive, negative), (negative, positive), and (negative, negative) for ($\delta_{l,-w}(x)$, $\delta_{g,-w}(x)$) differences.

\section*{Higher order enmity paradox}
\label{sec:high_order_enm_paradox}
We can also consider the paradoxes for higher-order neighbors. For example, in the global formulation, if the average degree of a random neighbor of a random node is greater than the average degree of a random node, is it also the case that the average degree of a random $2$-hop neighbor of a random node is also larger---or for any random $\ell$-hop neighbor of a random node (an $\ell$-hop neighbor is a neighbor we arrive through a random walk with length $\ell$). 

In this case, it is more specific to ask if there is a relationship between the paradox strength and the order of the neighbors. Here, the equation for global formulation of $\delta_g$ for an order of $\ell$ can be written as Eq.~\ref{eq:dif_global_higher_order},
\begin{align}
\label{eq:dif_global_higher_order}
\delta_g^{(\ell)} =& \frac{\one^TA\one \cdot \one^TA^{\ell}\one  - \one^TA^{\ell+1}\one\cdot\one^T\one}{\one^TA^\ell \one \cdot\one^T\one}\, ,
\end{align}
where, for $\ell=2$ can be written as follows.
\begin{align}
\label{eq:dif_global_higher_order_for_ell_2}
\delta_g^{(\ell=2)} =& \frac{\one^TA\one \cdot \one^TD^{2}\one  - \one^TDAD\one\cdot\one^T\one}{\one^TA^2 \one \cdot\one^T\one}\nonumber\\
=&\frac{\Tr{\left[A\left(\jdt D^2 \jdt - n D\jdt D\right)\right]}}{\one^TA^2\one\cdot \one^T\one}
\end{align}

Accordingly, we can see different scenarios under different conditions. If the degree of nodes that are connected have a positive correlation, i.e., $\nicefrac{1}{n}\sum_{i\sim j}d_id_j>\nicefrac{1}{n^2}\sum_i d_i \sum_j d_j^2$, we still have similar paradoxes, i.e., the average degree of a neighbor of a neighbor of a random node is larger than the average degree of a random node. If we have the opposite inequality, we see a paradox in a counterintuitive sense in comparison with the friendship paradox. And if we have equality, we do not see any paradox. However, the paradox is always valid for $\ell$ odd since the numerator of Eq.~\ref{eq:dif_global_higher_order} is always negative for $\ell$ odd due to Theorem 4.2 in Ref.~\cite{higham2019centrality} that was originally proposed in Ref.~\cite{lagarias1984inequality}. The theorem says that, given positive integers $r$ and $s$ such that $r+s$ is even, we have $\one^TA^r\one \cdot \one^TA^{s}\one  \leq \one^TA^{r+s}\one\cdot\one^T\one$. 

For a local definition, the equation can be reduced to Eq.~\ref{eq:dif_local_higher_order},
\begin{align}
\label{eq:dif_local_higher_order}
\delta_l^{(\ell)} =&  
\frac{\one^TA\one - \one^TD^{-1}A^{\ell}D\one}{\one^T\one}\, ,
\end{align}
where it can be both positive or negative under different circumstances. Statements similar to the relationship of $\delta_l$ and $\delta_g$ can be derived for this purpose. 

Hence, theory suggests that the enmity and friendship paradoxes for higher orders may not always be true and might only hold under some circumstances (e.g., particular sorts of networks, or particular regimes such as related to whether geodesic backsteps are allowed); while we note these mathematical observations here, we leave empirical investigation of these details to future work.

\section*{Methods}
\label{sec:Methods}
Despite the mathematical similarity of the enmity paradox formulation and the friendship paradox, the empirical existence of the enmity paradox in real networks is not guaranteed given the differences between the negative and positive environments. 
Positivity, for example, is characterized by a large clustering coefficient, reciprocity, and homophily~\cite{harrigan2017avoidance, jackson2010social}, while negativity can even seem like mere noise~\cite{harrigan2017avoidance, feng2022testing}.
Here, we characterize some of these structural characteristics of social networks before investigating the enmity paradox in comparison between the positive and negative ties. These noteworthy differences can be partly explained via the inherent avoidance between the sender and receiver of negative ties, which leads to low information transfer~\cite{harrigan2017avoidance}; in other words, people are more likely to know who likes them (because the other persons are more likely to so declare) than they are to know who dislikes them (which is information that is more often kept private)~\cite{isakov2019structure}. 
 
 Several distinct measures might play an important role in explaining the friendship paradox. (1) We denote the clustering coefficient using $T_l$, which is the local adaptation of the transitivity measure (Appendix B), and transitivity as $T_g$. (2) The inversity measure ($H_i$) (which is not the same as degree assortativity, with which is negatively correlated) has a direct effect on the 
 local friendship paradox strength~\cite{kumar2021interventions}.
(3) The degree assortativity ($H_a$) is also important~\cite{cantwell2021friendship}. (4) A heterogeneity index that captures starlike graphs---similar to the inversity measure---also seems to play an important role for the severity of friendship paradox. 
For degree assortativity, we can use extant measures~\cite{newman2003mixing}; and for starlike strength, we use a measure that represents topological heterogeneity in complex networks and is maximal for star graphs ($H_{\ast}$)~\cite{estrada2010quantifying}. We also consider (5) the variance of the degree ($H_{\rm{var}}$), 
and (6) a novel heterogeneity measure ($H_{\rm{deg-div}}$) that reflects degree diversity and has a close relationship with entropy~\cite{jacob2017measure}. To characterize other differences between positive and negative ties, we can also compare four more topological properties, including the reciprocity, the clustering coefficient, the homophily between the positive and negative ties, and the normalized betweenness-centrality (the mathematical definitions for all these measures are provided in Appendix B).

\subsection*{Maximum paradox strength}
\label{sub_sec:MSP}
The severity of the paradoxes depends on network structure. The global paradox strength is proportionate to the variance of its degree sequence divided by its average degree.
To maximize the global paradox strength given a constant average degree, we would need to maximize the degree variance by rewiring the edges so that the degree values oscillate between extreme values $1$ and $k_{max}$.
Similarly, a greedy rewiring algorithm can increase the local paradox strength~\cite{kumar2021interventions}. A cross-rewiring of paired edges that connect the nodes with extreme degrees---substituting connections of smallest degree nodes with the largest degree nodes instead of connection of mediocre degree nodes---can increase the strength of the paradox, with star networks having the maximum value.
Therefore, the enmity and friendship paradoxes would appear to have the greatest strength in both global and local formulations when connections have a star shape. 

\section*{Results}
We use data from a sociocentric network study of 24,687 people aged 11 to 93 years (with a mean age of 32) in 176 geographically isolated villages in western Honduras~\cite{shakya2017exploiting}. We first construct 176 binary directed signed networks. We use three name generators to determine (potentially overlapping) positive ties (“Who do you spend your free time with?” “Who is your closest friend?” and “Who do you discuss personal matters with?”). And we use one name generator for negative ties (“Who are the people with whom you do not get along well?”). 
Here, we focus on the enmity paradox for undirected (symmetrized) networks (an example village is illustrated in Fig.~\ref{fig:vill_67}). Results for undirected (reciprocated) and directed networks are in the Appendices C and D, respectively.

\begin{figure}[t!]
\begin{center}
  \includegraphics[width=0.8\textwidth]{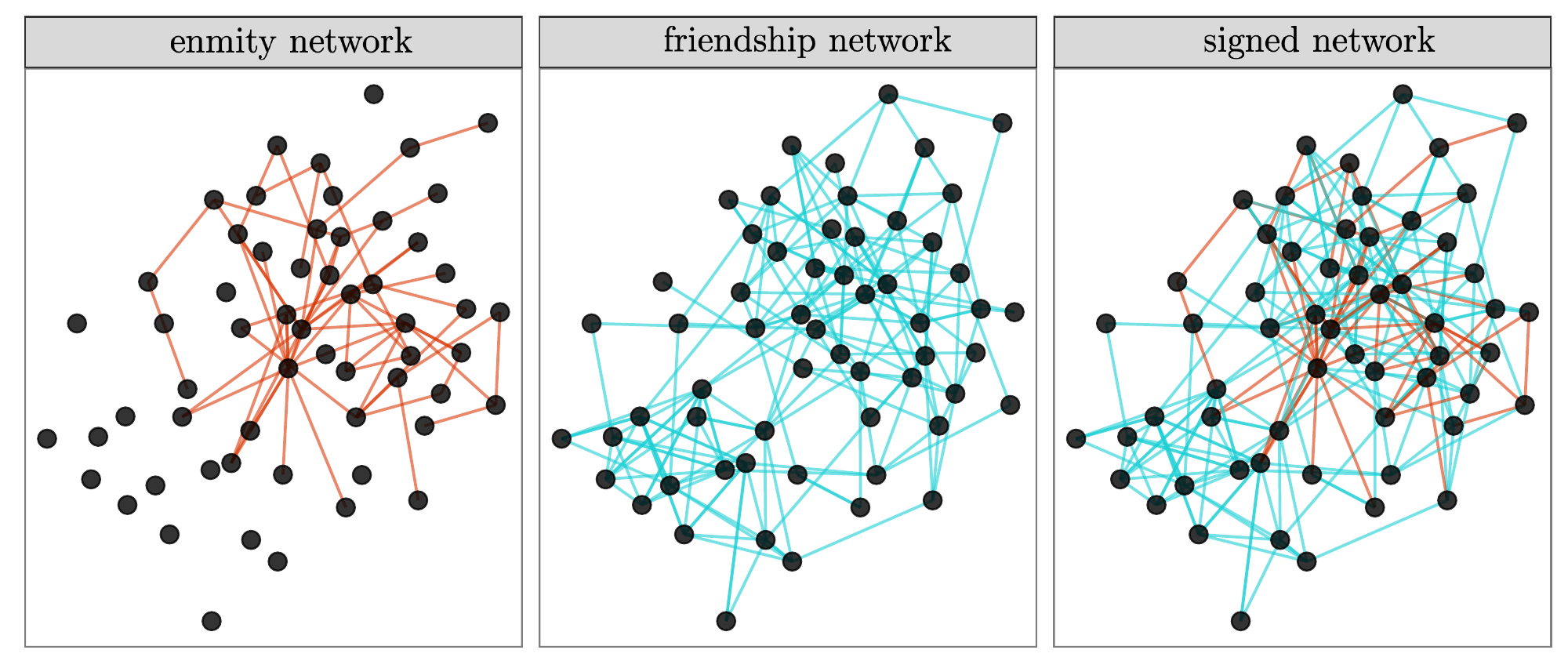}
    \end{center}
  \vspace{-8mm}
\caption{A visualization of the enmity, friendship, and signed networks within one Honduras village. The negative ties tend to be arranged in a star-shaped pattern, whereas the friendship ties contribute more towards larger-scale transitivity. Red indicates negative ties and blue indicates positive ties.}
\label{fig:vill_67}
\end{figure}

Across the whole dataset, nodes have an average of $6.89$ (${\rm{SD}} = 3.79$) friends, and friends have an average of $8.97$ (${\rm{SD}} = 4.4$) friends. In order to find the difference at the individual level, we must remove isolated nodes; and the average consequent difference is $-1.5$ (${\rm{SD}} = 3.57$).
For the enmity networks, a node has an average of $1.26$ (${\rm{SD}} = 1.70$) enemies, while an enemy has an average of $3.55$ (${\rm{SD}} = 2.58$) enemies. After removing the isolated nodes, the average difference between the number of enemies and the number of enemies of enemies is $-1.10$ (${\rm{SD}} = 2.57$). 
For the remaining analyses, we remove the isolated nodes from the constructed networks.

\begin{figure}[t!]
\begin{center}
  \includegraphics[width=0.65\textwidth]{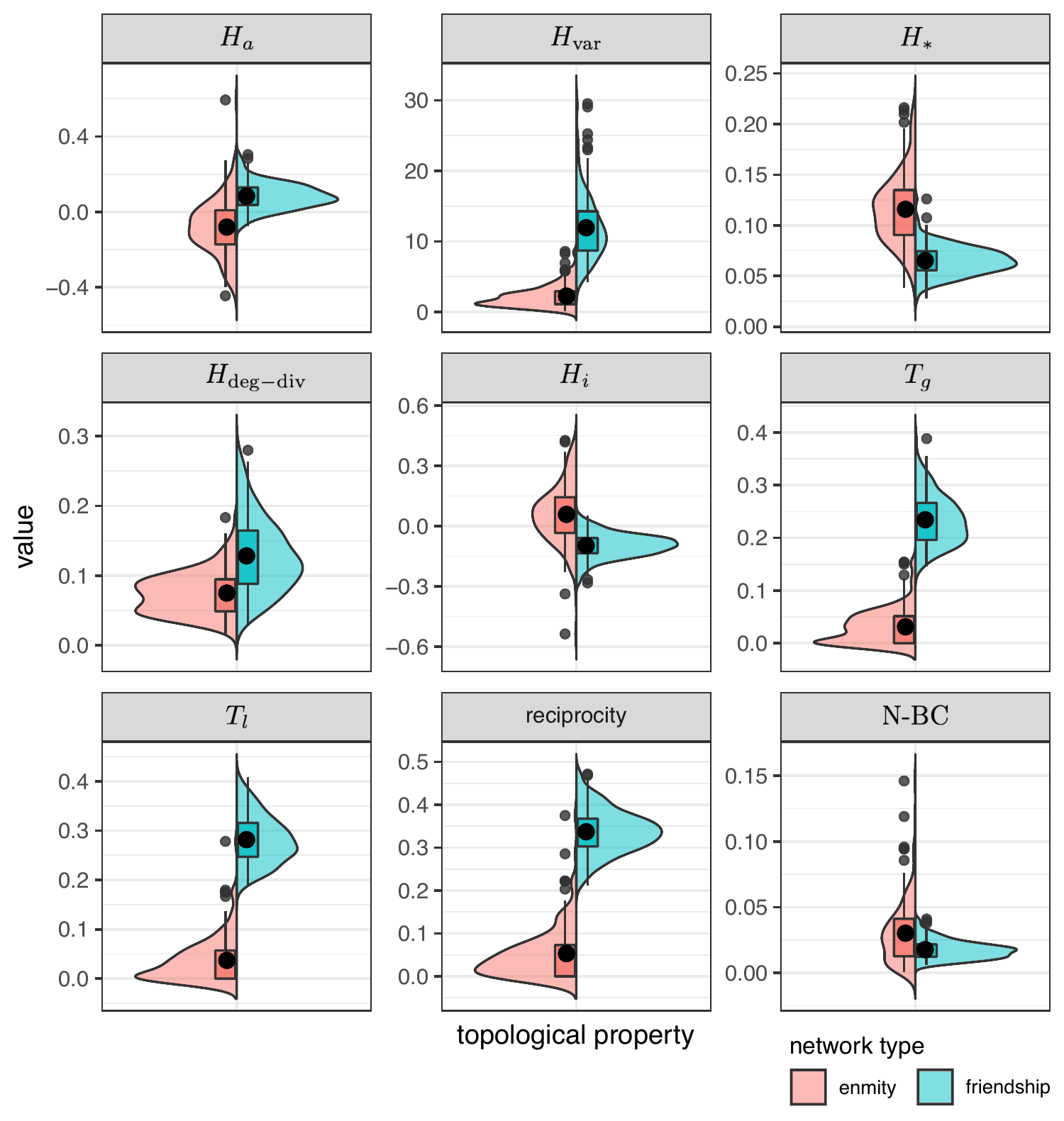}
    \end{center}
  \vspace{-8mm}
\caption{Comparison of topological properties between friendship and enmity networks. $H_a$, $H_{\rm{var}}$, $H_{\ast}$, $H_{\rm{deg-div}}$, $H_i$, $T_g$, $T_l$, and ${\rm{N-BC}}$ represents degree assortativity, variance, starlike strength, degree-diversity, inversity, transitivity, mean clustering coefficient, and normalized betweenness-centrality, respectively.}
\label{fig:stat_pos_neg}
\end{figure}

Core topological properties of friendship and enmity networks are presented in Fig.~\ref{fig:stat_pos_neg}. The reciprocity of positive edges is much larger than the reciprocity of negative edges, in part due to the relative scarcity of negative ties. Therefore, we expect the undirected enmity networks constructed from reciprocated edges to be much sparser compared to the friendship networks (Appendix C). The clustering coefficient of positive edges is much larger than the clustering coefficient of negative edges; that is, friendship networks have more triangles than enmity networks (Fig.~\ref{fig:vill_67}). The presence of starlike motifs is therefore expected to be more prevalent in negative environments (Fig.~\ref{fig:vill_67}). This is in turn aligned with the results for both $H_{\ast}$, and $H_i$. The variance of degree for positive networks is much larger than the negative networks. Thus, it appears that, among the different factors contributing to the strength of the enmity paradox, the starlike indices ($H_i$ and $H_{\ast}$) align with greater strength, and the lower $H_{\rm{var}}$ opposes it. Finally, we also plotted the difference between the normalized betweenness centrality (${\rm{N-BC}}$) of the positive and negative worlds to emphasize how the negative world has a more starlike shape as the maximum value of unnormalized ${\rm{BC}}$ is achieved by the central point in a star network~\cite{freeman2002centrality}. 

\begin{figure}[t!]
\centering
  \includegraphics[width=0.75\textwidth]{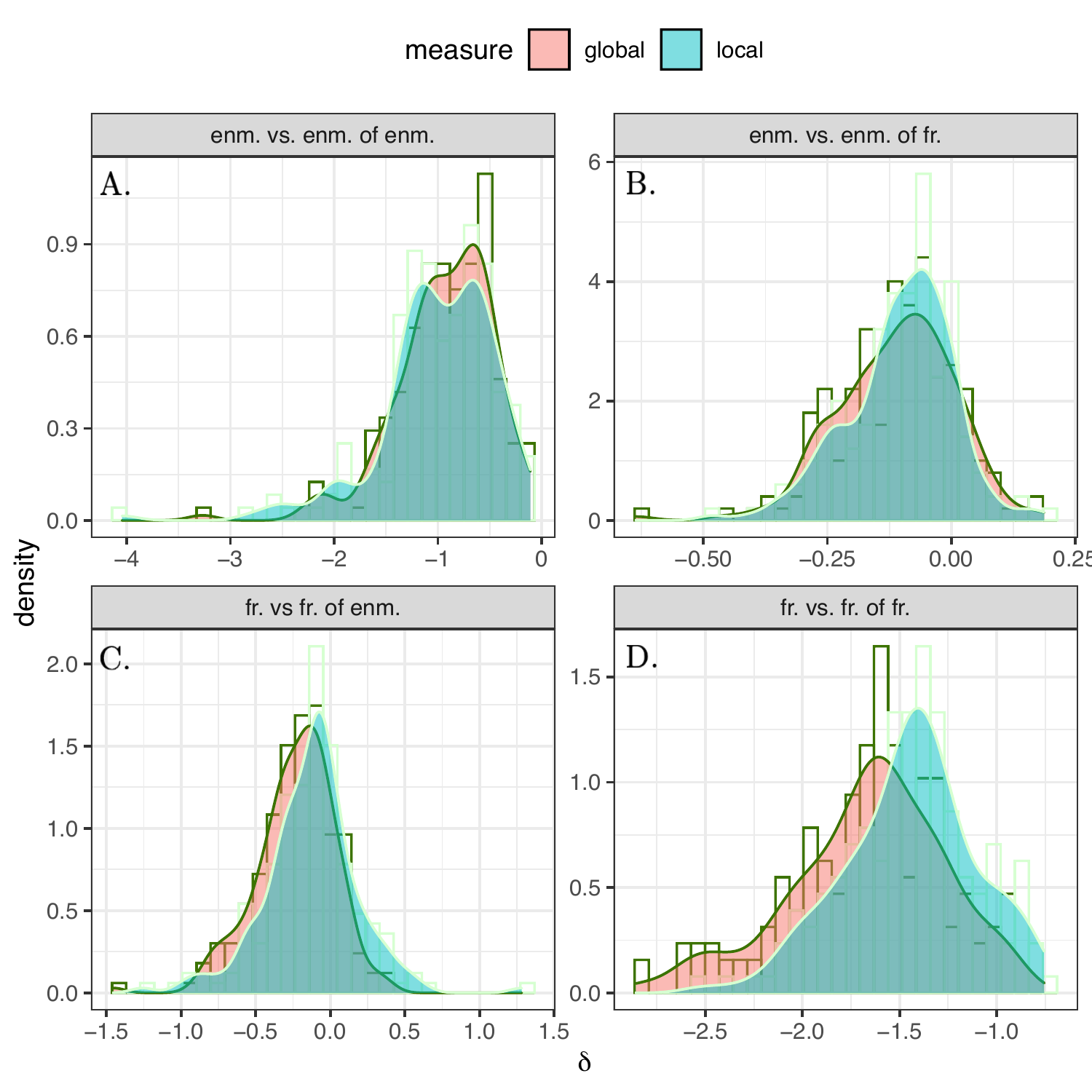}
\caption{Histograms of $\delta_g$ and $\delta_l$ for undirected networks among 176 village networks. The histogram of friendship and enmity paradoxes for undirected (symmetrized) networks is provided in panels A and D, respectively. Other panels represent the results for the mixed worlds. The histogram in panel B shows the global and local paradox distributions for the difference between the number of our enemies and the number of enemies of our friends, while panel C represents the difference between the number of our friends and the number of friends of our enemies.}
\label{fig:enm_fr_delta_us_ur}
\end{figure}

First, we study the global and local enmity paradox among the 176 villages and we present the histogram of these values in Fig.~\ref{fig:enm_fr_delta_us_ur}.
We observe very similar strengths in the enmity and friendship paradoxes in negative and positive worlds, respectively. Additionally, given slightly negative $H_i$ values for friendship networks, we expect the strength of local paradox to be greater than the global paradox for friendship networks as compared to enmity networks (Eq.~\ref{eq:delta_local_global}, Fig.~\ref{fig:enm_fr_delta_us_ur}, A and D, and Appendix, Fig. S12). 
The global and local paradoxes for enmity networks and also for the mixed worlds, however, are nearly equivalent, as expected due to their balanced $H_i$ values (Appendix, Fig. S12). 

The results for the mixed worlds are also presented in Fig.~\ref{fig:enm_fr_delta_us_ur}.
Although the paradoxes almost disappear in undirected (reciprocated) networks due to the sparsity of remaining negative edges
(Appendix, Fig. S1),
for undirected (symmetrized) networks, there is comparable strength in the paradoxes. In other words, in undirected networks using symmetrized edges, our friends are more likely to have enemies than we are (Fig.~\ref{fig:enm_fr_delta_us_ur}B) and our enemies are more likely to have friends than we are (Fig.~\ref{fig:enm_fr_delta_us_ur}C).

\begin{figure}[t!]
\begin{center}
  \includegraphics[width=0.65\textwidth]{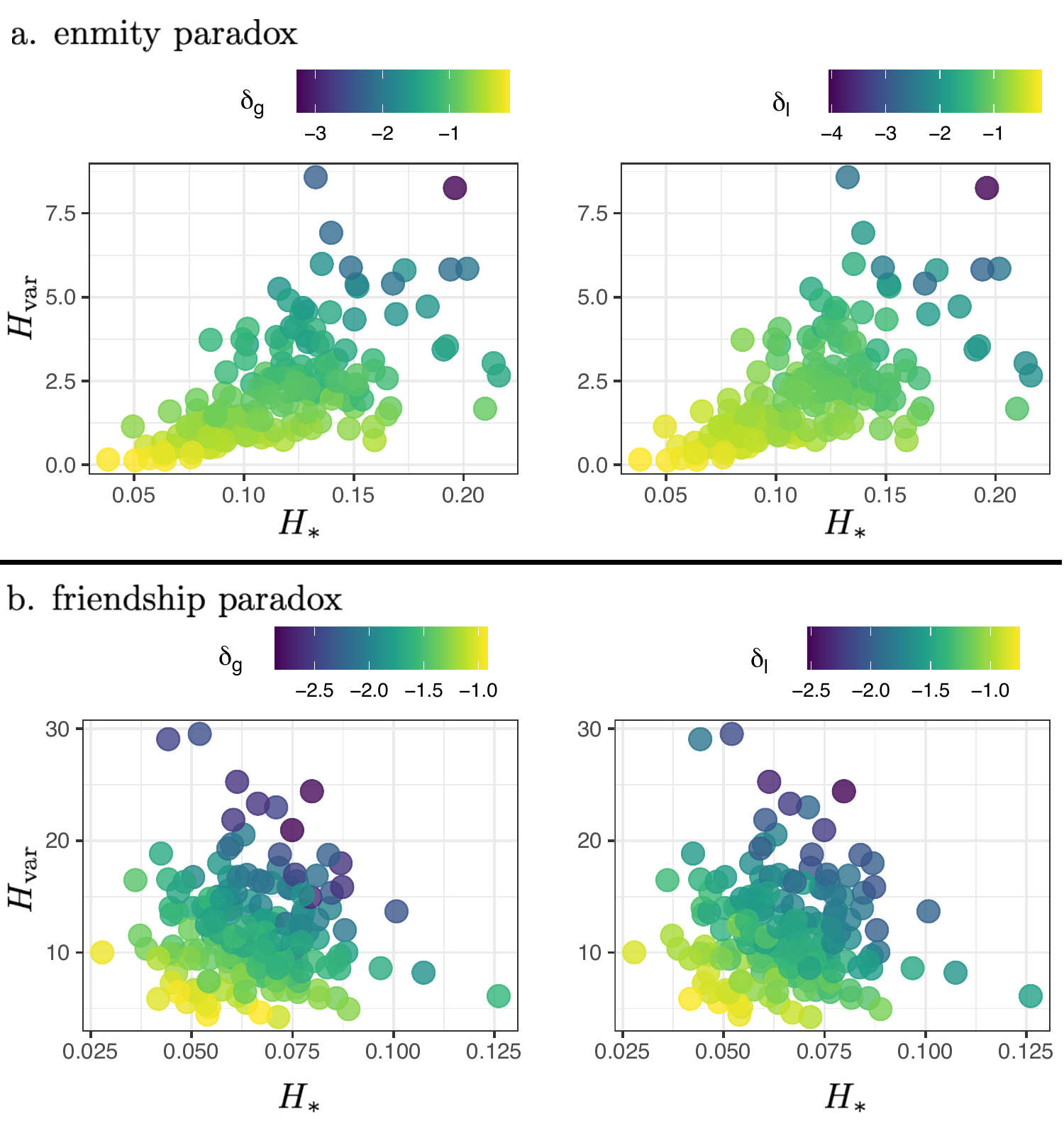}
    \end{center}
  \vspace{-8mm}
\caption{Critical features in enmity and friendship paradoxes. The heat maps show the global and local paradox strengths for undirected (symmetrized) networks as a function of two measures of degree variance $H_{\rm{var}}$ and the starlike embedding $H_{\ast}$. The larger the variance and the more the starlike embedding, the stronger the paradox.}
\label{fig:enm_fr_delta_us_vs_features}
\end{figure}

To understand the effect of different topological properties noted in Fig.~\ref{fig:stat_pos_neg} on the enmity paradox strength, we analyze the relationship between the local and global paradox measures and the topological properties. Using regression models, we can characterize the effect of various topological features on friendship and enmity paradox strengths (Appendix E).
Among the measures, the degree variance $H_{\rm{var}}$ and the starlike embedding $H_{\ast}$, besides the degree diversity $H_{\rm{deg-div}}$ and inversity $H_i$, can explain a significant portion of the intervillage variance in paradox strengths ($R^2\in[0.94,0.97]$). Furthermore, $H_{\rm{var}}$ and $H_{\ast}$ also have large effects. 
This relationship is illustrated in Fig.~\ref{fig:enm_fr_delta_us_vs_features} (the relationships between the strength and other heterogeneity measures are highlighted in Appendix, Figs. S5 and S6). 
As expected, the larger the variance and the more the starlike embedding, the stronger the paradox.

\begin{figure}[t!]
\begin{center}
\centering\includegraphics[width=0.75\textwidth]{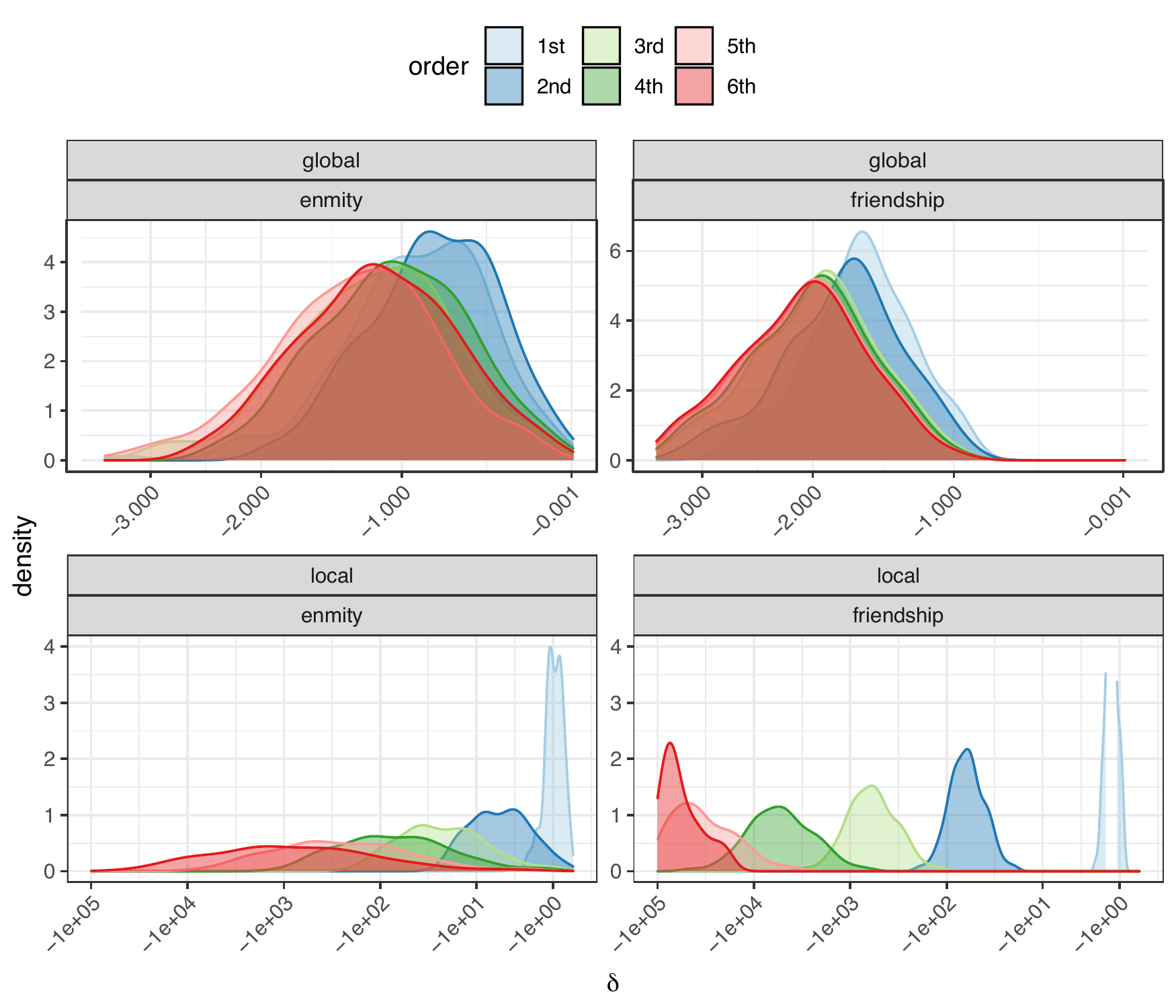}
    \end{center}
  \vspace{-8mm}
\caption{Higher order enmity and friendship paradox. Paradox strength increases with higher order compared to first order (the figure is plotted for first to sixth order). To better visualize the results, we employ the signed pseudo-logarithm transformation.}
\label{fig:higher_order_enm_paradox}
\end{figure}

 The higher-order enmity and friendship paradoxes are presented in Fig.~\ref{fig:higher_order_enm_paradox} (for the first to the sixth order). Because high-degree individuals are counted exponentially more than individuals with low degrees when increasing the order of walks, the higher-order enmity and friendship paradoxes are more severe than the typical enmity and friendship paradoxes. This phenomenon is much more severe in local paradoxes when we compare the degree of a random node with its higher-order neighbors (Fig.~\ref{fig:higher_order_enm_paradox}, bottom row).

\begin{figure}[t!]
\begin{center}
  \includegraphics[width=0.55\textwidth]{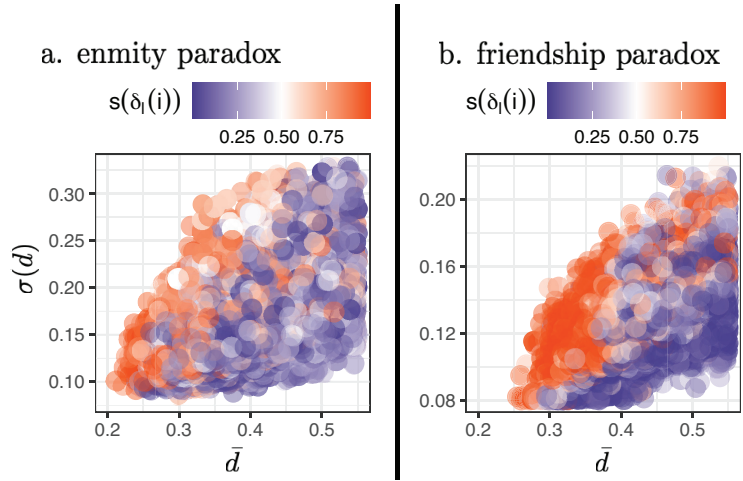}
      \end{center}
  \vspace{-8mm}
\caption{The relationship between a node's contribution in the local enmity and friendship paradoxes and the
location of the nodes within a network. The embedded points in the lower left of the figure correspond to central locations while points in the upper right are in peripheral locations.
Here, we use a sigmoid function $s(x)=\nicefrac{1}{1+e^{-x}}$ in order to increase the small positive and negative differences and reduce the larger differences. 
The peripheral nodes have a larger paradox strength compared with the central nodes.}
\label{fig:local_strength_vs_location_embedding}
\end{figure}

To investigate the relationship between a node's contribution in the local enmity and friendship paradoxes and its topological features, we plot the nodal contribution versus the location of the nodes in a network. The location of the nodes is defined by using a simple fact in networks, namely that the nodes in the center of a network have a smaller distance from other nodes while the nodes in the periphery have a larger distance from other nodes. Therefore, if we embed every node using its average and standard deviation of distances from other nodes in a network---normalized by the network diameter within each network component separately---we find that, in a network, peripheral nodes have a greater paradox strength than central nodes (Fig.~\ref{fig:local_strength_vs_location_embedding}).

Finally, the results for the generalized enmity paradox are provided in Appendix F, Tables~S3, and~S4, and Figs.~S8 and~S9. The results document the existence, for instance, of the generalized paradoxes in both enmity and friendship with respect to wealth. Due to the higher correlation between non-topological features and positive degree, the generalized friendship paradox is stronger than the generalized enmity paradox.

The results for undirected (reciprocated) and directed networks are provided in Appendix, Figs. S1 and S3, respectively. For undirected (reciprocated) networks, the enmity and mixed-world paradox strengths are much smaller compared to undirected (symmetrized) networks---due to the small reciprocity in enmity networks. For directed networks we almost have larger global enmity paradox strength. The local enmity and mixed-world paradox strengths can be explained using an inversity measure~\cite{alipourfard2020friendship} (Appendices D and G, and Fig. S13).

\section*{Discussion}
\label{sec:disc} 
Our perception of the world can be distorted by the sampling bias towards high-degree nodes that arise as a result of the friendship paradox~\cite{feld1991your}. In our minds, if not in reality, our friends have more friends, our collaborators have more collaborators, and our colleagues are wealthier or happier than we ourselves. Here, we showed that the same observations hold with respect to antagonistic ties. Yet, with a lower clustering coefficient and lower degree assortativity and with greater inversity, we expect even more starlike shapes in enmity networks. These factors all make the enmity paradox even more probable than the friendship paradox. Moreover, the enmity paradox obtains in a negative world when comparing the number of an ego’s enemies with their enemies’ enemies as well as in mixed worlds when we compare an ego’s friends with the ego’s enemies’ friends and an ego’s enemies with an ego’s friends’ enemies.

Some topological differences between enmity and friendship networks (e.g., the smaller variance and smaller reciprocity
in enmity networks) suggests possibly divergent perceptional biases. For example, since the paradox strength is proportional to the degree variance, it could be lower for enmity networks; however, we observe similar paradox patterns in both enmity networks and in a mixed world of friendship and antagonistic ties. 

There is also a fundamental connection between the generalized friendship paradox (involving non-topological features) and the enmity paradox. If we treat the number of enemies a person has as a non-topological attribute for the \textit{positive} network, then the enmity paradox follows from the generalized friendship paradox.
Looking at the correlation between positive and negative degrees reveals these observations (Appendix, Fig. S10). As a result of the positive correlation between positive and negative degrees in most of the village networks ($\textrm{Pearson’s correlation} = 0.18$, $p<10^{-16}$; in other words, the empirical reality that people who have many friends also tend to have more enemies), we are able to actually answer the question of why the enmity paradox exists in the first place through the lens of the generalized friendship paradox.

These paradoxes have further implications. Our understanding of social norms and of our social standing is influenced by our perceptions regarding those around us. For instance, the friendship paradox can help explain systematic biases in social perceptions such as regarding the prevalence of binge drinking and risky behaviors~\cite{jackson2019friendship, lerman2016majority}. Furthermore, the friendship paradoxes can explain why a given behavior in a society can be amplified. This can occur in two interwoven phases, where popular individuals act more intensely for activities associated with strategic complementarities, and those who are prone to certain behaviors interact more with other people who are also involved in that behavior, amplifying the effects of this behavior~\cite{jackson2019friendship}. As a result, perceptions of behavior increase and this could contribute to an increase in the behavior along the lines of the perception. Misperceptions about the habits of one’s enemies could act similarly. 

Such inter-personal influence, even if biased, can in turn be exploited to foster cascades, as field experiments have shown~\cite{kim2015social, shakya2017exploiting, alexander2022algorithms}. Indeed, as a result of phenomena like the friendship and enmity paradoxes, we could further perfect network targeting algorithms that exploit the friendship paradox; and taking into account a person’s enemies could make network targeting even more effective. 

Biases associated with our antagonistic ties could be consequential in still another way. For instance, the friendship paradox may intensify homophilous patterns in network formation due to the misperception~\cite{jackson2019friendship, evtushenko2021paradox}. Because of these paradoxes and the misperceptions they can give rise to, people might form a miscalbirated perception of their own attributes and thus seek out connections with people different than they might otherwise truly wish or ``deserve.'' The enmity paradox might similarly change our perceptions of reality and may function as a deterrent force in the formation of network ties between a person and the social connections around a person’s antagonists. When people judge whom to either befriend or avoid, they may be biased in their perceptions.

\begin{acknowledgements}
This work was supported in part by the National Science Foundation under Grant No.~2030859 to the Computing Research Association for the CIFellows Project (A.G.), and also by the NOMIS Foundation. Also, we thank the Yale Center for Research Computing for guidance and use of the research computing infrastructure. Replication code is provided at \url{https://github.com/Aghasemian/EnmityParadox}.
\end{acknowledgements}

\clearpage
\newpage
\appendix
\counterwithout{equation}{section}
\renewcommand{\theequation}{S\arabic{equation}}
\setcounter{equation}{0}
\renewcommand{\thetable}{S\arabic{table}}
\setcounter{table}{0}
\renewcommand{\thefigure}{S\arabic{figure}}
\setcounter{figure}{0}

\section{Data}
Our data come from wave 1 of a sociocentric network study of 24,687 people aged 11 to 93 years, with a mean of around 32 and a median of around 28 years old in 176 geographically isolated villages in western Honduras~\cite{shakya2017exploiting}. Using this empirical data, we first construct 176 binary directed signed networks (with no multi-edges
or self-loops). We use three name generators to determine (overlapping) positive ties (“Who do you spend your free time with?” “Who is your closest friend?” and “Who do you discuss personal matters with?”) to construct the positive world, and one name generator for negative ties (“Who are the people with whom you do not get along well?”) to construct the negative world.  
Due to the village-based nature of our analysis, we also exclude the rare connections reported outside of the villages. 

The reciprocity of negative ties in the analyzed dataset is much smaller than that of positive ties, which may influence the findings. There are several factors contributing to this, including self-censorship by individuals who do not share sensitive information in order to avoid embarrassment, reduce the risk of social norm violation, and protect their privacy~\cite{tourangeau2000psychology, shoemaker2002item}. In addition, there can be a lack of information flow due to the avoidance of any interaction with the receiver of the antagonistic tie by the sender, which may result in the receiver being unaware of the tie and thus unable to ``reciprocate'' it~\cite{harrigan2017avoidance}.   
Nevertheless, the amount of reciprocity depends on the question, and studying undirected networks in Honduras dataset can be justified based on the fact that negative ties are created from the
question “Who are the people with whom
you do not get along well?” that is a symmetric relationship in practice. If an individual $i$ does not get along with another individual $j$, it may be reasonable, as a first approximation, to assume that the individual $j$ also does not get along with $i$. Furthermore, we turned the directed networks into undirected ones by either removing unreciprocated edges or by symmetrizing them. However, alternative approaches are described below. 

\section{Measures}
\label{sec:measures}
\begin{enumerate}
    \item Inversity measure ($H_i$): this measure is proposed in~\cite{kumar2021interventions} where it has been shown it determines the relative effectiveness of the local and global strategies. Actually, the authors have shown that $\mu_L = \mu_G + H_i \Psi(\kappa_{-1},\kappa_1, \kappa_2, \kappa_3)$, where, $\mu_G$ and $\mu_L$ are the global and local mean number of neighbor's enemies/friends, respectively, $\kappa_m=\nicefrac{1}{n} \sum_i k_i^m$ is the $m$-th moment degree, and $H_i=Cor(k_i, \nicefrac{1}{k_j})\,, \forall(i,j) \in E$, i.e., the edge-based correlation of the degree of an endpoint of an edge $E$ with the inverse degree of another endpoint $j$ of that edge is the inversity measure.
    \item Degree assortativity ($H_a$): this measure of homophily measures the tendency for vertices with similar degrees to connect with each other. In~\cite{cantwell2021friendship}, the authors have shown that this measure has a key role in the strength of the local definition of the friendship paradox. For degree assortativity, we use the definition in~\cite{newman2003mixing}, i.e., $H_a = \nicefrac{\sum_{j,k}jk(e_{jk}-q(j)q(k))}{\sigma_q^2}$, where $q(k)$ is the probability of a random endpoint of a random edge has degree $k$, and $e_{jk}$ is the probability of a random edge has endpoints of degree $j$ and $k$. This definition is provided for undirected networks and it can be easily generalized for directed networks.
    \item Starlike strength ($H_\ast$): this measure is introduced by Estrada in Ref.~\cite{estrada2010quantifying} and it represents topological heterogeneity in complex networks; it is maximal for star graphs. The normalized heterogeneity index can be written as $H_{\ast}=\nicefrac{\sum_{(i,j)\in E}\left(k_i^{-1/2}-k_j^{-1/2}\right)}{\left(n-2\sqrt{n-1}\right)}$, where $n$ is the number of nodes in a network.
    \item Variance of the degree ($H_{\rm{var}}$): this typical degree heterogeneity measure is the variance of the degree distribution in a network.
    \item  Degree-diversity ($H_{\rm{deg-div}}$): this is a novel degree heterogeneity measure proposed in~\cite{jacob2017measure} that reflects the degree diversity and has a close relationship with the entropy of the degree distribution. Given a network $G$, the authors define a heterogeneity index $h$ for a network of $n$ nodes as $h^2 = \nicefrac{1}{n}\sum_{k_{\rm{min}}}^{k_{\rm{max}}}\left(1-P(k)\right)^2$, where $P(k)$ is the probability of a random node with degree $k$, $k_{\rm{min}}$ and $k_{\rm{max}}$ are the minimum and maximum degrees, respectively, and the summation is only for $k$, which $P(k)\neq 0$. Then, for a completely homogeneous network of $k$-regular network, it is the case that $h_{\rm{hom}}=0$, and for the most heterogenous case, where the $P(k)$ is uniform, $h_{\rm{het}} = 1-\nicefrac{3}{n}+\nicefrac{(n+2)}{n^3}$. Therefore, the degree diversity is the normalized version of $h$ as $H_{\rm{deg-div}} = \nicefrac{h}{h_{\rm{het}}}$. 
    \item Transitivity $T_g$ and clustering coefficient $T_l$: transitivity is one of the two most common topological features that is designed to measure the commonness of triangles in social networks. This global measure computes the fraction of transitive triads that are subgraphs of triangles~\cite{latora2017complex}. Another topological feature, the clustering coefficient, which is very similar to transitivity, computes the fraction of triads that are subgraphs of triangles for each node $i$, and then averages these values. As a result, we can consider the clustering coefficient as the local adaptation of the transitivity measure, and we refer to the clustering coefficient as $T_l$, and transitivity as $T_g$. In most graphs, these two measures take similar values; however, they are rare occasions that they disagree.
    \item Normalized betweenness-centrality (${\rm{N-BC}}$): the number of shortest paths traversing a node $v$ represents its importance in information flow in a network and is called betweenness centrality (${\rm{BC}}$). Freeman~\cite{freeman2002centrality} proved that the maximum value of this measure is achieved by the central point in a star, and it is $\nicefrac{n^2-3n+2}{2}$. Therefore, the ${\rm{N-BC}}$ is a normalization of ${\rm{BC}}$ by this maximum value.
\end{enumerate}
\section{Enmity paradox in undirected networks}
\label{sec:enm_par_und_nets}
\subsection*{The generalized inversity in mixed worlds} 
Based on the inversity measure for undirected networks, it is possible to formalize the differences between global and local paradoxes in four different scenarios introduced in the main text: 1. the enmity paradox; 2. the friendship paradox; and the two mixed worlds of 3. when we compare the number of our friends with the number of our enemies' friends; and 4. when we compare the number of our enemies with the number of our friends' enemies. 

Here, we summarize these equations in the following manner. (Detailed proof of the generalization of the inversity measure for the mixed worlds (Eqs.~\ref{eq:inv_nwp} and~\ref{eq:inv_pwn}) are provided in Section~\ref{sec:gen_inv}.)
\begin{enumerate}
    \item \begin{align}
    \label{eq:inv_nwn}
        \delta_{g, -w}(-)-\delta_{l, -w}(-) \propto cor(k_{(-),i}, \nicefrac{1}{k_{(-),j}} | (i,j) \in E_{(-)})
    \end{align}
    \item \begin{align}
    \label{eq:inv_pwp}
        \delta_{g, +w}(+)-\delta_{l, +w}(+) \propto cor(k_{(+),i}, \nicefrac{1}{k_{(+),j}} | (i,j) \in E_{(+)})
    \end{align}
    \item \begin{align}
    \label{eq:inv_nwp}
        \delta_{g, -w}(+)-\delta_{l, -w}(+) \propto cor(k_{(+),i}, \nicefrac{1}{k_{(-),j}} | (i,j) \in E_{(-)})
    \end{align}
    \item \begin{align}
    \label{eq:inv_pwn}
        \delta_{g, +w}(-)-\delta_{l, +w}(-) \propto cor(k_{(-),i}, \nicefrac{1}{k_{(+),j}} | (i,j) \in E_{(+)})
    \end{align}
\end{enumerate}
\subsection*{Enmity paradox in undirected (reciprocated) networks}
\begin{figure}[t!]
\centering
  \includegraphics[width=0.75\textwidth]{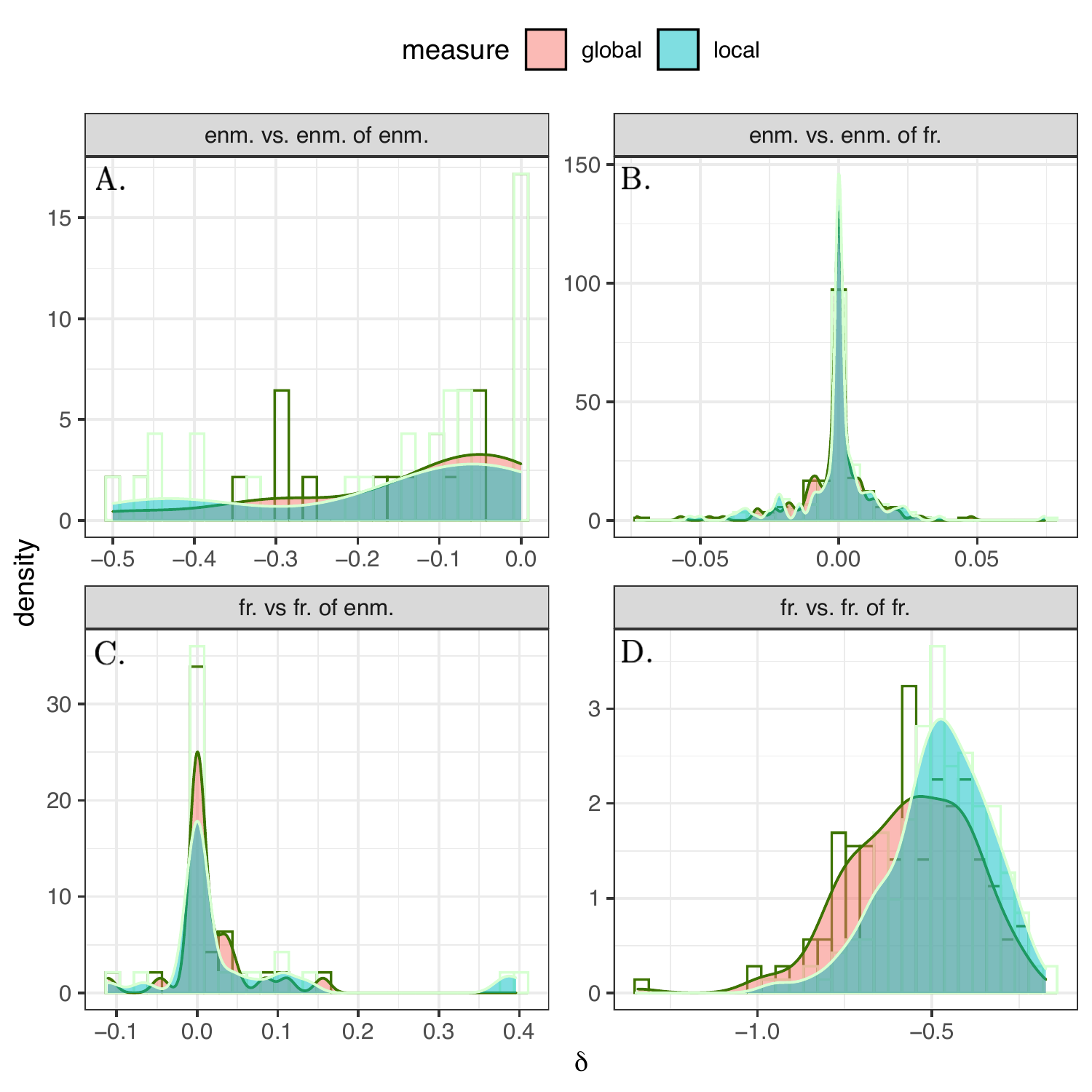}
\caption{Histograms of $\delta_g$ and $\delta_l$ for undirected (reciprocated) networks. The histogram of enmity and friendship paradoxes is provided in panels A and D, respectively. Other panels represent the results for the mixed worlds. The histogram in panel B shows the global and local paradox distributions for the difference between the number of our enemies and the number of enemies of our friends, while panel C represents the difference between the number of our friends and the number of friends of our enemies.
These paradoxes are minimized in undirected (reciprocated) networks when compared to undirected (symmetrized) networks (see the main text), especially in mixed worlds due to the sparsity of negative edges that cause the networks created to be very small after removing isolated nodes (only $27$ out of $176$ networks were left in our analysis).}
\label{fig:enm_fr_delta_ur}
\end{figure}

The results for undirected (symmetrized) networks have been provided in the main text. Here, the results for undirected networks constructed by only keeping the reciprocated edges are presented (see Fig.~\ref{fig:enm_fr_delta_ur}). 
Once the isolated nodes are removed, many undirected (reciprocated) networks disappear or become very small, and the results are limited to networks with at least $8$ nodes ($27$ out of $176$ networks).
Additionally, due to the small reciprocity in the negative world, the enmity paradox for undirected enmity networks constructed with reciprocated edges is much smaller than the friendship paradox for friendship networks constructed similarly (see Fig.~\ref{fig:enm_fr_delta_ur}, A versus D). Essentially, this observation reflects the sparsity of the networks created as a result of antagonistic ties being infrequently reciprocal compared to positive ones. 
The amount of reciprocity varies from question to question, and we believe that questions such as “Who are the people with whom you do not get along well” should be accompanied by greater reciprocity. Therefore, it is possible that measurement errors and self-censorship lead to smaller reciprocity, further justifying symmetrized undirected networks in our study.

The results for the mixed worlds in undirected (reciprocated) networks are also presented in Fig.~\ref{fig:enm_fr_delta_ur}. As with the enmity paradox of undirected networks constructed with reciprocated edges, we do not observe the paradoxes in the mixed worlds with reciprocated edges due to the sparsity of remaining negative edges; however, for undirected (symmetrized) networks (main text, Fig. 4) we see comparable strengths of the paradoxes.

\section{Enmity paradox in directed networks}
\label{sec:enm_par_dir_nets}

Here, we derive the equations of the enmity paradox for directed networks. We follow a similar convention as in Ref.~\cite{alipourfard2020friendship} and use distinguishing names for those who introduce us as antagonistic/friendship connections and those we introduce as antagonistic/friendship connections. We name whom we report as antagonistic/friendship connection as ``enemy''/``friend'' and a person who introduce us as antagonistic/friendship connections as ``hater''/``liker.''

Then, we have four possibilities that we should consider (Fig.~\ref{fig:dir_cases}): 1. our enemies/friends have more haters/likers than we do; 2. our haters/likers have more enemies/friends than we do; 3. our enemies/friends have more enemies/friends than we do; and 4. our haters/likers have more haters/likers than we do. Similar to what the authors in Ref.~\cite{alipourfard2020friendship} have shown for the positive world of followers/followees, the first two statements are correct mathematically, but for the last two, a positive correlation between the in-degree and out-degree is required.

The global formulation of the enmity paradox for directed networks is a comparison between the average in-degree/out-degree of a random node $i$ and the average in-degree/out-degree of its neighbors' in-degrees/out-degrees when the neighbor is chosen between the enemies and haters as described below. Here, $A_{i,j}=1$ means node $j$ is an enemy of node $i$ and node $i$ is a hater of node $j$. In other words node $i$ reports node $j$ as its antagonistic connection. As a result of the symmetry, the equations for the directed friendship paradoxes are similar. To simplify the notation, we use $A$ instead of $A_{(-)}$. As we did for the undirected networks in the main text, these equations can be generalized to the mixed world when positive and negative ties coexist. 

\begin{enumerate}
\item the average number of enemies of a random node versus the average number of enemies of a random hater~\footnote{In this notation, the $in-w$ and $out-w$ indicate the direction of one's neighbor, as one's in-neighbor and out-neighbor, respectively. The $(in)$ and $(out)$ denote the type of comparison as in-degree or out-degree.}.
\begin{align}
\label{eq:dif_global_num_enms_vs_num_enms_haters}
\delta_{g, in-w} (out) =& \frac{(\one^TA\one)^2 - \one^TA^TA\one\cdot\one^T\one}{\one^TA\one \cdot\one^T\one}
\end{align}
\item the average number of haters of a random node versus the average number of haters of a random enemy.
\begin{align}
\label{eq:dif_global_num_htrs_vs_num_htrs_enemies}
\delta_{g, out-w} (in) =& \frac{(\one^TA\one)^2 - \one^TAA^T\one\cdot\one^T\one}{\one^TA\one \cdot\one^T\one}
\end{align}
\item the average number of enemies of a random node versus the average number of enemies of a random enemy.
\begin{align}
\label{eq:dif_global_num_enms_vs_num_enms_enmiess}
\delta_{g, out-w} (out) =& \frac{(\one^TA\one)^2 - \one^TA^2\one\cdot\one^T\one}{\one^TA\one \cdot\one^T\one}
\end{align}
\item the average number of haters of a random node versus the average number of haters of a random hater.
\begin{align}
\label{eq:dif_global_num_htrs_vs_num_htrs_haters}
\delta_{g, in-w} (in) =& \frac{(\one^TA\one)^2 - \one^T{A}^2\one\cdot\one^T\one}{\one^TA\one \cdot\one^T\one}
\end{align}
\end{enumerate}
The global formulation of the last two scenarios is the same due to symmetry. It is not different when we compute the average number of haters of our haters or the average number of enemies of our enemies globally. 

However, it matters when we have the local formulation. 
For the local formulation, we have again four possibilities (Fig.~\ref{fig:dir_cases}). The in-degree and out-degree vectors are denoted as $\din$ and $\dout$, while the diagonal in-degree and out-degree matrices are denoted as $\Din$ and $\Dout$, respectively.  Therefore, ${\dout}_{,i}$ is the $i$-th entry of the out-degree vector $\dout$, and, similarly, it can be defined for the in-degree vector $\din$.

\begin{enumerate}
\item the average difference of the number of enemies of a random node with the average number of enemies of its haters.
\begin{align}
\label{eq:ave_dif_local_num_enms_vs_num_enms_haters}
\delta_{l, in-w} (out) =& 
\frac{\one^TA\one - \one^T\Din^{-1}A^T\Dout\one}{\one^T\one} 
\end{align}
\item the average difference of the number of haters of a random node with the average number of haters of its enemies.
\begin{align}
\label{eq:ave_dif_local_num_htrs_vs_num_htrs_enemies}
\delta_{l, out-w} (in) =& 
\frac{\one^TA^T\one - \one^T\Dout^{-1}A\Din\one}{\one^T\one} 
\end{align}
\item the average difference of the number of enemies of a random node with the average number of enemies of its enemies.
\begin{align}
\label{eq:ave_dif_local_num_enms_vs_num_enms_enemies}
\delta_{l, out-w} (out) =& 
\frac{\one^TA\one - \one^T\Dout^{-1}A\Dout\one}{\one^T\one} 
\end{align}
\item the average difference of the number of haters of a random node with the average number of haters of its haters.
\begin{align}
\label{eq:ave_dif_local_num_htrs_vs_num_htrs_haters}
\delta_{l, in-w} (in) =& 
\frac{\one^TA^T\one - \one^T\Din^{-1}A^T\Din\one}{\one^T\one} 
\end{align}
\end{enumerate}
In the local formulation, $\one^TA\one = \one^TA^T\one = \sum_i {\dout}_{,i} = \sum_i {\din}_{,i}$. However, to compute the individual values of the vector of average difference we need to write it this way.

\begin{enumerate}
\item the difference between the number of enemies of a random node with the average number of enemies of its haters.
\begin{align}
\label{eq:dif_local_num_enms_vs_num_enms_haters}
\Delta_{l, in-w}(out) =& (A - \Din^{-1}A^T\Dout)\one
\end{align}
\item the difference between the number of haters of a random node with the average number of haters of its enemies.
\begin{align}
\label{eq:dif_local_num_htrs_vs_num_htrs_enemies}
\Delta_{l, out-w}(in) =& 
(A^T - \Dout^{-1}A\Din)\one
\end{align}
\item the difference between the number of enemies of a random node with the average number of enemies of its enemies.
\begin{align}
\label{eq:dif_local_num_enms_vs_num_enms_enemies}
\Delta_{l, out-w}(out) =& 
(A - \Dout^{-1}A\Dout)\one 
\end{align}
\item the difference between the number of haters of a random node with the average number of haters of its haters.
\begin{align}
\label{eq:dif_local_num_htrs_vs_num_htrs_haters}
\Delta_{l, in-w}(in) =& 
(A^T - \Din^{-1}A^T\Din)\one 
\end{align}
\end{enumerate}
\subsection*{The generalized inversity in directed networks} 
Based on the inversity measure expanded for directed networks, it is possible to formalize the differences between the four pairs of global and local paradoxes in directed networks~\cite{alipourfard2020friendship}. Here, we summarize these equations in the following manner. (Detailed proof of the generalization of the inversity measure for directed networks is provided in Section~\ref{sec:gen_inv}.)
\begin{enumerate}
    \item \begin{align}
    \label{eq:inv_inwout}
        \delta_{g, in-w}(out)-\delta_{l, in-w}(out) \propto cor(k_{i, out}, \nicefrac{1}{k_{j, in}} | (i,j) \in E)
    \end{align}
    \item \begin{align}
    \label{eq:inv_outwin}
        \delta_{g, out-w}(in)-\delta_{l, out-w}(in) \propto cor(\nicefrac{1}{k_{i, out}}, k_{j, in} | (i,j) \in E)
    \end{align}
    \item \begin{align}
    \label{eq:inv_outwout}
        \delta_{g, out-w}(out)-\delta_{l, out-w}(out) \propto cor(\nicefrac{1}{k_{i, out}}, k_{j, out} | (i,j) \in E)
    \end{align}
    \item \begin{align}
    \label{eq:inv_inwin}
        \delta_{g, in-w}(in)-\delta_{l, in-w}(in) \propto cor(k_{i, in}, \nicefrac{1}{k_{j, in}} | (i,j) \in E)
    \end{align}
\end{enumerate}

Similarly, we can easily generalize the equations corresponding to the local and global formulations of enmity and friendship paradoxes in directed networks to the mixed world of directed networks, when both antagonistic and friendship interactions coexist. For example, we can compare the number of enemies of a random node with the number of enemies of a random liker in the global and local formulations as $\delta_g = \nicefrac{\left(\one^TA_{(-)}\one\cdot\one^TA_{(+)}\one - \one^TA^T_{(+)}A_{(-)}\one\cdot\one^T\one\right)}{\one^TA_{(+)}\one\cdot\one^T\one}$ and $\delta_l = \nicefrac{\left(\one^TA_{(-)}\one - \one^TD_{(+),{\rm{in}}}^{-1}A^T_{(+)}D_{(-),{\rm{out}}}\one\right)}{\one^T\one}$, respectively.
However, a careful study of these mixed worlds for directed networks is out of scope of this study, though it could be examined in future work. 

Also for the generalized enmity paradox, we can expand the equations for directed networks. The global formulation of the generalized paradox for directed networks can be written for enemies in Eq.~\ref{eq:dif_global_gen_enemy}, and for haters in Eq.~\ref{eq:dif_global_gen_hater}.
\begin{align}
\label{eq:dif_global_gen_enemy}
\delta_g(x) =& \frac{\one^TD_x\one\one^TA_{(-)}\one - \one^TA_{(-)}D_x\one\cdot\one^T\one}{\one^TA_{(-)}\one \cdot\one^T\one}  
\end{align}
\begin{align}
\label{eq:dif_global_gen_hater}
\delta_g(x) =& \frac{\one^TD_x\one\one^TA_{(-)}\one - \one^TA_{(-)}^TD_x\one\cdot\one^T\one}{\one^TA_{(-)}\one \cdot\one^T\one} \,. 
\end{align}
Similarly, for the local formulation of the generalized paradox, we can write the equations for enemies and haters in Eqs.~\ref{eq:dif_local_gen_enemy} and~\ref{eq:dif_local_gen_hater}, respectively.
\begin{align}
\label{eq:dif_local_gen_enemy}
\delta_l(x) =& 
\frac{\one^TD_x\one - \one^TD_{(-),{\rm{out}}}^{-1}A_{(-)}D_x\one}{\one^T\one}
\end{align}
\begin{align}
\label{eq:dif_local_gen_hater}
\delta_l(x) =& 
\frac{\one^TD_x\one - \one^TD_{(-),{\rm{in}}}^{-1}A_{(-)}^TD_x\one}{\one^T\one}\,.
\end{align}
 
\begin{figure}[t!]
\centering
  \includegraphics[width=0.7\textwidth]{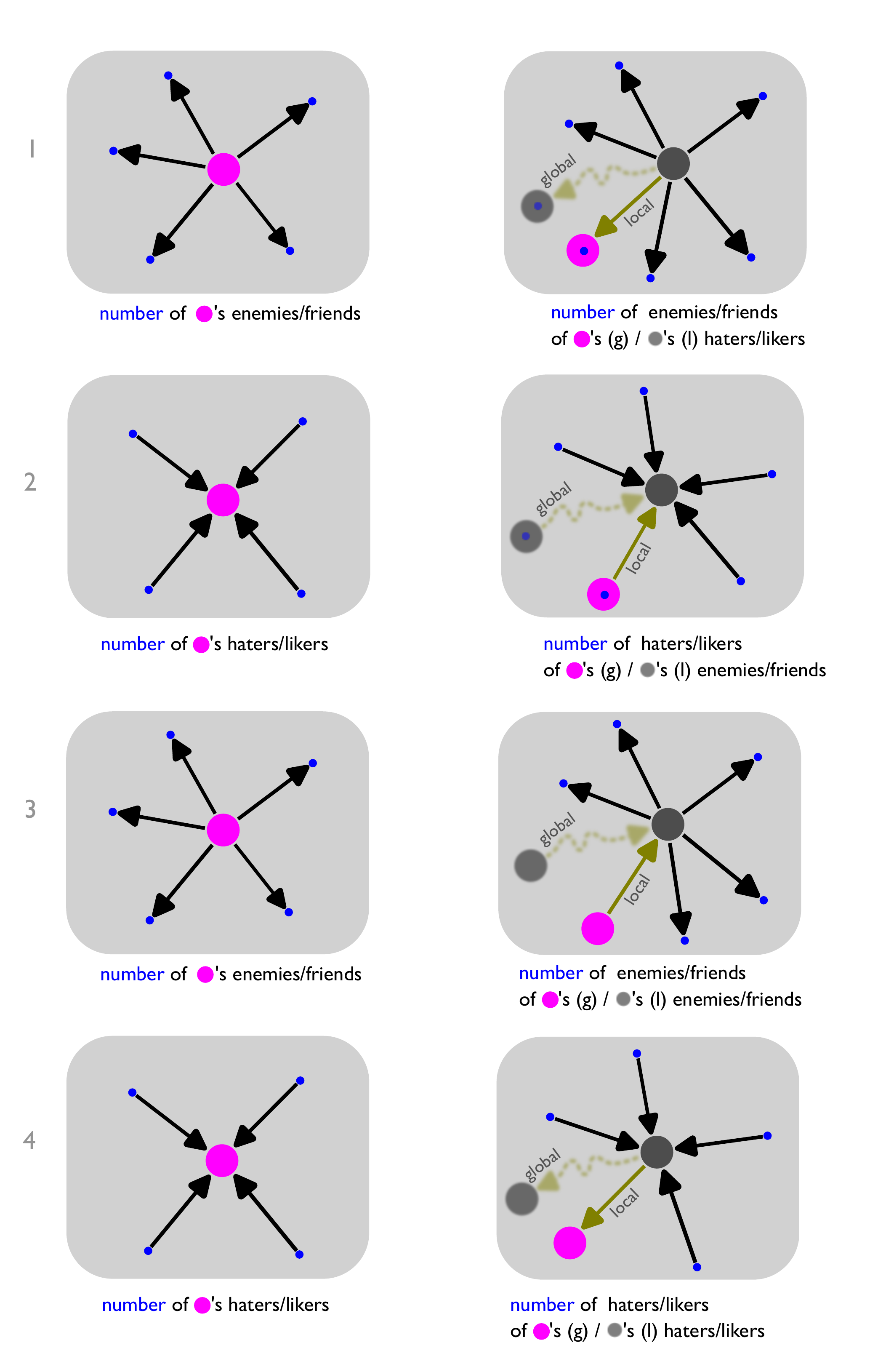}
\caption{A schematic of four possibilities in the global and local formulations of the enmity and friendship paradoxes. For local paradoxes, we compare a person's number of in-/out-neighbors with their in-/out-neighbor's average number of in-/out-neighbors. For the global paradoxes, we compare a person's number of in-/out-neighbors with the average number of in-/out-neighbors of a random endpoint of an edge.}
\label{fig:dir_cases}
\end{figure}

As part of an analysis of both directed and undirected village networks in western Honduras, we compared the average degrees of all nodes with the average degrees of their neighbors in Table~\ref{tab:statistics_degree}. We define neighbors in directed networks in two ways: in-neighbors and out-neighbors. A comparison of the average degrees of the nodes is made with their neighbors' average degrees. We consider four cases for directed networks corresponding to the scenarios presented in Fig.~\ref{fig:dir_cases}.

\begin{figure}[t!]
\centering
  \includegraphics[width=0.7\textwidth]{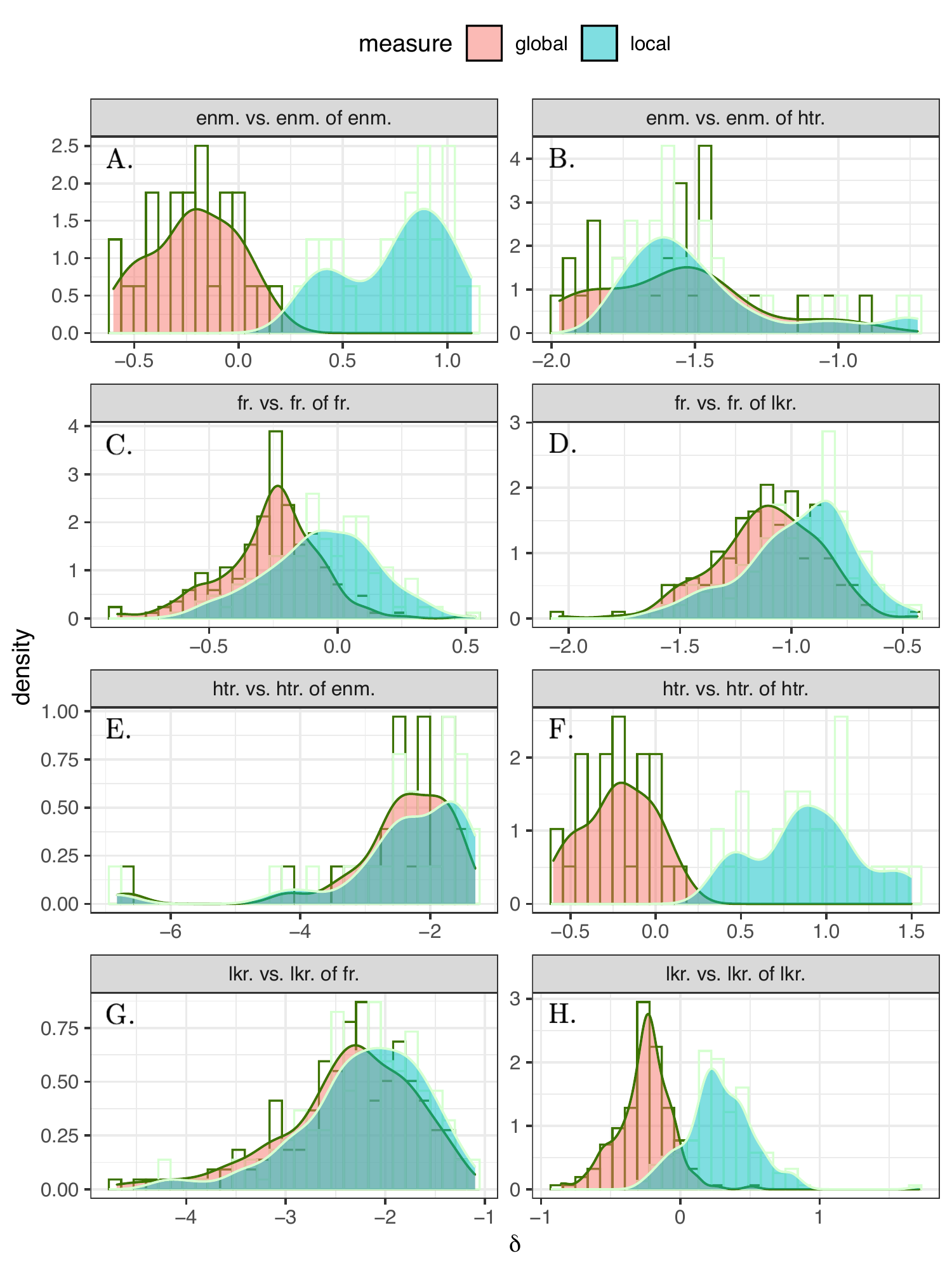}
\caption{Histograms of $\delta_g$ and $\delta_l$ for directed networks among 176 network villages. The difference between the global and local paradoxes can be explained using the generalized inversity measures for directed networks. These (correlations, p-values) for the whole dataset can be summarized as $(-0.03, 4.2e^{-5})$, $(-0.01, 0.2)$, $(-0.06, 2.32e^{-12})$, and $(-0.09, <2.2e^{-16})$ for the enmity world and $(-0.15,<2.2e^{-16})$, $(-0.14,<2.2e^{-16})$, $(-0.19,<2.2e^{-16})$, and $(-0.16,<2.2e^{-16})$ for the friendship world. The order of these correlations is for the paradox strengths comparing the number of enemies/friends with the  number of enemies/friends of haters/likers (panels B and D); the number of haters/likers with the  number of haters/likers of enemies/friends (panels E and G); the  number of enemeis/friends with the  number of enemies/friends of enemeis/friends (panels A and C); and the  number of haters/likers with the  number of haters/likers of haters/likers (panels F and H). A detailed analysis of these correlations is provided in Fig.~\ref{fig:inversity_directed}.}
\label{fig:enm_fr_delta_directed}
\end{figure}

The empirical results for the aforementioned enmity paradox in directed networks, introduced by four different formulations in Eqs.~\ref{eq:dif_global_num_enms_vs_num_enms_haters}-\ref{eq:ave_dif_local_num_htrs_vs_num_htrs_haters}, and their corresponding friendship paradoxes have been provided in Fig.~\ref{fig:enm_fr_delta_directed}. 
The results indicate that we see global paradoxes for both enmity and friendship directed networks. Also, the strength of these paradoxes is maximum when comparing our number of haters with the number of haters of our enemies and our number of likers with the number of likers of our friends, as these two are mathematically supported. Thus, our enemies have more haters than we do, and our friends have more likers. As with the previous two mathematical facts, our likers have more friends than we do, and our haters have more enemies than we do, but the strength of these two mathematical paradoxes is smaller than those from the previous two facts. In order for the other four global paradoxes to be valid, the in- and out-degrees must be positively correlated~\cite{alipourfard2020friendship}. In view of the positive correlation between in- and out-degrees (Fig.~\ref{fig:in-out-deg-corr_pos_neg}), we expect the other four global paradoxes to be satisfied as shown in Fig.~\ref{fig:enm_fr_delta_directed}. Alternatively, while there are local paradoxes when comparing our number of haters/likers with the number of haters/likers of our enemies/friends and also comparing our number of enemies/friends with the number of enemies/friends of our haters/likers, they do not exist or exist in a counterintuitive sense when we compare our number of enemies/friends with the number of enemies/friends of our enemies/friends, or when we compare our number of haters/likers with the number of haters/likers of our haters/likers. 
Generally, the difference between global and local paradoxes can be explained using the generalized inversity measures for directed networks. The correlations and p-values for the whole dataset are summarized for both enmity and friendship networks in the caption of Fig.~\ref{fig:enm_fr_delta_directed}. Also, a detailed analysis of these correlations has been provided in Fig.~\ref{fig:inversity_directed}.
Several applications can benefit from exploring the enmity paradox in directed mixed networks including both positive and negative ties. Future research can examine the details of the enmity paradox in such an environment, as well as the relationship between the paradox strength and the topological features of directed networks.

\begin{table}[t!]
\centering
\begin{tabular}[t]{lrrrr}
\toprule
\multicolumn{1}{c}{ } & \multicolumn{2}{c}{enmity} & \multicolumn{2}{c}{friendship}\\
\cmidrule(l{2pt}){2-3}\cmidrule(l{2pt}){4-5} 
value &  node  &  neighbor   &  node  &  neighbor     \\
\midrule
\bf{(undirected)} & ~ & ~ & ~ & ~\\
~~ degree & 1.26 & 3.55& 6.89 &8.97\\
\addlinespace
\bf{(directed)} & ~ & ~& ~ & ~\\
~~ 1. out-degree (in-neighbor) & 0.65 &2.48 &4.13&5.47\\
~~ 2. in-degree (out-neighbor) & 0.65 & 3.06 &4.13&6.83\\
~~ 3. out-degree (out-neighbor) & 0.65 &0.93 &4.13&4.61\\
~~ 4. in-degree (in-neighbor) & 0.65 & 0.93&4.13&4.61\\
\addlinespace
  \bottomrule
\end{tabular}
\caption{Comparing the average degrees of nodes in 176 village networks in western Honduras with the average degrees of their neighbors for both directed and undirected networks. Directed networks define neighbors as either in- or out-neighbors. We compare the average degree of the nodes with the average degree of their neighbors. For directed networks, four scenarios are considered, corresponding to those presented in Fig.~\ref{fig:dir_cases}.}
\label{tab:statistics_degree}
\vspace{-5mm}
\end{table}

\clearpage
\newpage
\section{Relationship of topological features with paradox strength}
\begin{figure}[t!]
\centering
\vspace{-10mm}
\subfloat[enmity]{\includegraphics[width=0.6\textwidth]{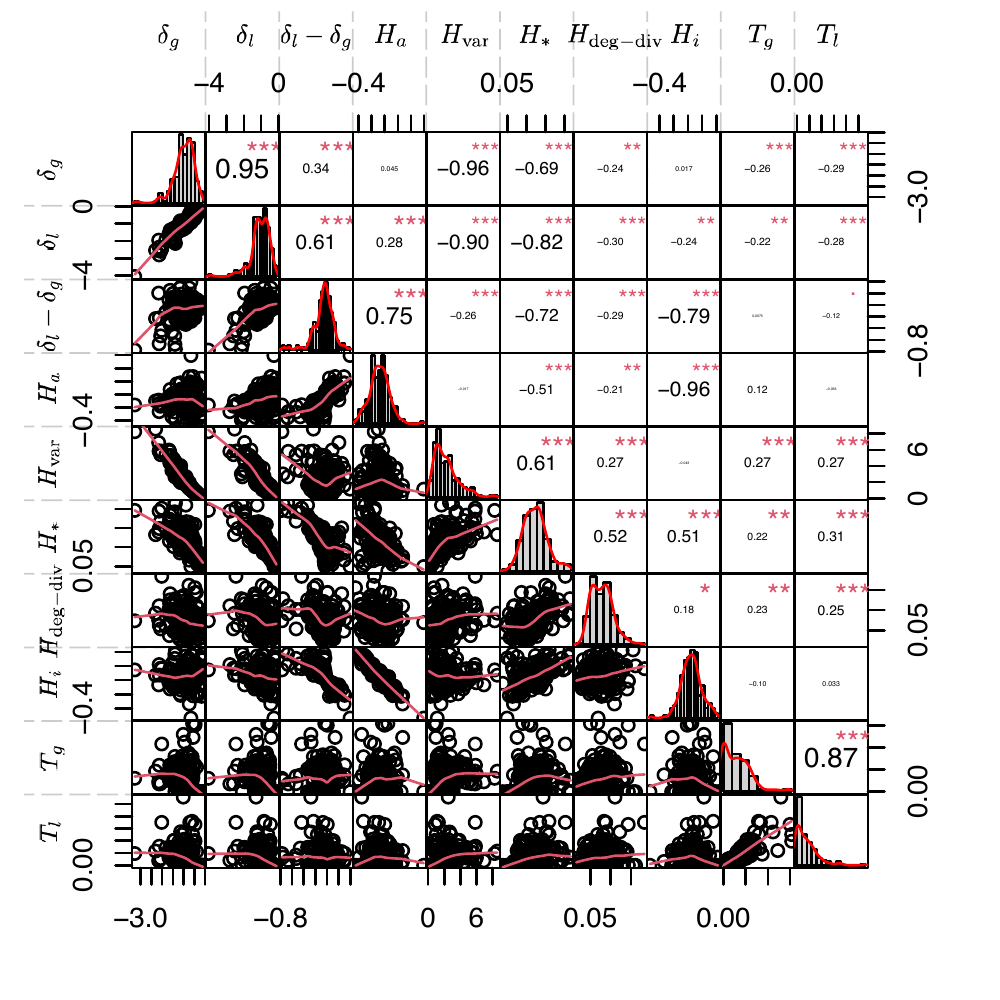}
  \vspace{-8mm}
  \label{fig:correlation_matrix_enm_data}}

\subfloat[friendship]{\includegraphics[width=0.6\textwidth]{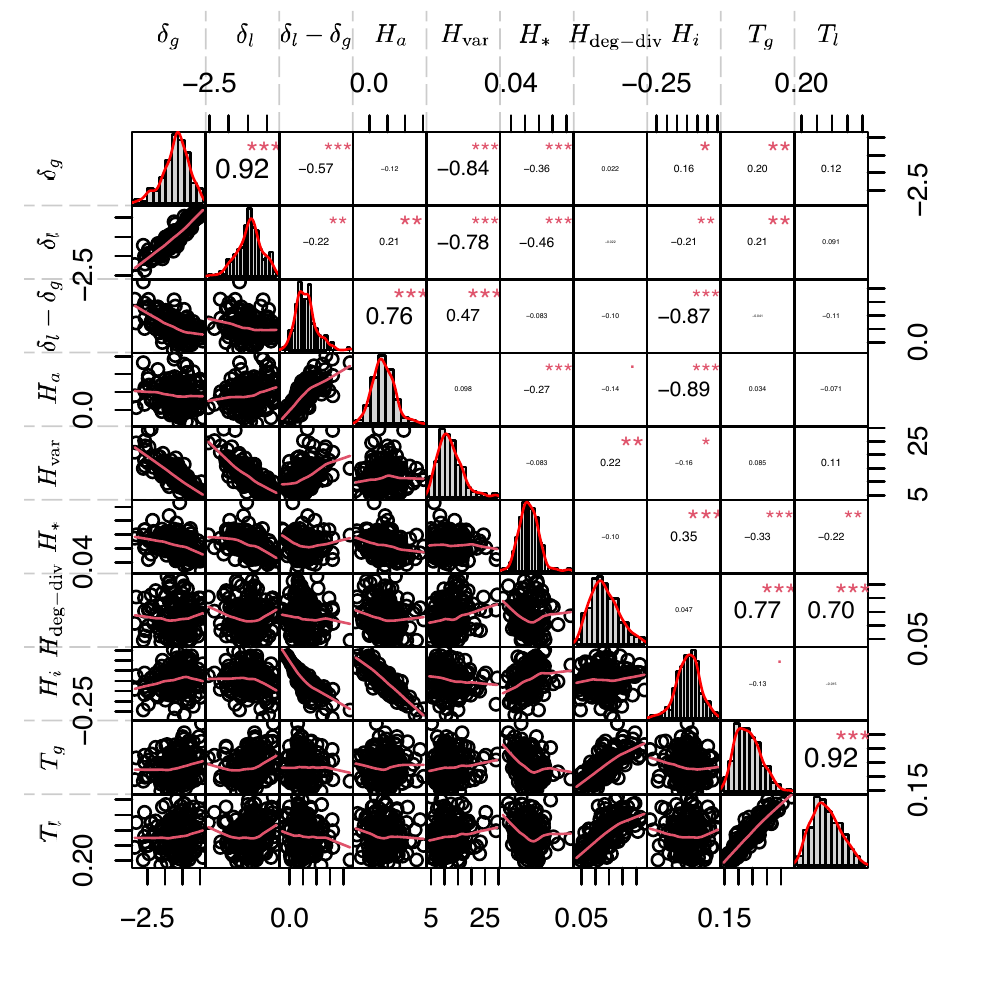}
  \vspace{-8mm}
  \label{fig:correlation_matrix_fr_data}}
\caption{The chart of correlation matrices for enmity and friendship data frames.}
\label{fig:correlation_matrix_enm_fr_data}
\end{figure}

Our study examines the relationship of various topological features with friendship and enmity paradox strength using regression modeling. The correlation matrix for both enmity and friendship paradox data frames is shown in Fig.~\ref{fig:correlation_matrix_enm_fr_data}. Our regression analysis considers only a subset of these topological features due to their significant correlation, including $H_{\rm{var}}$, $H_{\ast}$, $H_{\rm{deg-div}}$, $H_i$, and $T_g$. The effect of $H_a$ is almost the same as the effect of $H_i$ with a negative sign since they are highly negatively correlated, whereas the effect of $T_l$ is almost the same as that of $T_g$ since they are highly positively correlated.
The effect of regression of $\delta_l$ and $\delta_g$ on different topological features is in Table~\ref{tab:modeling_delta_g_l_en_fr}.

\begin{table}[!ht]
\footnotesize
\begin{center}
\begin{tabular}{l c c c c}
\hline
 & Model $\delta_g$ (enmity paradox) & Model $\delta_l$ (enmity paradox) & Model $\delta_g$ (friendship paradox)  & Model $\delta_l$ (friendship paradox) \\
\hline
$H_{\rm{var}}$           & $-0.76^{***}$ & $-0.66^{***}$ & $-0.88^{***}$ & $-0.88^{***}$ \\
                & $(0.02)$      & $(0.03)$      & $(0.02)$      & $(0.02)$      \\
$H_{\ast}$        & $-0.39^{***}$ & $-0.46^{***}$ & $-0.47^{***}$ & $-0.44^{***}$ \\
                & $(0.03)$      & $(0.04)$      & $(0.02)$      & $(0.02)$      \\
$H_{\rm{deg-div}}$     & $0.14^{***}$  & $0.13^{***}$  & $0.12^{***}$  & $0.10^{***}$  \\
                & $(0.02)$      & $(0.02)$      & $(0.03)$      & $(0.03)$      \\
$H_i$            & $0.16^{***}$  & $-0.05$       & $0.18^{***}$  & $-0.20^{***}$ \\
                & $(0.02)$      & $(0.03)$      & $(0.02)$      & $(0.02)$      \\
$T_g$ & $0.01$        & $0.03$        & $0.05$        & $0.04$        \\
                & $(0.02)$      & $(0.02)$      & $(0.03)$      & $(0.03)$      \\
\hline
R$^2$           & $0.97$        & $0.94$        & $0.95$        & $0.95$        \\
Adj. R$^2$      & $0.96$        & $0.93$        & $0.95$        & $0.94$        \\
Num. obs.       & $176$         & $176$         & $176$         & $176$         \\
\hline
\multicolumn{5}{l}{\scriptsize{$^{***}p<0.001$; $^{**}p<0.01$; $^{*}p<0.05$}}
\end{tabular}
\caption{Regression analysis of $\delta_g$ and $\delta_l$ of the enmity paradox and $\delta_g$ and $\delta_l$ of the friendship paradox with respect to different measures. The model considered in our analysis is summarized as $\delta \sim H_{\rm{var}} + H_{\ast} + H_{\rm{deg-div}} + H_i + T_g $.}
\label{tab:modeling_delta_g_l_en_fr}
\end{center}
\end{table}

\begin{figure}[t!]
\centering
  \includegraphics[width=1\textwidth]{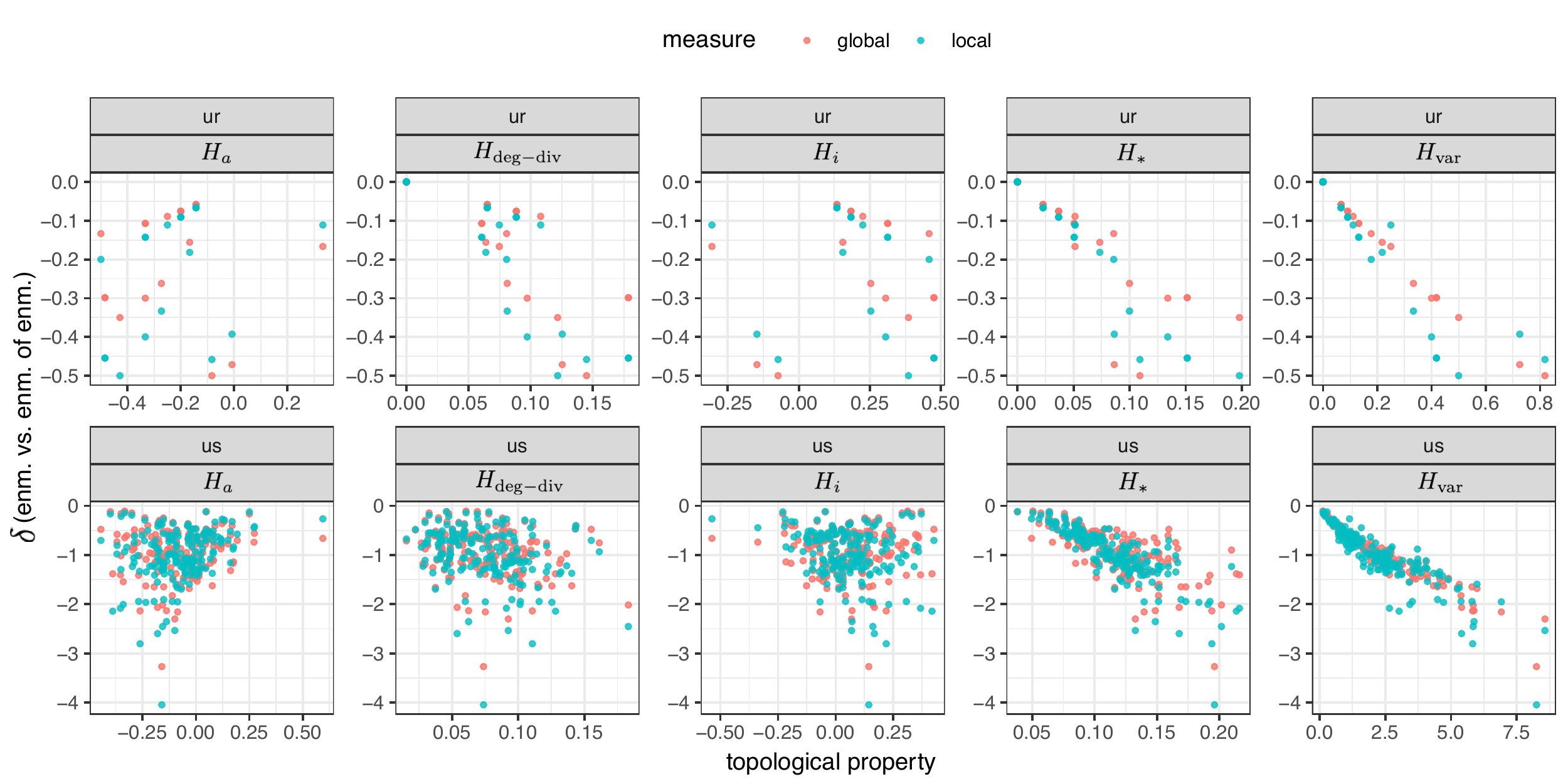}
\caption{The strength of the global and local enmity paradox for undirected networks created by reciprocated (${\rm{ur}}$) and symmetrized edges (${\rm{us}}$) changes with different topological measures.}
\label{fig:en_vs_en_of_en_feats}
\end{figure}

\begin{figure}[t!]
\centering
  \includegraphics[width=1\textwidth]{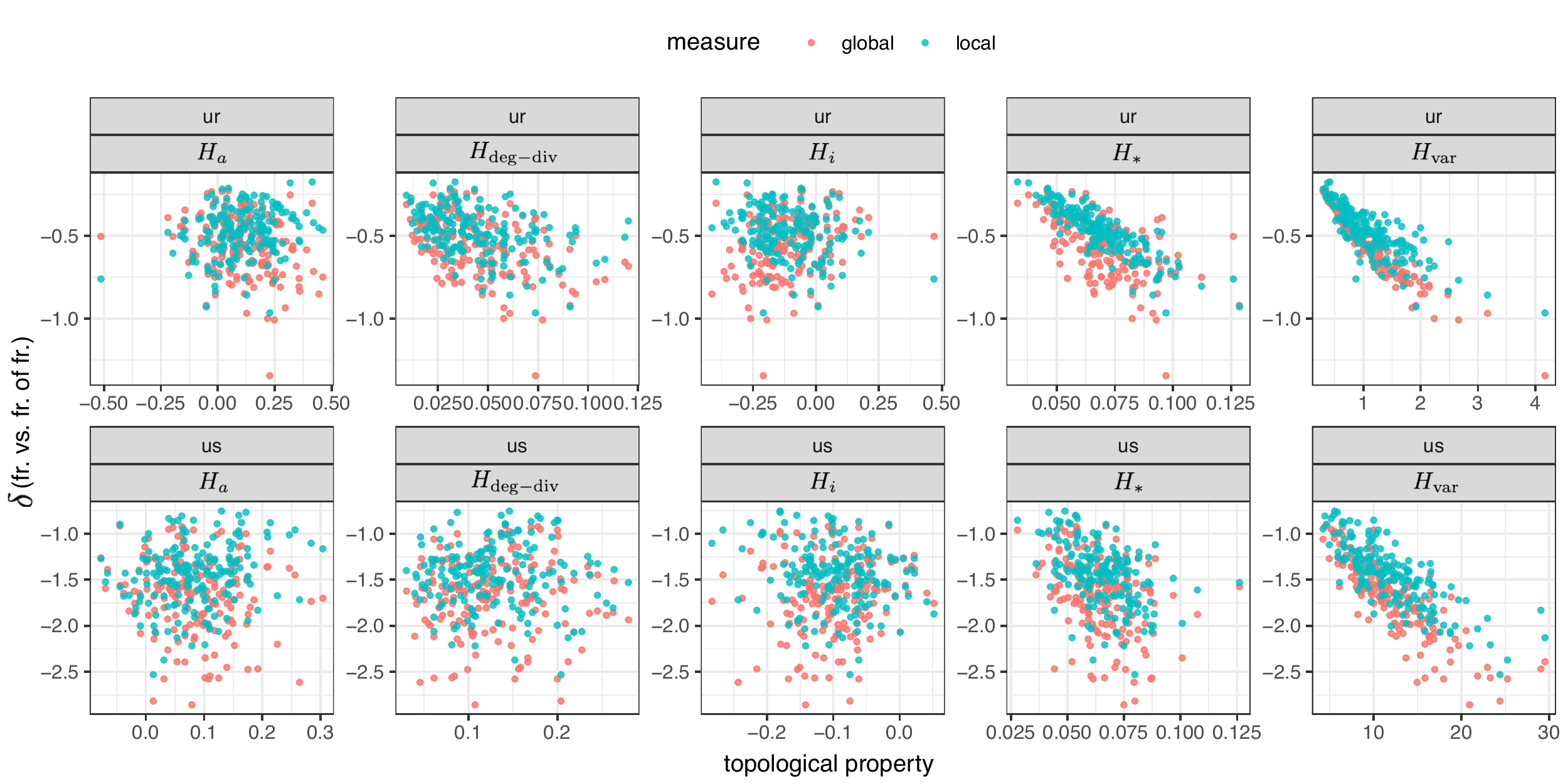}
\caption{The strength of the global and local friendship paradox for undirected networks created by reciprocated (${\rm{ur}}$) and symmetrized edges (${\rm{us}}$) changes with different topological measures.}
\label{fig:fr_vs_fr_of_fr_feats}
\end{figure}

\begin{figure}[t!]
\centering
  \includegraphics[width=0.65\textwidth]{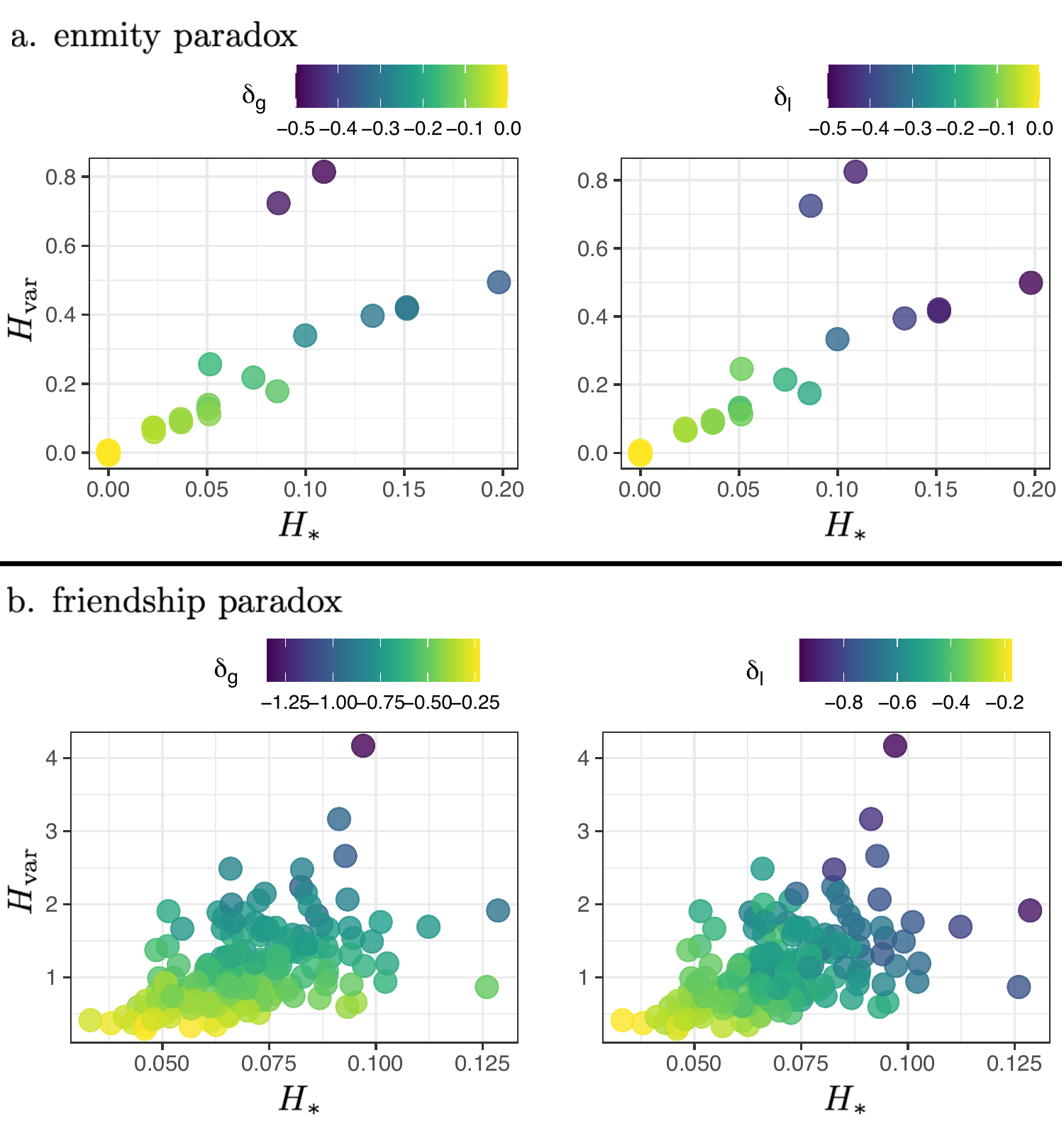}
\caption{Critical features in enmity and friendship paradoxes. The heat maps show the global and local paradox strengths for undirected (reciprocated) networks as a function of two measures: degree variance $H_{\rm{var}}$ and the starlike embedding $H_{\ast}$. The larger the variance and the more the starlike embedding, the stronger the paradox.}
\label{fig:enm_fr_delta_ur_vs_features}
\end{figure}

Figs.~\ref{fig:en_vs_en_of_en_feats}-\ref{fig:fr_vs_fr_of_fr_feats} illustrate how the global and local enmity and friendship paradox strengths for undirected networks created by reciprocated (${\rm{ur}}$) and symmetrized edges (${\rm{us}}$) changes with different topological measures. Among the different topological features, $H_{\ast}$ and $H_{\rm{var}}$ have a negative correlation with the strength of the paradoxes. In these figures, $H_a$ shows a positive correlation with strength, and also $H_i$ and $H_{\rm{deg-div}}$ show a negative correlation with strength. Among the various measures, $H_{\rm{var}}$ and $H_{\ast}$ are the most significant, with large effects for both undirected (symmetrized) and undirected (reciprocated) networks (see Figs. 5 [main text] and~\ref{fig:enm_fr_delta_ur_vs_features}). 
It is expected that the paradoxes become stronger as the variance and starlike embedding increase.


\clearpage
\newpage
\section{Generalized enmity and friendship paradoxes}
\label{sec:gen_enm_fr_par}
\begin{figure}[t!]
\centering
  \includegraphics[width=0.85\textwidth]{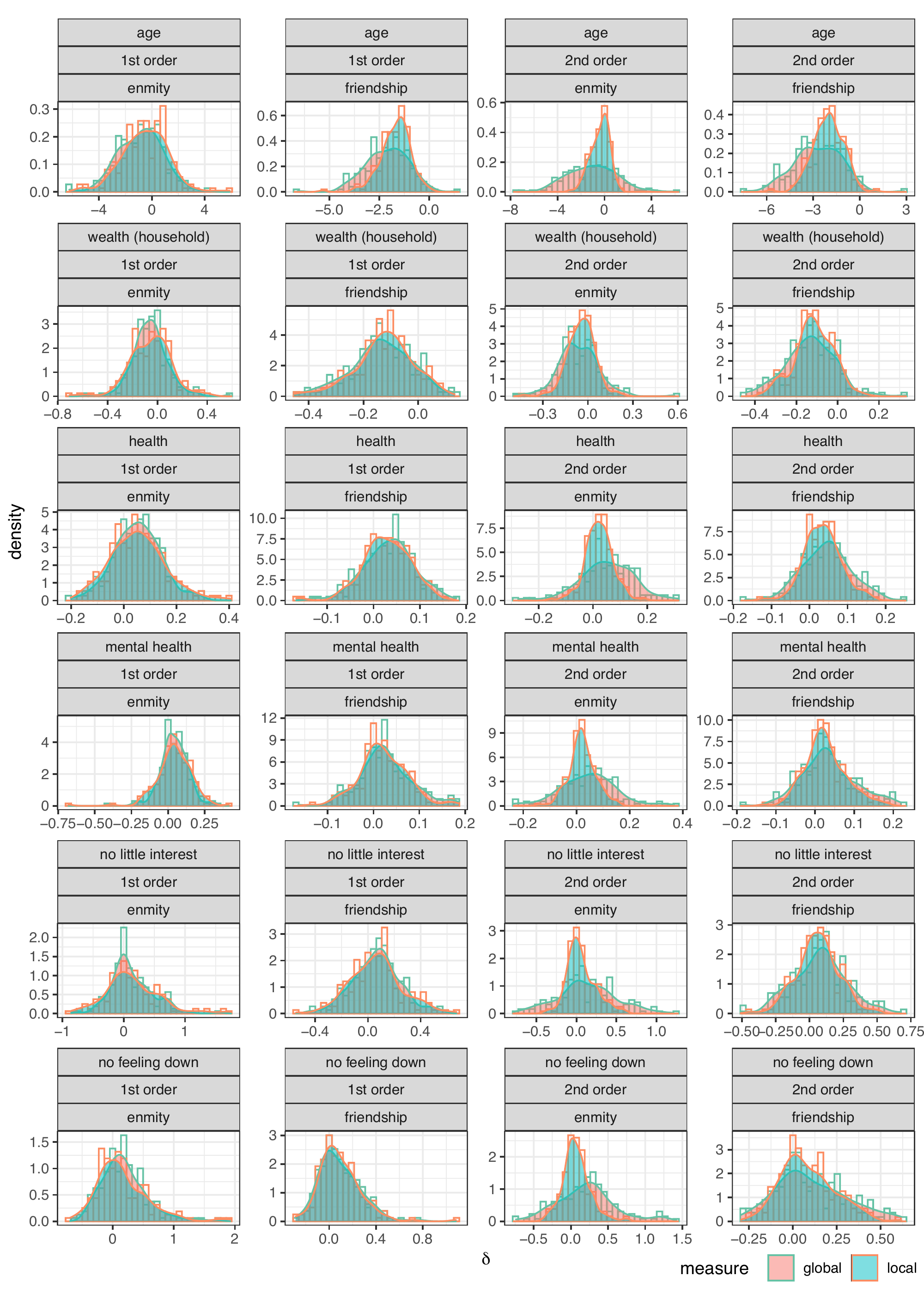}
\caption{Histograms of global and local measures, $\delta_g$ and $\delta_l$, for the generalized enmity and friendship paradoxes in undirected (symmetrized) networks. Here, we provide these measures for both the first and second order neighbors.}
\label{fig:hist_global_local_generalized_paradoxes}
\end{figure}

\begin{figure}[t!]
\centering
  \includegraphics[width=0.95\textwidth]{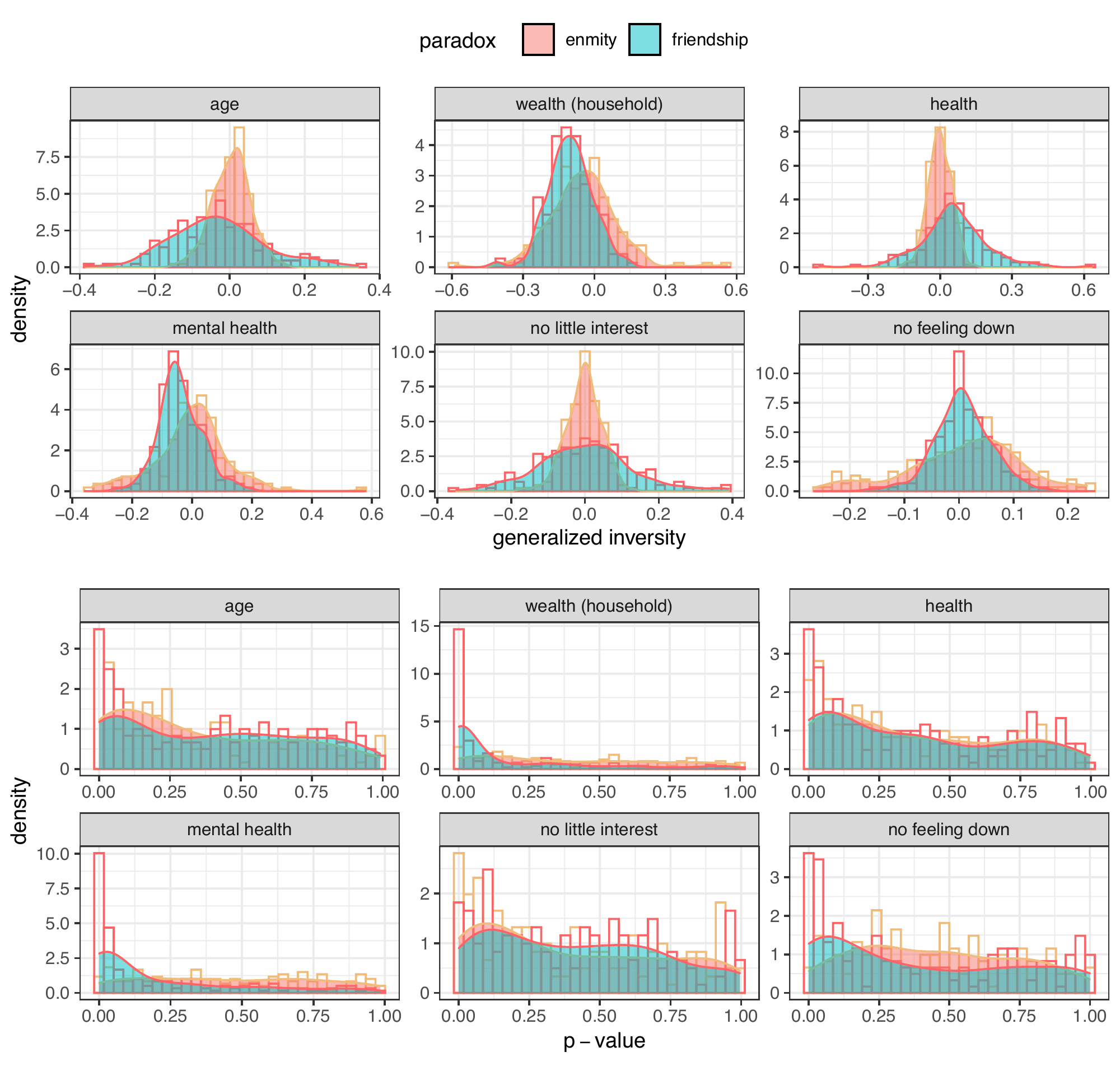}
\caption{The village-level inversity distribution for undirected (symmetrized) networks. 
These inversities are aligned with the inversity measures defined in order to explain the differences between global and local measures of generalized enmity and friendship paradoxes.
The distribution of these correlations over $176$ village networks in the Honduras dataset is represented in the two upper rows, while the p-values of these correlation tests are shown in the two bottom rows.}
\label{fig:generalized_inversity_for_generalized_paradoxes}
\end{figure}

The results for global and local measures of the generalized enmity and friendship paradoxes for undirected (symmetrized) networks are provided in Tables~\ref{tab:gen_en_paradox_us_ur} and~\ref{tab:gen_fr_paradox_us_ur}. The distribution of these measures over $176$ village networks in the Honduras dataset is represented in Fig.~\ref{fig:hist_global_local_generalized_paradoxes}. The difference between the global and local measures can be explained through the inversity measure defined 
as the following correlation:
$\rho_{-w/+w, (x)} = \cor(x_i,\nicefrac{1}{k_{(-/+),j}}|(i,j)\in E_{(-/+)})$, i.e., $\delta_{g,-w/+w}(x)-\delta_{l,-w/+w}(x) \propto \rho_{-w/+w, (x)}$. The distribution of these inversity measures and the p-values for $176$ village networks in the Honduras dataset is provided in Fig.~\ref{fig:generalized_inversity_for_generalized_paradoxes}.
 The results show the generalized paradoxes in both enmity and friendship paradoxes (e.g., wealth). Due to the higher correlation between topological features and positive degrees, generalized friendship paradoxes are stronger than generalized enmity paradoxes. We see an agreement in the existence of the generalized paradoxes for different attributes between the enmity and friendship networks. A further investigation of the generalized enmity paradox is left for future study.

\begin{table}[!ht]
\footnotesize
\centering
\begin{tabular}[t]{lrrrrr}
\toprule
\multicolumn{1}{c}{ } & \multicolumn{1}{c}{random ppl.} & \multicolumn{2}{c}{enmity} & \multicolumn{2}{c}{hypothesis: $\delta_x<0$}\\
\cmidrule(l{2pt}){3-4}\cmidrule(l{2pt}){5-6} 
\multicolumn{1}{c}{  }  & \multicolumn{1}{c}{in enmity} & \multicolumn{1}{c}{enemy of} & \multicolumn{1}{c}{enemy of enemy} & \multicolumn{2}{c}{estimate (significant)} \\
attribute &  network (n = 2857) &  of random ppl. (n = 2816)   & of random ppl. (n = 2816) & (1st order) & (2nd order)    \\
\midrule
\bf{(global)} & ~ & ~ & ~ & ~ & ~\\
~~ age & 33.33 $\pm$ 2.81 & 34.19 $\pm$ 3.34 & 34.41 $\pm$ 3.62 & -0.86 (****) & -1.08 (****)\\
~~ wealth (household) & 2.88 $\pm$ 0.86 & 2.94 $\pm$ 0.88 & 2.95 $\pm$ 0.88 & -0.06 (****) & -0.06 (****)\\
~~ health & 2.68 $\pm$ 0.23 & 2.62 $\pm$ 0.26 & 2.62 $\pm$ 0.27 & 0.05 & 0.05\\
~~ mental health & 2.76 $\pm$ 0.25 & 2.72 $\pm$ 0.29 & 2.72 $\pm$ 0.29 & 0.04 & 0.04\\
~~ no little interest & 11.35 $\pm$ 0.80 & 11.21 $\pm$ 0.94 & 11.22 $\pm$ 0.95 & 0.14 & 0.13\\
~~ no feeling down & 11.58 $\pm$ 0.71 & 11.40 $\pm$ 0.85 & 11.38 $\pm$ 0.86 & 0.19 & 0.21\\

\addlinespace
\bf{(local)} & ~ & ~ & ~ & ~ & ~\\
~~ age & 33.33 $\pm$ 2.81 & 33.85 $\pm$ 3.32 & 33.71 $\pm$ 2.95 & -0.52 (****) & -0.38 (****)\\
~~ wealth (household) & 2.88 $\pm$ 0.86 & 2.94 $\pm$ 0.87 & 2.92 $\pm$ 0.87 & -0.06 (****) & -0.04 (****)\\
~~ health & 2.68 $\pm$ 0.23 & 2.62 $\pm$ 0.26 & 2.65 $\pm$ 0.24 & 0.05 & 0.02\\
~~ mental health & 2.76 $\pm$ 0.25 & 2.72 $\pm$ 0.30 & 2.74 $\pm$ 0.26 & 0.04 & 0.02\\
~~ no little interest & 11.34 $\pm$ 0.80 & 11.21 $\pm$ 1.00 & 11.29 $\pm$ 0.85 & 0.14 & 0.06\\
~~ no feeling down & 11.58 $\pm$ 0.71 & 11.43 $\pm$ 0.86 & 11.51 $\pm$ 0.74 & 0.16 & 0.07\\
\addlinespace
  \bottomrule
  \multicolumn{5}{l}{\scriptsize{$^{(****)}p<0.001$; $^{(***)}p<0.001$; $^{(**)}p<0.01$; $^{(*)}p<0.05$}}
\end{tabular}
\caption{The empirical validation for global and local generalized enmity paradox for undirected (symmetrized) networks. The average of demographic qualities of random people in comparison with the average of demographics for neighbors of random people and neighbors' neighbors of random people. The ``no little interest'' and ``no feeling down'' are measured over the most recent 2 weeks.}
\label{tab:gen_en_paradox_us_ur}
\vspace{-5mm}
\end{table}

\begin{table}[!ht]
\footnotesize
\centering
\begin{tabular}[t]{lrrrrr}
\toprule
\multicolumn{1}{c}{ } & \multicolumn{1}{c}{random ppl.} & \multicolumn{2}{c}{friendship} & \multicolumn{2}{c}{hypothesis: $\delta_x<0$} \\
\cmidrule(l{2pt}){3-4} \cmidrule(l{2pt}){5-6} 
\multicolumn{1}{c}{  }  & \multicolumn{1}{c}{in friendship} & \multicolumn{1}{c}{friend of} & \multicolumn{1}{c}{friend of friend} & \multicolumn{2}{c}{estimate (significant)} \\
attribute &  network (n = 2821) &  of random ppl. (n = 2816)   & of random ppl. (n = 2816) & (1st order) & (2nd order)       \\
\midrule
\bf{global} & ~ & ~ & ~ & ~ & ~\\
~~ age & 32.60 $\pm$ 2.25 & 34.70 $\pm$ 2.67 & 35.30 $\pm$ 2.95 & -2.1 (****) & -2.7 (****)\\
~~ wealth (household) & 2.85 $\pm$ 0.85 & 2.98 $\pm$ 0.89 & 2.98 $\pm$ 0.89 & -0.13 (****) & -0.13 (****)\\
~~ health & 2.73 $\pm$ 0.19 & 2.69 $\pm$ 0.21 & 2.68 $\pm$ 0.21 & 0.03 & 0.04\\
~~ mental health & 2.84 $\pm$ 0.21 & 2.81 $\pm$ 0.23 & 2.81 $\pm$ 0.24 & 0.03 & 0.03\\
~~ no little interest & 11.56 $\pm$ 0.72 & 11.49 $\pm$ 0.74 & 11.48 $\pm$ 0.76 & 0.06 & 0.07\\
~~ no feeling down & 11.82 $\pm$ 0.59 & 11.73 $\pm$ 0.67 & 11.72 $\pm$ 0.68 & 0.09 & 0.1\\

\addlinespace
\bf{(local)} & ~ & ~ & ~ & ~ & ~\\
~~ age & 32.60 $\pm$ 2.25 & 34.29 $\pm$ 2.42 & 34.64 $\pm$ 2.58 & -1.7 (****) & -2.04 (****)\\
~~ wealth (household) & 2.85 $\pm$ 0.85 & 2.97 $\pm$ 0.88 & 2.96 $\pm$ 0.88 & -0.12 (****) & -0.11 (****)\\
~~ health & 2.73 $\pm$ 0.19 & 2.70 $\pm$ 0.21 & 2.69 $\pm$ 0.21 & 0.03 & 0.03\\
~~ mental health & 2.84 $\pm$ 0.21 & 2.82 $\pm$ 0.23 & 2.81 $\pm$ 0.23 & 0.02 & 0.03\\
~~ no little interest & 11.56 $\pm$ 0.72 & 11.50 $\pm$ 0.74 & 11.50 $\pm$ 0.74 & 0.06 & 0.06\\
~~ no feeling down & 11.82 $\pm$ 0.59 & 11.73 $\pm$ 0.66 & 11.74 $\pm$ 0.65 & 0.09 & 0.08\\
\addlinespace
  \bottomrule
  \multicolumn{5}{l}{\scriptsize{$^{(****)}p<0.001$; $^{(***)}p<0.001$; $^{(**)}p<0.01$; $^{(*)}p<0.05$}}
\end{tabular}
\caption{The empirical validation for global and local generalized friendship paradox for undirected (symmetrized) networks. The average of demographic qualities of random people in comparison with the average of demographics for neighbors of random people and neighbors' neighbors of random people. The ``no little interest'' and ``no feeling down'' are measured over the most recent 2 weeks.}
\label{tab:gen_fr_paradox_us_ur}
\vspace{-5mm}
\end{table}

\clearpage
\section{The generalized inversity}
\label{sec:gen_inv}
To find the relationships between the global and local paradox strengths in the mixed worlds, i.e., $\delta_{g,-w}(+) - \delta_{l,-w}(+)$ and $\delta_{g,+w}(-) - \delta_{l,+w}(-)$,
we generalize the inversity measure originally proposed for a friendship network in Ref.~\cite{kumar2021interventions}. 
The difference between $\delta_{g,-w}(+)$ and $\delta_{l,-w}(+)$ can be written as Eq.~\ref{eq:global_local_diff_in_mixed}.

\begin{align}
\label{eq:global_local_diff_in_mixed}
\delta_{g,-w}(+) - \delta_{l,-w}(+) &= \frac{\one^T D_{(-)}^{-1} A_{(-)} D_{(+)}\one\cdot\one^TA_{(-)}\one- \one^TA_{(+)}A_{(-)}\one\cdot\one^T\one}{\one^TA_{(-)}\one \cdot\one^T\one}\, \\
&= \frac{\sum_{(i,j)\in E_{(-)}}\nicefrac{k_{(+),j}}{k_{(-),i}}}{n} - \frac{\sum_i k_{(+),i}k_{(-),i}}{2|E_{(-)}|}
\end{align}

The generalized inversity for the mixed world of $-w$ is defined as the correlation between the positive degree of node $i$ and the inverse negative degree of node $j$ on the two endpoints of a random negative edge $(i,j)\in E_{(-)}$, which is derivable as follows: Using a similar convention as in Ref.~\cite{kumar2021interventions}, we denote the positive degree variable corresponding to one endpoint as $k_{D,(+)}$ and the inverse negative degree of the other endpoint as $k_{ID,(-)}$. The generalized inversity in this mixed world can be written as the following equation,

\begin{align}
    \label{proof:gen_inversity}
    \rho_{\rm{mixed},D(+),ID(-)} &= \frac{1}{2|E_{(-)}|\sigma_{D,(+)}\sigma_{ID,(-)}} \sum_{(i,j)\in E_{(-)}}\left(k_{(+),i} - \mu_{D,(+)}\right)\left(\nicefrac{1}{k_{(-),j}} - \mu_{ID,(-)}\right) \nonumber \\ 
    &= \frac{1}{2|E_{(-)}|\sigma_{D,(+)}\sigma_{ID,(-)}} \sum_{(i,j)\in E_{(-)}}\left(\nicefrac{k_{(+),i}}{k_{(-),j}} - k_{(+),i}\mu_{ID,(-)} - \nicefrac{\mu_{D,(+)}}{k_{(-),j}} + 
    \mu_{D,(+)}\mu_{ID,(-)}\right) \nonumber   \\
    &= \frac{1}{2|E_{(-)}|\sigma_{D,(+)}\sigma_{ID,(-)}} \sum_{(i,j)\in E_{(-)}}\left(\nicefrac{k_{(+),i}}{k_{(-),j}} - k_{(+),i}\mu_{ID,(-)}\right) \nonumber\\
    &= \frac{n}{2|E_{(-)}|\sigma_{D,(+)}\sigma_{ID,(-)}} \left(\frac{\sum_{(i,j)\in E_{(-)}}\nicefrac{k_{(+),i}}{k_{(-),j}}}{n} - \frac{\sum_{i} k_{(+),i}k_{(-),i}}{2|E_{(-)}|}\right) \, ,
\end{align}
where the first two moments of $k_{D,(+)}$ and $k_{ID,(-)}$ in Eq.~\ref{proof:gen_inversity} are computed as follows~\footnote{In Eq. S29 and other equations regarding the generalized inversity for the mixed world of $-w$, $n$ denotes the size of nodes with non-zero negative degrees. For the mixed world of $+w$, $n$ denotes the size of nodes with non-zero positive degrees.}:
\begin{align}
    \mu_{D,(+)} & = \frac{\sum_{(i,j)\in E_{(-)}} k_{(+),i}}{2|E_{(-)}|} = \frac{\sum_{i} k_{(+),i}k_{(-),i}}{2|E_{(-)}|} \\
    \mu_{ID,(-)} & = \frac{\sum_{(i,j)\in E_{(-)}} \nicefrac{1}{k_{(-),j}}}{2|E_{(-)}|} = \frac{n}{2|E_{(-)}|}\\
    \sigma_{D,(+)} & = \frac{\sum_{(i,j)\in E_{(-)}} (k_{(+),i}-\mu_{D,(+)})^2}{2|E_{(-)}|} = \frac{\sum_{i} (k_{(+),i}-\mu_{D,(+)})^2k_{(-),i}}{2|E_{(-)}|}\\
    \sigma_{ID,(-)} & = \frac{\sum_{(i,j)\in E_{(-)}} (\nicefrac{1}{k_{(-),j}}-\mu_{ID,(-)})^2}{2|E_{(-)}|} = \frac{\sum_{j: k_{(-),j}>0}\nicefrac{1}{k_{(-),j}}-2n\mu_{ID,(-)}+2|E_{(-)}|\mu_{ID,(-)}^2}{2|E_{(-)}|}\, .
\end{align}

Therefore, Eq.~\ref{proof:gen_inversity} can be written as Eq.~\ref{Eq:generalized_relation_local_global_strength_nwp},
\begin{align}
\label{Eq:generalized_relation_local_global_strength_nwp}
    \delta_{g,-w}(+) - \delta_{l,-w}(+) &= \rho_{\rm{mixed},D(+),ID(-)} \sigma_{D,(+)}\sigma_{ID,(-)}\overline{ k_{(-)}} \, ,
\end{align}
where, $\overline{ k_{(-)}} = \nicefrac{2|E_{(-)}|}{n}$ and $\rho_{\rm{mixed},D(+),ID(-)} = cor(k_{(+),i}, \nicefrac{1}{k_{(-),j}} | (i,j) \in E_{(-)})$.
Similarly, the difference $\delta_{l,+w}(-)$ and $\delta_{g,+w}(-)$ can be written as Eq.~\ref{eq:global_local_diff_in_mixed_pwn}.

\begin{align}
\label{eq:global_local_diff_in_mixed_pwn}
\delta_{g,+w}(-) - \delta_{l,+w}(-) &= \frac{\one^T D_{(+)}^{-1} A_{(+)} D_{(-)}\one\cdot\one^TA_{(+)}\one- \one^TA_{(+)}A_{(-)}\one\cdot\one^T\one}{\one^TA_{(+)}\one \cdot\one^T\one}\, \\
&= \frac{\sum_{(i,j)\in E_{(+)}}\nicefrac{k_{(-),j}}{k_{(+),i}}}{n} - \frac{\sum_i k_{(+),i}k_{(-),i}}{2|E_{(+)}|}
\end{align}
And, through similar derivation, it can be shown that 
\begin{align}
\label{Eq:generalized_relation_local_global_strength_pwn}
    \delta_{g,+w}(-) - \delta_{l,+w}(-) &= \rho_{\rm{mixed},D(-),ID(+)} \sigma_{D,(-)}\sigma_{ID,(+)} \overline{ k_{(+)}} \, ,
\end{align}
where, $\overline{ k_{(+)}} = \nicefrac{2|E_{(+)}|}{n}$ and $\rho_{\rm{mixed},D(-),ID(+)} = cor(k_{(-),i}, \nicefrac{1}{k_{(+),j}} | (i,j) \in E_{(+)})$.

For directed networks, we can similarly derive the equations corresponding to the inversity measure as follows. We only present the proof for the difference between $\delta_{g, in-w} (out)$ and $\delta_{l, in-w} (out)$. 

\begin{align}
    \label{eq:global_local_diff_in_directed_inwout}
\delta_{g, in-w} (out) - \delta_{l, in-w} (out) &= \frac{\one^T\Din^{-1}A^T\Dout\one\cdot\one^TA\one- \one^TA^TA\one\cdot\one^T\one}{\one^TA\one \cdot\one^T\one}\, \\
&= \frac{\sum_{(i,j)\in E}\nicefrac{k_{i, out}}{k_{j, in}}}{n} - \frac{\sum_i k_{i, out}^2}{|E|}
\end{align}

The inversity in this scenario is defined as the correlation between the out-degree of node $i$ and the inverse in-degree of node $j$ on the two endpoints of a random edge $(i,j)\in E$, which can be derived as follows. We denote the out-degree variable corresponding to one endpoint as $k_{D,out}$ and the inverse in-degree of the other endpoint as $k_{ID,in}$. The inversity measure can be expanded for this directed network via the following equation,

\begin{align}
    \label{proof:gen_inversity_directed}
    \rho_{\rm{mixed},D(out),ID(in)} &= \frac{1}{|E|\sigma_{D,(out)}\sigma_{ID,(in)}} \sum_{(i,j)\in E}\left(k_{i, out} - \mu_{D,(out)}\right)\left(\nicefrac{1}{k_{j, in}} - \mu_{ID,(in)}\right) \nonumber \\ 
    &= \frac{1}{|E|\sigma_{D,(out)}\sigma_{ID,(in)}} \sum_{(i,j)\in E}\left(\nicefrac{k_{i, out}}{k_{j, in}} - k_{i, out}\mu_{ID,(in)} - \nicefrac{\mu_{D,(out)}}{k_{j, in}} + 
    \mu_{D,(out)}\mu_{ID,(in)}\right) \nonumber   \\
    &= \frac{1}{|E|\sigma_{D,(out)}\sigma_{ID,(in)}} \sum_{(i,j)\in E}\left(\nicefrac{k_{i, out}}{k_{j, in}} - k_{i, out}\mu_{ID,(in)}\right) \nonumber\\
    &= \frac{n}{|E|\sigma_{D,(out)}\sigma_{ID,(in)}} \left(\frac{\sum_{(i,j)\in E}\nicefrac{k_{i, out}}{k_{j, in}}}{n} - \frac{\sum_{i: k_{i, in}>0} k_{i, out}k_{i, in}}{|E|}\right) \, ,
\end{align}
where the first two moments of $k_{D,(out)}$ and $k_{ID,(in)}$ in Eq.~\ref{proof:gen_inversity_directed} are computed as follows:
\begin{align}
    \mu_{D,(out)} & = \frac{\sum_{(i,j)\in E} k_{i, out}}{|E|} = \frac{\sum_{i} k_{i, out}^2}{|E|} \\
    \mu_{ID,(in)} & = \frac{\sum_{(i,j)\in E} \nicefrac{1}{k_{j, in}}}{|E|} = \frac{n}{|E|}\\
    \sigma_{D,(out)} & = \frac{\sum_{(i,j)\in E} (k_{i, out}-\mu_{D,(out)})^2}{|E|} = \frac{\sum_{i} (k_{i, out}-\mu_{D,(out)})^2 k_{i,out}}{|E|}\\
    \sigma_{ID,(in)} & = \frac{\sum_{(i,j)\in E} (\nicefrac{1}{k_{j, in}}-\mu_{ID,(in)})^2}{|E|} = \frac{\sum_{j: k_{j, in}>0}\nicefrac{1}{k_{j, in}}-2n\mu_{ID,(in)}+2|E|\mu_{ID,(in)}^2}{|E|}\, .
\end{align}

\begin{align}
\label{Eq:generalized_relation_local_global_strength_directed_inwout}
    \delta_{g, in-w} (out) - \delta_{l, in-w} (out) &= \rho_{\rm{mixed},D(out),ID(in)} \sigma_{D,(out)}\sigma_{ID,(in)} \overline{k} \, ,
\end{align}
where, $\overline{k} = \nicefrac{2|E|}{n}$ and $\rho_{\rm{mixed},D(out),ID(in)} = cor(k_{i,out}, \nicefrac{1}{k_{j,in}} | (i,j) \in E)$.

\clearpage
\newpage
\begin{figure}[t!]
\centering
  \includegraphics[width=0.75\textwidth]{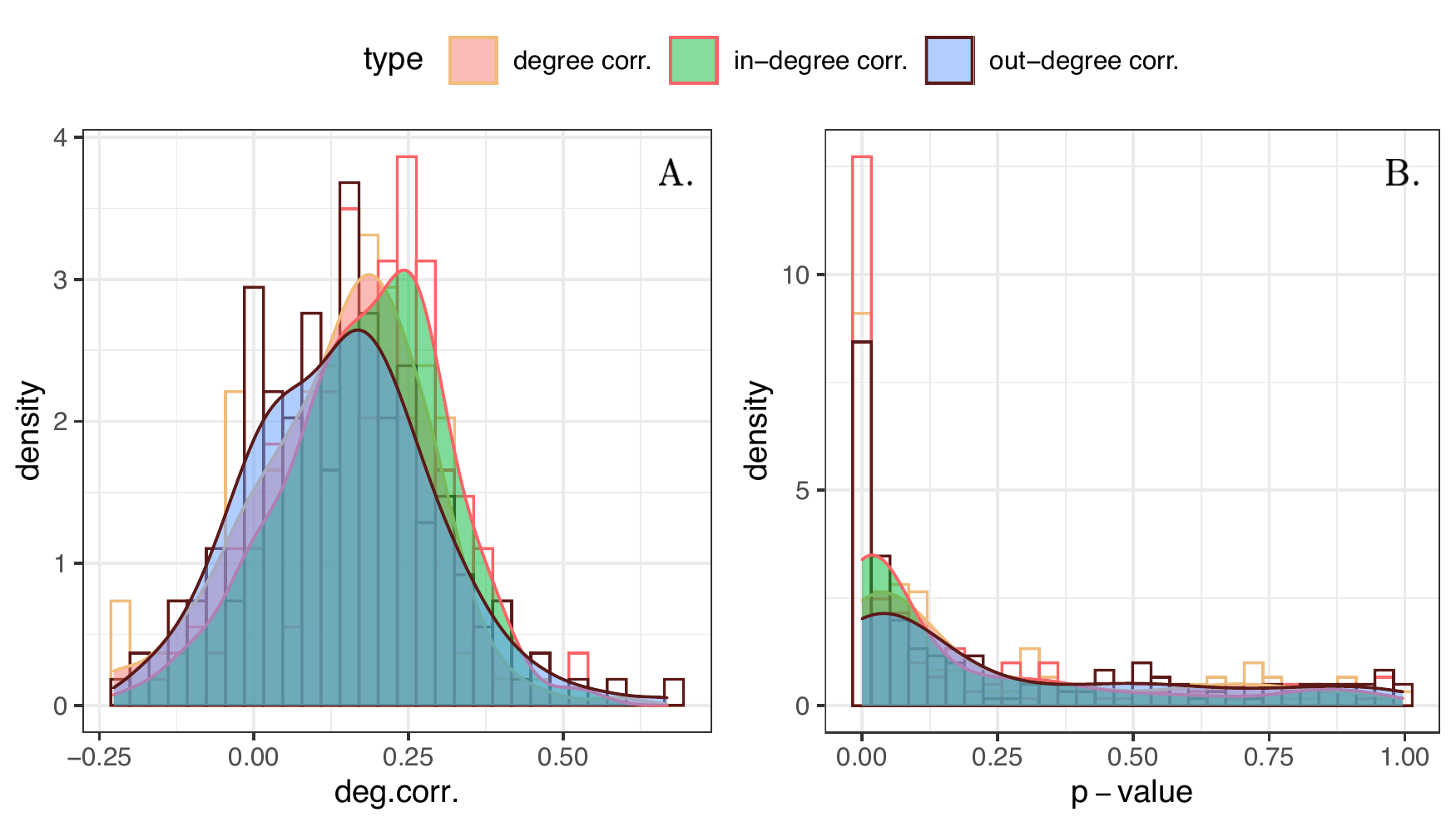}
\caption{Correlation between positive and negative degree. The distribution of correlation between positive and negative degrees over $176$ village networks in the Honduras dataset is represented in panel A, while the p-values of these correlation tests for these networks are represented in panel B.}
\label{fig:pos-neg-corr}
\end{figure}

\begin{figure}[t!]
\centering
  \includegraphics[width=0.75\textwidth]{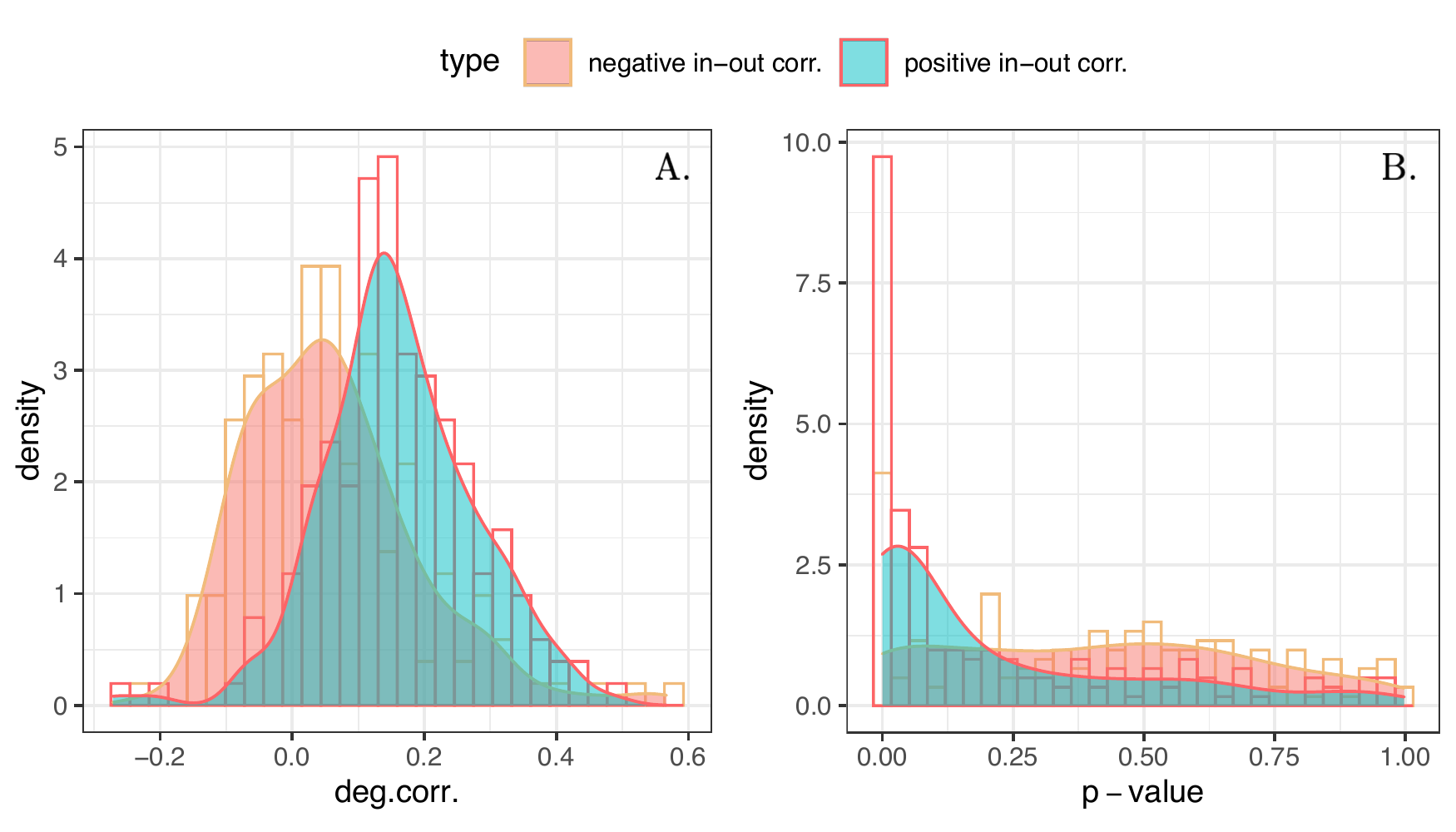}
\caption{Correlation between in- and out-degrees for positive and negative networks. The distribution of correlation between in- and out-degrees over $176$ enmity and friendship village networks in the Honduras dataset is represented in panel A, while the p-values of these correlation tests for these networks are represented in panel B.}
\label{fig:in-out-deg-corr_pos_neg}
\end{figure}

\begin{figure}[t!]
\centering
  \includegraphics[width=0.75\textwidth]{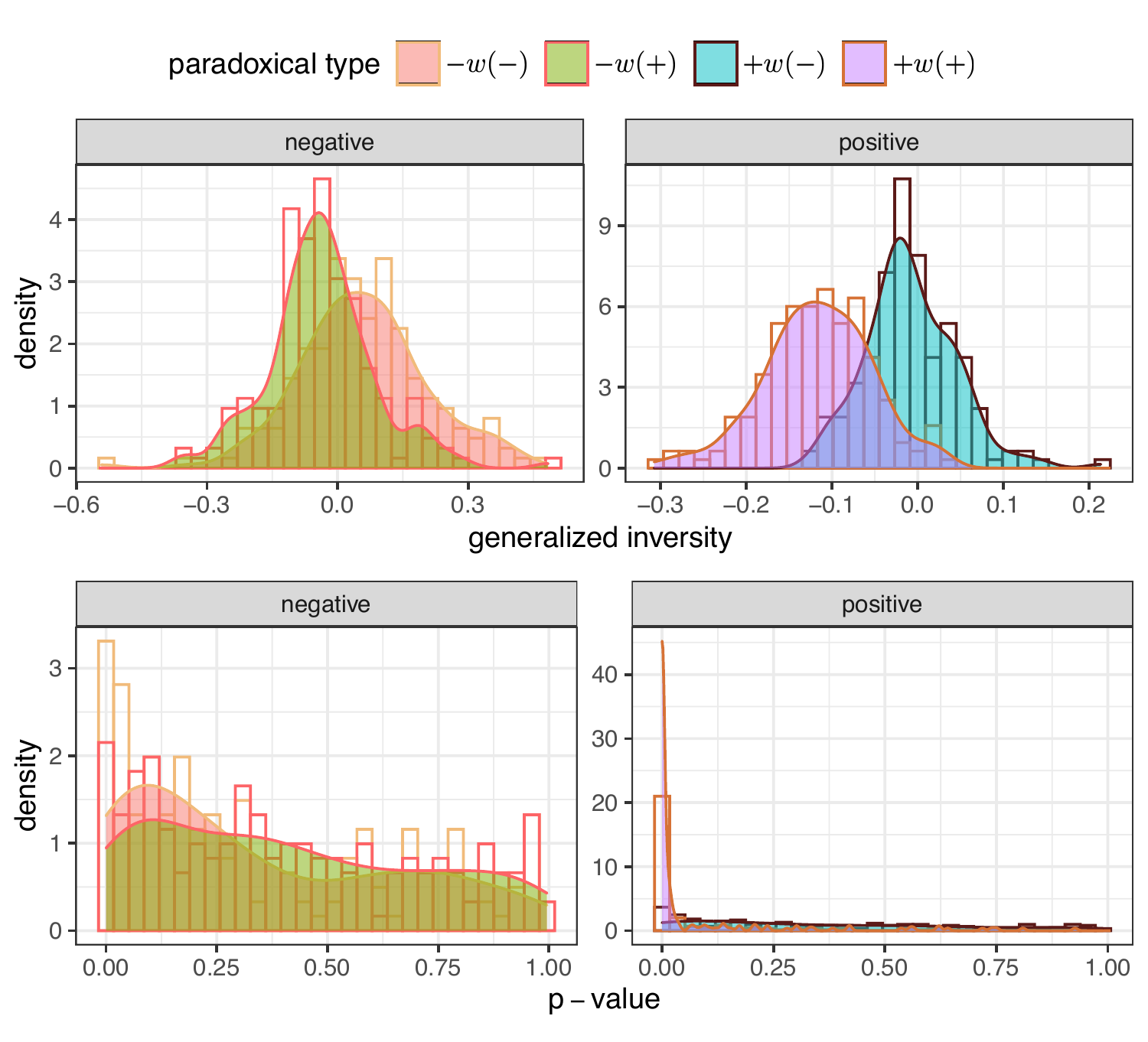}
\caption{The village-level inversity distribution for undirected (symmetrized) networks. Here, we use the same notation we used in the main text. In this notation, the $-w$ and $+w$ indicate the type of one's neighbor, as one's enemy and friend, respectively. The $(-)$ and $(+)$ denote the type of comparison as enemies or friends.
These inversities are aligned with four cases: the enmity paradox in the negative world  $-w(-)$, the friendship paradox in the positive world $+w(+)$, and the two mixed worlds explained in the main text $-w(+)$/$+w(-)$.
The distribution of these correlations over $176$ village networks in the Honduras dataset is represented in the upper row, while the p-values of these correlation tests for these networks are represented in the bottom row. These (correlations, p-values) for the whole dataset for the above four cases can be summarized as $(-0.09,<2.2e^{-16})$, $(-0.21, <2.2e^{-16})$ for enmity $-w(-)$ and friendship $+w(+)$ paradoxes, and $(-0.09,<2.2e^{-16})$/$(-0.05, <2.2e^{-16})$ for the two mixed worlds, when we compare one's number of friends/enemies with the number of friends/enemies of their enemies/friends $-w(+)$/$+w(-)$. For the friendship paradox $+w(+)$, the distribution of inversity is predominantly distributed over negative values, which results in $\delta_g<\delta_l$, whereas the other distributions are around $0$, making the local and global strengths almost identical (main text, Fig. 4).
}
\label{fig:undirected_inversity}
\end{figure}

\begin{figure}[t!]
\centering
  \includegraphics[width=0.75\textwidth]{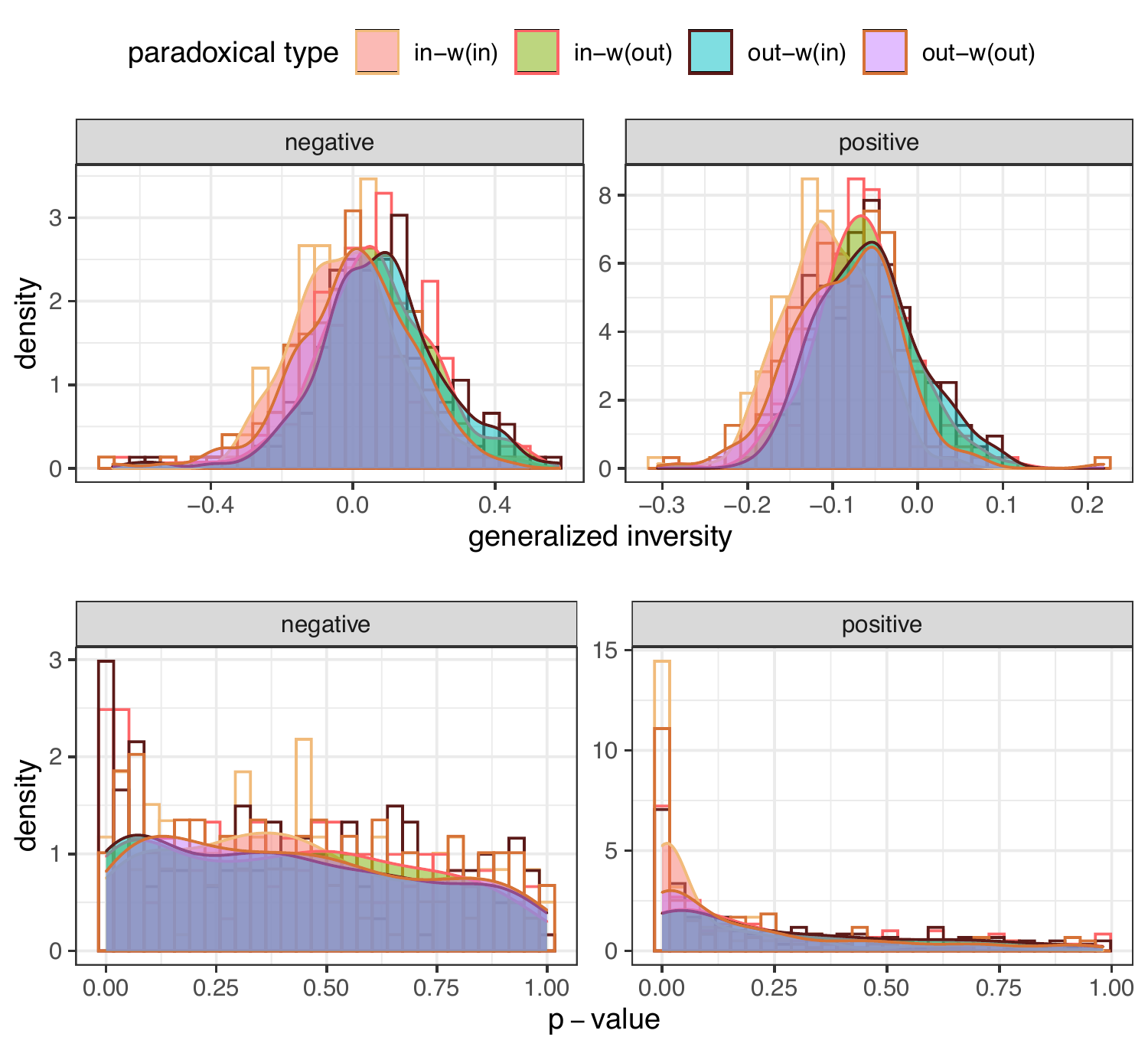}
\caption{The village-level inversities associated with four cases of enmity and friendship paradoxes in directed networks.
These inversities are aligned with the four cases introduced in Section~\ref{sec:enm_par_dir_nets}, i.e., $cor(k_{i, out}, \nicefrac{1}{k_{j, in}} | (i,j) \in E)$, 
$cor(\nicefrac{1}{k_{i, out}}, k_{j, in} | (i,j) \in E)$, $cor(\nicefrac{1}{k_{i, out}}, k_{j, out} | (i,j) \in E)$, and 
$cor(k_{i, in}, \nicefrac{1}{k_{j, in}} | (i,j) \in E)$ in order.
The distribution of these correlations over $176$ enmity and friendship village networks in the Honduras dataset is represented in the upper row, while the p-values of these correlation tests for these networks are represented in the bottom row. These (correlations, p-values) for the whole dataset can be summarized as $(-0.03, 4.2e^{-5})$, $(-0.01, 0.2)$, $(-0.06, 2.32e^{-12})$, and $(-0.09, <2.2e^{-16})$ for enmity world and $(-0.15,<2.2e^{-16})$, $(-0.14,<2.2e^{-16})$, $(-0.19,<2.2e^{-16})$, and $(-0.16,<2.2e^{-16})$ for friendship world.}
\label{fig:inversity_directed}
\end{figure}

\clearpage
\newpage

%


\begin{thebibliography}{32}%
\makeatletter
\providecommand \@ifxundefined [1]{%
 \@ifx{#1\undefined}
}%
\providecommand \@ifnum [1]{%
 \ifnum #1\expandafter \@firstoftwo
 \else \expandafter \@secondoftwo
 \fi
}%
\providecommand \@ifx [1]{%
 \ifx #1\expandafter \@firstoftwo
 \else \expandafter \@secondoftwo
 \fi
}%
\providecommand \natexlab [1]{#1}%
\providecommand \enquote  [1]{``#1''}%
\providecommand \bibnamefont  [1]{#1}%
\providecommand \bibfnamefont [1]{#1}%
\providecommand \citenamefont [1]{#1}%
\providecommand \href@noop [0]{\@secondoftwo}%
\providecommand \href [0]{\begingroup \@sanitize@url \@href}%
\providecommand \@href[1]{\@@startlink{#1}\@@href}%
\providecommand \@@href[1]{\endgroup#1\@@endlink}%
\providecommand \@sanitize@url [0]{\catcode `\\12\catcode `\$12\catcode
  `\&12\catcode `\#12\catcode `\^12\catcode `\_12\catcode `\%12\relax}%
\providecommand \@@startlink[1]{}%
\providecommand \@@endlink[0]{}%
\providecommand \url  [0]{\begingroup\@sanitize@url \@url }%
\providecommand \@url [1]{\endgroup\@href {#1}{\urlprefix }}%
\providecommand \urlprefix  [0]{URL }%
\providecommand \Eprint [0]{\href }%
\providecommand \doibase [0]{https://doi.org/}%
\providecommand \selectlanguage [0]{\@gobble}%
\providecommand \bibinfo  [0]{\@secondoftwo}%
\providecommand \bibfield  [0]{\@secondoftwo}%
\providecommand \translation [1]{[#1]}%
\providecommand \BibitemOpen [0]{}%
\providecommand \bibitemStop [0]{}%
\providecommand \bibitemNoStop [0]{.\EOS\space}%
\providecommand \EOS [0]{\spacefactor3000\relax}%
\providecommand \BibitemShut  [1]{\csname bibitem#1\endcsname}%
\let\auto@bib@innerbib\@empty
\bibitem [{\citenamefont {Feld}(1991)}]{feld1991your}%
  \BibitemOpen
  \bibfield  {author} {\bibinfo {author} {\bibfnamefont {S.~L.}\ \bibnamefont
  {Feld}},\ }\bibfield  {title} {\bibinfo {title} {Why your friends have more
  friends than you do},\ }\href@noop {} {\bibfield  {journal} {\bibinfo
  {journal} {American journal of sociology}\ }\textbf {\bibinfo {volume}
  {96}},\ \bibinfo {pages} {1464} (\bibinfo {year} {1991})}\BibitemShut
  {NoStop}%
\bibitem [{\citenamefont {Eom}\ and\ \citenamefont
  {Jo}(2014)}]{eom2014generalized}%
  \BibitemOpen
  \bibfield  {author} {\bibinfo {author} {\bibfnamefont {Y.-H.}\ \bibnamefont
  {Eom}}\ and\ \bibinfo {author} {\bibfnamefont {H.-H.}\ \bibnamefont {Jo}},\
  }\bibfield  {title} {\bibinfo {title} {Generalized friendship paradox in
  complex networks: The case of scientific collaboration},\ }\href@noop {}
  {\bibfield  {journal} {\bibinfo  {journal} {Scientific reports}\ }\textbf
  {\bibinfo {volume} {4}},\ \bibinfo {pages} {1} (\bibinfo {year}
  {2014})}\BibitemShut {NoStop}%
\bibitem [{\citenamefont {Nettasinghe}\ and\ \citenamefont
  {Krishnamurthy}(2019)}]{nettasinghe2019friendship}%
  \BibitemOpen
  \bibfield  {author} {\bibinfo {author} {\bibfnamefont {B.}~\bibnamefont
  {Nettasinghe}}\ and\ \bibinfo {author} {\bibfnamefont {V.}~\bibnamefont
  {Krishnamurthy}},\ }\bibfield  {title} {\bibinfo {title} {The friendship
  paradox: Implications in statistical inference of social networks},\ }in\
  \href@noop {} {\emph {\bibinfo {booktitle} {2019 IEEE 29th International
  Workshop on Machine Learning for Signal Processing (MLSP)}}}\ (\bibinfo
  {organization} {IEEE},\ \bibinfo {year} {2019})\ pp.\ \bibinfo {pages}
  {1--6}\BibitemShut {NoStop}%
\bibitem [{\citenamefont {Grund}(2014)}]{grund2014your}%
  \BibitemOpen
  \bibfield  {author} {\bibinfo {author} {\bibfnamefont {T.}~\bibnamefont
  {Grund}},\ }\bibfield  {title} {\bibinfo {title} {Why your friends are more
  important and special than you think},\ }\href@noop {} {\bibfield  {journal}
  {\bibinfo  {journal} {Sociological Science}\ }\textbf {\bibinfo {volume}
  {1}},\ \bibinfo {pages} {128} (\bibinfo {year} {2014})}\BibitemShut {NoStop}%
\bibitem [{\citenamefont {Higham}(2019)}]{higham2019centrality}%
  \BibitemOpen
  \bibfield  {author} {\bibinfo {author} {\bibfnamefont {D.~J.}\ \bibnamefont
  {Higham}},\ }\bibfield  {title} {\bibinfo {title} {Centrality-friendship
  paradoxes: when our friends are more important than us},\ }\href@noop {}
  {\bibfield  {journal} {\bibinfo  {journal} {Journal of Complex Networks}\ }
  (\bibinfo {year} {2019})}\BibitemShut {NoStop}%
\bibitem [{\citenamefont {Alipourfard}\ \emph {et~al.}(2020)\citenamefont
  {Alipourfard}, \citenamefont {Nettasinghe}, \citenamefont {Abeliuk},
  \citenamefont {Krishnamurthy},\ and\ \citenamefont
  {Lerman}}]{alipourfard2020friendship}%
  \BibitemOpen
  \bibfield  {author} {\bibinfo {author} {\bibfnamefont {N.}~\bibnamefont
  {Alipourfard}}, \bibinfo {author} {\bibfnamefont {B.}~\bibnamefont
  {Nettasinghe}}, \bibinfo {author} {\bibfnamefont {A.}~\bibnamefont
  {Abeliuk}}, \bibinfo {author} {\bibfnamefont {V.}~\bibnamefont
  {Krishnamurthy}},\ and\ \bibinfo {author} {\bibfnamefont {K.}~\bibnamefont
  {Lerman}},\ }\bibfield  {title} {\bibinfo {title} {Friendship paradox biases
  perceptions in directed networks},\ }\href@noop {} {\bibfield  {journal}
  {\bibinfo  {journal} {Nature communications}\ }\textbf {\bibinfo {volume}
  {11}},\ \bibinfo {pages} {1} (\bibinfo {year} {2020})}\BibitemShut {NoStop}%
\bibitem [{\citenamefont {Galesic}\ \emph {et~al.}(2021)\citenamefont
  {Galesic}, \citenamefont {Bruine~de Bruin}, \citenamefont {Dalege},
  \citenamefont {Feld}, \citenamefont {Kreuter}, \citenamefont {Olsson},
  \citenamefont {Prelec}, \citenamefont {Stein},\ and\ \citenamefont {van
  Der~Does}}]{galesic2021human}%
  \BibitemOpen
  \bibfield  {author} {\bibinfo {author} {\bibfnamefont {M.}~\bibnamefont
  {Galesic}}, \bibinfo {author} {\bibfnamefont {W.}~\bibnamefont {Bruine~de
  Bruin}}, \bibinfo {author} {\bibfnamefont {J.}~\bibnamefont {Dalege}},
  \bibinfo {author} {\bibfnamefont {S.~L.}\ \bibnamefont {Feld}}, \bibinfo
  {author} {\bibfnamefont {F.}~\bibnamefont {Kreuter}}, \bibinfo {author}
  {\bibfnamefont {H.}~\bibnamefont {Olsson}}, \bibinfo {author} {\bibfnamefont
  {D.}~\bibnamefont {Prelec}}, \bibinfo {author} {\bibfnamefont {D.~L.}\
  \bibnamefont {Stein}},\ and\ \bibinfo {author} {\bibfnamefont
  {T.}~\bibnamefont {van Der~Does}},\ }\bibfield  {title} {\bibinfo {title}
  {Human social sensing is an untapped resource for computational social
  science},\ }\href@noop {} {\bibfield  {journal} {\bibinfo  {journal}
  {Nature}\ }\textbf {\bibinfo {volume} {595}},\ \bibinfo {pages} {214}
  (\bibinfo {year} {2021})}\BibitemShut {NoStop}%
\bibitem [{\citenamefont {Christakis}\ and\ \citenamefont
  {Fowler}(2010)}]{christakis2010social}%
  \BibitemOpen
  \bibfield  {author} {\bibinfo {author} {\bibfnamefont {N.~A.}\ \bibnamefont
  {Christakis}}\ and\ \bibinfo {author} {\bibfnamefont {J.~H.}\ \bibnamefont
  {Fowler}},\ }\bibfield  {title} {\bibinfo {title} {Social network sensors for
  early detection of contagious outbreaks},\ }\href@noop {} {\bibfield
  {journal} {\bibinfo  {journal} {PloS one}\ }\textbf {\bibinfo {volume} {5}},\
  \bibinfo {pages} {e12948} (\bibinfo {year} {2010})}\BibitemShut {NoStop}%
\bibitem [{\citenamefont {Garcia-Herranz}\ \emph {et~al.}(2014)\citenamefont
  {Garcia-Herranz}, \citenamefont {Moro}, \citenamefont {Cebrian},
  \citenamefont {Christakis},\ and\ \citenamefont {Fowler}}]{garcia2014using}%
  \BibitemOpen
  \bibfield  {author} {\bibinfo {author} {\bibfnamefont {M.}~\bibnamefont
  {Garcia-Herranz}}, \bibinfo {author} {\bibfnamefont {E.}~\bibnamefont
  {Moro}}, \bibinfo {author} {\bibfnamefont {M.}~\bibnamefont {Cebrian}},
  \bibinfo {author} {\bibfnamefont {N.~A.}\ \bibnamefont {Christakis}},\ and\
  \bibinfo {author} {\bibfnamefont {J.~H.}\ \bibnamefont {Fowler}},\ }\bibfield
   {title} {\bibinfo {title} {Using friends as sensors to detect global-scale
  contagious outbreaks},\ }\href@noop {} {\bibfield  {journal} {\bibinfo
  {journal} {PloS one}\ }\textbf {\bibinfo {volume} {9}},\ \bibinfo {pages}
  {e92413} (\bibinfo {year} {2014})}\BibitemShut {NoStop}%
\bibitem [{\citenamefont {Kim}\ \emph {et~al.}(2015)\citenamefont {Kim},
  \citenamefont {Hwong}, \citenamefont {Stafford}, \citenamefont {Hughes},
  \citenamefont {O'Malley}, \citenamefont {Fowler},\ and\ \citenamefont
  {Christakis}}]{kim2015social}%
  \BibitemOpen
  \bibfield  {author} {\bibinfo {author} {\bibfnamefont {D.~A.}\ \bibnamefont
  {Kim}}, \bibinfo {author} {\bibfnamefont {A.~R.}\ \bibnamefont {Hwong}},
  \bibinfo {author} {\bibfnamefont {D.}~\bibnamefont {Stafford}}, \bibinfo
  {author} {\bibfnamefont {D.~A.}\ \bibnamefont {Hughes}}, \bibinfo {author}
  {\bibfnamefont {A.~J.}\ \bibnamefont {O'Malley}}, \bibinfo {author}
  {\bibfnamefont {J.~H.}\ \bibnamefont {Fowler}},\ and\ \bibinfo {author}
  {\bibfnamefont {N.~A.}\ \bibnamefont {Christakis}},\ }\bibfield  {title}
  {\bibinfo {title} {Social network targeting to maximise population behaviour
  change: a cluster randomised controlled trial},\ }\href@noop {} {\bibfield
  {journal} {\bibinfo  {journal} {The Lancet}\ }\textbf {\bibinfo {volume}
  {386}},\ \bibinfo {pages} {145} (\bibinfo {year} {2015})}\BibitemShut
  {NoStop}%
\bibitem [{\citenamefont {Shakya}\ \emph {et~al.}(2017)\citenamefont {Shakya},
  \citenamefont {Stafford}, \citenamefont {Hughes}, \citenamefont {Keegan},
  \citenamefont {Negron}, \citenamefont {Broome}, \citenamefont {McKnight},
  \citenamefont {Nicoll}, \citenamefont {Nelson}, \citenamefont {Iriarte} \emph
  {et~al.}}]{shakya2017exploiting}%
  \BibitemOpen
  \bibfield  {author} {\bibinfo {author} {\bibfnamefont {H.~B.}\ \bibnamefont
  {Shakya}}, \bibinfo {author} {\bibfnamefont {D.}~\bibnamefont {Stafford}},
  \bibinfo {author} {\bibfnamefont {D.~A.}\ \bibnamefont {Hughes}}, \bibinfo
  {author} {\bibfnamefont {T.}~\bibnamefont {Keegan}}, \bibinfo {author}
  {\bibfnamefont {R.}~\bibnamefont {Negron}}, \bibinfo {author} {\bibfnamefont
  {J.}~\bibnamefont {Broome}}, \bibinfo {author} {\bibfnamefont
  {M.}~\bibnamefont {McKnight}}, \bibinfo {author} {\bibfnamefont
  {L.}~\bibnamefont {Nicoll}}, \bibinfo {author} {\bibfnamefont
  {J.}~\bibnamefont {Nelson}}, \bibinfo {author} {\bibfnamefont
  {E.}~\bibnamefont {Iriarte}}, \emph {et~al.},\ }\bibfield  {title} {\bibinfo
  {title} {Exploiting social influence to magnify population-level behaviour
  change in maternal and child health: study protocol for a randomised
  controlled trial of network targeting algorithms in rural honduras},\
  }\href@noop {} {\bibfield  {journal} {\bibinfo  {journal} {BMJ open}\
  }\textbf {\bibinfo {volume} {7}},\ \bibinfo {pages} {e012996} (\bibinfo
  {year} {2017})}\BibitemShut {NoStop}%
\bibitem [{\citenamefont {Kumar}\ \emph {et~al.}(2021)\citenamefont {Kumar},
  \citenamefont {Krackhardt},\ and\ \citenamefont
  {Feld}}]{kumar2021interventions}%
  \BibitemOpen
  \bibfield  {author} {\bibinfo {author} {\bibfnamefont {V.}~\bibnamefont
  {Kumar}}, \bibinfo {author} {\bibfnamefont {D.}~\bibnamefont {Krackhardt}},\
  and\ \bibinfo {author} {\bibfnamefont {S.}~\bibnamefont {Feld}},\ }\bibfield
  {title} {\bibinfo {title} {Interventions with inversity in unknown networks
  can help regulate contagion},\ }\href@noop {} {\bibfield  {journal} {\bibinfo
   {journal} {preprint arXiv:2105.08758}\ } (\bibinfo {year}
  {2021})}\BibitemShut {NoStop}%
\bibitem [{\citenamefont {Alexander}\ \emph {et~al.}(2022)\citenamefont
  {Alexander}, \citenamefont {Forastiere}, \citenamefont {Gupta},\ and\
  \citenamefont {Christakis}}]{alexander2022algorithms}%
  \BibitemOpen
  \bibfield  {author} {\bibinfo {author} {\bibfnamefont {M.}~\bibnamefont
  {Alexander}}, \bibinfo {author} {\bibfnamefont {L.}~\bibnamefont
  {Forastiere}}, \bibinfo {author} {\bibfnamefont {S.}~\bibnamefont {Gupta}},\
  and\ \bibinfo {author} {\bibfnamefont {N.~A.}\ \bibnamefont {Christakis}},\
  }\bibfield  {title} {\bibinfo {title} {Algorithms for seeding social networks
  can enhance the adoption of a public health intervention in urban india},\
  }\href@noop {} {\bibfield  {journal} {\bibinfo  {journal} {Proceedings of the
  National Academy of Sciences}\ }\textbf {\bibinfo {volume} {119}},\ \bibinfo
  {pages} {e2120742119} (\bibinfo {year} {2022})}\BibitemShut {NoStop}%
\bibitem [{\citenamefont {Cantwell}\ \emph {et~al.}(2021)\citenamefont
  {Cantwell}, \citenamefont {Kirkley},\ and\ \citenamefont
  {Newman}}]{cantwell2021friendship}%
  \BibitemOpen
  \bibfield  {author} {\bibinfo {author} {\bibfnamefont {G.~T.}\ \bibnamefont
  {Cantwell}}, \bibinfo {author} {\bibfnamefont {A.}~\bibnamefont {Kirkley}},\
  and\ \bibinfo {author} {\bibfnamefont {M.~E.~J.}\ \bibnamefont {Newman}},\
  }\bibfield  {title} {\bibinfo {title} {The friendship paradox in real and
  model networks},\ }\href@noop {} {\bibfield  {journal} {\bibinfo  {journal}
  {Journal of Complex Networks}\ }\textbf {\bibinfo {volume} {9}},\ \bibinfo
  {pages} {cnab011} (\bibinfo {year} {2021})}\BibitemShut {NoStop}%
\bibitem [{\citenamefont {Shirado}\ \emph {et~al.}(2019)\citenamefont
  {Shirado}, \citenamefont {Iosifidis}, \citenamefont {Tassiulas},\ and\
  \citenamefont {Christakis}}]{shirado2019resource}%
  \BibitemOpen
  \bibfield  {author} {\bibinfo {author} {\bibfnamefont {H.}~\bibnamefont
  {Shirado}}, \bibinfo {author} {\bibfnamefont {G.}~\bibnamefont {Iosifidis}},
  \bibinfo {author} {\bibfnamefont {L.}~\bibnamefont {Tassiulas}},\ and\
  \bibinfo {author} {\bibfnamefont {N.~A.}\ \bibnamefont {Christakis}},\
  }\bibfield  {title} {\bibinfo {title} {Resource sharing in technologically
  defined social networks},\ }\href@noop {} {\bibfield  {journal} {\bibinfo
  {journal} {Nature Communications}\ }\textbf {\bibinfo {volume} {10}},\
  \bibinfo {pages} {1079} (\bibinfo {year} {2019})}\BibitemShut {NoStop}%
\bibitem [{\citenamefont {Jackson}(2019)}]{jackson2019friendship}%
  \BibitemOpen
  \bibfield  {author} {\bibinfo {author} {\bibfnamefont {M.~O.}\ \bibnamefont
  {Jackson}},\ }\bibfield  {title} {\bibinfo {title} {The friendship paradox
  and systematic biases in perceptions and social norms},\ }\href@noop {}
  {\bibfield  {journal} {\bibinfo  {journal} {Journal of Political Economy}\
  }\textbf {\bibinfo {volume} {127}},\ \bibinfo {pages} {777} (\bibinfo {year}
  {2019})}\BibitemShut {NoStop}%
\bibitem [{\citenamefont {Lagarias}\ \emph {et~al.}(1984)\citenamefont
  {Lagarias}, \citenamefont {Mazo}, \citenamefont {Shepp},\ and\ \citenamefont
  {McKay}}]{lagarias1984inequality}%
  \BibitemOpen
  \bibfield  {author} {\bibinfo {author} {\bibfnamefont {J.~C.}\ \bibnamefont
  {Lagarias}}, \bibinfo {author} {\bibfnamefont {J.~E.}\ \bibnamefont {Mazo}},
  \bibinfo {author} {\bibfnamefont {L.~A.}\ \bibnamefont {Shepp}},\ and\
  \bibinfo {author} {\bibfnamefont {B.}~\bibnamefont {McKay}},\ }\bibfield
  {title} {\bibinfo {title} {An inequality for walks in a graph},\ }\href@noop
  {} {\bibfield  {journal} {\bibinfo  {journal} {SIAM Review}\ }\textbf
  {\bibinfo {volume} {26}},\ \bibinfo {pages} {580} (\bibinfo {year}
  {1984})}\BibitemShut {NoStop}%
\bibitem [{\citenamefont {Harrigan}\ and\ \citenamefont
  {Yap}(2017)}]{harrigan2017avoidance}%
  \BibitemOpen
  \bibfield  {author} {\bibinfo {author} {\bibfnamefont {N.}~\bibnamefont
  {Harrigan}}\ and\ \bibinfo {author} {\bibfnamefont {J.}~\bibnamefont {Yap}},\
  }\bibfield  {title} {\bibinfo {title} {Avoidance in negative ties: Inhibiting
  closure, reciprocity, and homophily},\ }\href@noop {} {\bibfield  {journal}
  {\bibinfo  {journal} {Social Networks}\ }\textbf {\bibinfo {volume} {48}},\
  \bibinfo {pages} {126} (\bibinfo {year} {2017})}\BibitemShut {NoStop}%
\bibitem [{\citenamefont {Jackson}(2010)}]{jackson2010social}%
  \BibitemOpen
  \bibfield  {author} {\bibinfo {author} {\bibfnamefont {M.~O.}\ \bibnamefont
  {Jackson}},\ }\href@noop {} {\emph {\bibinfo {title} {Social and economic
  networks}}}\ (\bibinfo  {publisher} {Princeton university press},\ \bibinfo
  {year} {2010})\BibitemShut {NoStop}%
\bibitem [{\citenamefont {Feng}\ \emph {et~al.}(2022)\citenamefont {Feng},
  \citenamefont {Altmeyer}, \citenamefont {Stafford}, \citenamefont
  {Christakis},\ and\ \citenamefont {Zhou}}]{feng2022testing}%
  \BibitemOpen
  \bibfield  {author} {\bibinfo {author} {\bibfnamefont {D.}~\bibnamefont
  {Feng}}, \bibinfo {author} {\bibfnamefont {R.}~\bibnamefont {Altmeyer}},
  \bibinfo {author} {\bibfnamefont {D.}~\bibnamefont {Stafford}}, \bibinfo
  {author} {\bibfnamefont {N.~A.}\ \bibnamefont {Christakis}},\ and\ \bibinfo
  {author} {\bibfnamefont {H.~H.}\ \bibnamefont {Zhou}},\ }\bibfield  {title}
  {\bibinfo {title} {Testing for balance in social networks},\ }\href@noop {}
  {\bibfield  {journal} {\bibinfo  {journal} {Journal of the American
  Statistical Association}\ }\textbf {\bibinfo {volume} {117}},\ \bibinfo
  {pages} {156} (\bibinfo {year} {2022})}\BibitemShut {NoStop}%
\bibitem [{\citenamefont {Isakov}\ \emph {et~al.}(2019)\citenamefont {Isakov},
  \citenamefont {Fowler}, \citenamefont {Airoldi},\ and\ \citenamefont
  {Christakis}}]{isakov2019structure}%
  \BibitemOpen
  \bibfield  {author} {\bibinfo {author} {\bibfnamefont {A.}~\bibnamefont
  {Isakov}}, \bibinfo {author} {\bibfnamefont {J.~H.}\ \bibnamefont {Fowler}},
  \bibinfo {author} {\bibfnamefont {E.~M.}\ \bibnamefont {Airoldi}},\ and\
  \bibinfo {author} {\bibfnamefont {N.~A.}\ \bibnamefont {Christakis}},\
  }\bibfield  {title} {\bibinfo {title} {The structure of negative social ties
  in rural village networks},\ }\href@noop {} {\bibfield  {journal} {\bibinfo
  {journal} {Sociological science}\ }\textbf {\bibinfo {volume} {6}},\ \bibinfo
  {pages} {197} (\bibinfo {year} {2019})}\BibitemShut {NoStop}%
\bibitem [{\citenamefont {Newman}(2003)}]{newman2003mixing}%
  \BibitemOpen
  \bibfield  {author} {\bibinfo {author} {\bibfnamefont {M.~E.~J.}\
  \bibnamefont {Newman}},\ }\bibfield  {title} {\bibinfo {title} {Mixing
  patterns in networks},\ }\href@noop {} {\bibfield  {journal} {\bibinfo
  {journal} {Physical review E}\ }\textbf {\bibinfo {volume} {67}},\ \bibinfo
  {pages} {026126} (\bibinfo {year} {2003})}\BibitemShut {NoStop}%
\bibitem [{\citenamefont {Estrada}(2010)}]{estrada2010quantifying}%
  \BibitemOpen
  \bibfield  {author} {\bibinfo {author} {\bibfnamefont {E.}~\bibnamefont
  {Estrada}},\ }\bibfield  {title} {\bibinfo {title} {Quantifying network
  heterogeneity},\ }\href@noop {} {\bibfield  {journal} {\bibinfo  {journal}
  {Physical Review E}\ }\textbf {\bibinfo {volume} {82}},\ \bibinfo {pages}
  {066102} (\bibinfo {year} {2010})}\BibitemShut {NoStop}%
\bibitem [{\citenamefont {Jacob}\ \emph {et~al.}(2017)\citenamefont {Jacob},
  \citenamefont {Harikrishnan}, \citenamefont {Misra},\ and\ \citenamefont
  {Ambika}}]{jacob2017measure}%
  \BibitemOpen
  \bibfield  {author} {\bibinfo {author} {\bibfnamefont {R.}~\bibnamefont
  {Jacob}}, \bibinfo {author} {\bibfnamefont {K.}~\bibnamefont {Harikrishnan}},
  \bibinfo {author} {\bibfnamefont {R.}~\bibnamefont {Misra}},\ and\ \bibinfo
  {author} {\bibfnamefont {G.}~\bibnamefont {Ambika}},\ }\bibfield  {title}
  {\bibinfo {title} {Measure for degree heterogeneity in complex networks and
  its application to recurrence network analysis},\ }\href@noop {} {\bibfield
  {journal} {\bibinfo  {journal} {Royal Society open science}\ }\textbf
  {\bibinfo {volume} {4}},\ \bibinfo {pages} {160757} (\bibinfo {year}
  {2017})}\BibitemShut {NoStop}%
\bibitem [{\citenamefont {Freeman}(2002)}]{freeman2002centrality}%
  \BibitemOpen
  \bibfield  {author} {\bibinfo {author} {\bibfnamefont {L.~C.}\ \bibnamefont
  {Freeman}},\ }\bibfield  {title} {\bibinfo {title} {Centrality in social
  networks: Conceptual clarification},\ }\href@noop {} {\bibfield  {journal}
  {\bibinfo  {journal} {Social network: critical concepts in sociology.
  Londres: Routledge}\ }\textbf {\bibinfo {volume} {1}},\ \bibinfo {pages}
  {238} (\bibinfo {year} {2002})}\BibitemShut {NoStop}%
\bibitem [{\citenamefont {Lerman}\ \emph {et~al.}(2016)\citenamefont {Lerman},
  \citenamefont {Yan},\ and\ \citenamefont {Wu}}]{lerman2016majority}%
  \BibitemOpen
  \bibfield  {author} {\bibinfo {author} {\bibfnamefont {K.}~\bibnamefont
  {Lerman}}, \bibinfo {author} {\bibfnamefont {X.}~\bibnamefont {Yan}},\ and\
  \bibinfo {author} {\bibfnamefont {X.-Z.}\ \bibnamefont {Wu}},\ }\bibfield
  {title} {\bibinfo {title} {The" majority illusion" in social networks},\
  }\href@noop {} {\bibfield  {journal} {\bibinfo  {journal} {PloS one}\
  }\textbf {\bibinfo {volume} {11}},\ \bibinfo {pages} {e0147617} (\bibinfo
  {year} {2016})}\BibitemShut {NoStop}%
\bibitem [{\citenamefont {Evtushenko}\ and\ \citenamefont
  {Kleinberg}(2021)}]{evtushenko2021paradox}%
  \BibitemOpen
  \bibfield  {author} {\bibinfo {author} {\bibfnamefont {A.}~\bibnamefont
  {Evtushenko}}\ and\ \bibinfo {author} {\bibfnamefont {J.}~\bibnamefont
  {Kleinberg}},\ }\bibfield  {title} {\bibinfo {title} {The paradox of
  second-order homophily in networks},\ }\href@noop {} {\bibfield  {journal}
  {\bibinfo  {journal} {Scientific Reports}\ }\textbf {\bibinfo {volume}
  {11}},\ \bibinfo {pages} {1} (\bibinfo {year} {2021})}\BibitemShut {NoStop}%
\bibitem [{\citenamefont {Tourangeau}\ \emph {et~al.}(2000)\citenamefont
  {Tourangeau}, \citenamefont {Rips},\ and\ \citenamefont
  {Rasinski}}]{tourangeau2000psychology}%
  \BibitemOpen
  \bibfield  {author} {\bibinfo {author} {\bibfnamefont {R.}~\bibnamefont
  {Tourangeau}}, \bibinfo {author} {\bibfnamefont {L.~J.}\ \bibnamefont
  {Rips}},\ and\ \bibinfo {author} {\bibfnamefont {K.}~\bibnamefont
  {Rasinski}},\ }\href@noop {} {\emph {\bibinfo {title} {The psychology of
  survey response}}}\ (\bibinfo  {publisher} {Cambridge University Press},\
  \bibinfo {year} {2000})\BibitemShut {NoStop}%
\bibitem [{\citenamefont {Shoemaker}\ \emph {et~al.}(2002)\citenamefont
  {Shoemaker}, \citenamefont {Eichholz},\ and\ \citenamefont
  {Skewes}}]{shoemaker2002item}%
  \BibitemOpen
  \bibfield  {author} {\bibinfo {author} {\bibfnamefont {P.~J.}\ \bibnamefont
  {Shoemaker}}, \bibinfo {author} {\bibfnamefont {M.}~\bibnamefont
  {Eichholz}},\ and\ \bibinfo {author} {\bibfnamefont {E.~A.}\ \bibnamefont
  {Skewes}},\ }\bibfield  {title} {\bibinfo {title} {Item nonresponse:
  Distinguishing between don't know and refuse},\ }\href@noop {} {\bibfield
  {journal} {\bibinfo  {journal} {International Journal of Public Opinion
  Research}\ }\textbf {\bibinfo {volume} {14}},\ \bibinfo {pages} {193}
  (\bibinfo {year} {2002})}\BibitemShut {NoStop}%
\bibitem [{\citenamefont {Latora}\ \emph {et~al.}(2017)\citenamefont {Latora},
  \citenamefont {Nicosia},\ and\ \citenamefont {Russo}}]{latora2017complex}%
  \BibitemOpen
  \bibfield  {author} {\bibinfo {author} {\bibfnamefont {V.}~\bibnamefont
  {Latora}}, \bibinfo {author} {\bibfnamefont {V.}~\bibnamefont {Nicosia}},\
  and\ \bibinfo {author} {\bibfnamefont {G.}~\bibnamefont {Russo}},\
  }\href@noop {} {\emph {\bibinfo {title} {Complex networks: principles,
  methods and applications}}}\ (\bibinfo  {publisher} {Cambridge University
  Press},\ \bibinfo {year} {2017})\BibitemShut {NoStop}%
\bibitem [{Note1()}]{Note1}%
  \BibitemOpen
  \bibinfo {note} {In this notation, the $in-w$ and $out-w$ indicate the
  direction of one's neighbor, as one's in-neighbor and out-neighbor,
  respectively. The $(in)$ and $(out)$ denote the type of comparison as
  in-degree or out-degree.}\BibitemShut {Stop}%
\bibitem [{Note2()}]{Note2}%
  \BibitemOpen
  \bibinfo {note} {In Eq. S29 and other equations regarding the generalized
  inversity for the mixed world of $-w$, $n$ denotes the size of nodes with
  non-zero negative degrees. For the mixed world of $+w$, $n$ denotes the size
  of nodes with non-zero positive degrees.}\BibitemShut {Stop}%
\end{thebibliography}

\end{document}